\documentclass[fleqn,usenatbib,useAMS]{mnras}

\usepackage{multirow}
\usepackage{makecell}

\usepackage{graphicx}	% Including figure files
\usepackage{amsmath}	% Advanced maths commands
\usepackage{amssymb}	% Extra maths symbols
\usepackage{multicol}        % Multi-column entries in tables
\usepackage{bm}		% Bold maths symbols, including upright Greek
\usepackage{pdflscape}	% Landscape pages
\usepackage{comment}

\usepackage[normalem]{ulem}
\newcommand{\nzradd}[1]{{#1}}
\newcommand{\nzrremove}[1]{}
\newcommand{\nzraddone}[1]{{#1}}
\newcommand{\nzrremoveone}[1]{}

\usepackage[T1]{fontenc}
\usepackage{ae,aecompl}

\usepackage{newtxtext,newtxmath}
\title[Strong magnetogravity waves]{Gravity waves in strong magnetic fields}

\author[N. Z. Rui and J. Fuller]{
Nicholas Z. Rui,$^{1}$\thanks{E-mail: nrui@caltech.edu}
Jim Fuller$^{1}$
\\
$^{1}$TAPIR, California Institute of Technology, Pasadena, CA 91125, USA}

\date{Last updated XXX; in original form YYY}

\pubyear{2020}

\begin{document}
\label{firstpage}
\pagerange{\pageref{firstpage}--\pageref{lastpage}}
\maketitle

\begin{abstract}
Strong magnetic fields in the cores of stars are expected to significantly modify the behavior of gravity waves\nzrremove{, which}\nzradd{: this} is likely the origin of suppressed dipole modes observed in many red giants.
However, a detailed understanding of how such fields alter the spectrum and spatial structure of magnetogravity waves has been elusive.
%, especially for non-axisymmetric pulsations.
For a dipole field, we analytically characterize the horizontal eigenfunctions of magnetogravity modes, assuming that the wavevector is primarily radial.
%For a dipole field, we apply a radial WKB approximation to analytically characterize the horizontal eigenfunctions of magnetogravity modes.
%for general values of their horizontal wavenumbers $\ell$ and $m$ in the formal absence of dissipation.
For axisymmetric modes ($m=0$), the magnetogravity wave eigenfunctions become Hough functions, and they have a radial turning point for sufficiently strong magnetic fields.
For non-axisymmetric modes ($m\neq0$), the interaction between the discrete $g$ mode spectrum and a continuum of Alfv\'en \nzrremoveone{resonances}\nzraddone{waves} produces nearly discontinuous features in the fluid displacements at critical latitudes associated with a singularity in the fluid equations.
%Hence, even vanishingly small amounts of dissipation can greatly affect the mode structure, as confirmed by our numerical calculations including dissipative effects.
We find that magnetogravity modes cannot propagate in regions with sufficiently strong magnetic fields, instead becoming evanescent.
When encountering strong magnetic fields, ingoing gravity waves are likely refracted into outgoing \nzrremoveone{Alfv\'en-like}\nzraddone{slow magnetic} waves.
These outgoing waves approach infinite radial wavenumbers, which are likely to be damped efficiently.
However, it may be possible for a small fraction of the wave power to escape the stellar core as pure Alfv\'en waves or magnetogravity waves confined to a very narrow equatorial band.
%It is possible that some wave energy may escape back to the surface through a very narrow equatorial band.
%When encountering strong magnetic fields, ingoing gravity waves are likely converted into outgoing Alfv\'en-like waves which are damped in the stellar interior due to their large radial wavenumbers. %and efficient coupling to the Alfv\'en continuum.
The artificially sharp features in the WKB-separated solutions suggest the need for global mode solutions which include small terms neglected in our analysis.
%We discuss how non-WKB and/or dissipative terms will smooth the wavefunctions near the critical latitudes.

\end{abstract}

%we present numerical calculations of these horizontal eigenfunctions in the presence of dissipation.
%even under a radial magnetic field with realistic spatial dependence,
%When encountering strong magnetic fields, ingoing gravity waves are likely converted into outgoing magnetosonic waves that obtain very large radial wave numbers and damp within the star's interior, in agreement with prior work.


%%%%%%%%%%%%%%%%%%%%%%%%%%%%%%%%%%%%%%%%%%%%%%%%%%

%%%%%%%%%%%%%%%%% BODY OF PAPER %%%%%%%%%%%%%%%%%%

%\tableofcontents

\section{Introduction}

Stellar magnetism is a highly impactful, but often neglected, property of many main sequence stars \citep{ferrario2009origin,vidotto2014stellar}, red giants \nzraddone{\citep{garcia2014study,stello2016suppression,stello2016prevalence,fuller2015asteroseismology}}, white dwarfs \citep{angel1977magnetism,wickramasinghe2000magnetism,liebert2003true}, and neutron stars \citep{thompson1993neutron,kulkarni1998star,levin2006qpos} alike.
In stars, such magnetic fields are expected to originate from dynamo mechanisms \citep{baliunas1996dynamo,spruit2002dynamo,maeder2005stellar,brun2017magnetism}, as fossils leftover from the star's formation \citep{braithwaite2004fossil,dudorov2015theory,ferrario2015magnetic}, or from stellar mergers \citep{ferrario2009origin,tutukov2010possible,wickramasinghe2014most,schneider2019stellar}.
Despite the importance and ubiquity of strong stellar magnetism, our understanding of oscillations in such highly magnetized stars remains incomplete, even at the qualitative level.

Interest in the influence of magnetic fields on nonradial stellar oscillations has been reignited in the past few years by the discovery of suppressed dipole ($\ell=1$) and quadrupole ($\ell=2$) oscillation modes in a family of red giants \nzraddone{\citep{mosser2012characterization,garcia2014study,stello2016suppression,stello2016prevalence,mosser2017dipole}}.
It is largely believed that the origin of this phenomenon is magnetic in nature, with recent work suggesting that ingoing gravity waves can damp out after either being trapped inside the core \citep[the ``magnetic greenhouse effect,''][]{fuller2015asteroseismology}, refracted into high-wavenumber oscillations \citep{lecoanet2016conversion}, or dissipated by Alfv\'en waves \citep{loi2017torsional}.
\nzrremoveone{At the same time}\nzraddone{In parallel}, \citet{gang2022topology} have made the first-ever constraints on the interior magnetic field \textit{topology}---the recent development of such new powerful observational tools further demands proportionate advances in our theoretical understanding of internal magnetogravity waves.

Efforts to understand the impact of magnetic fields on stellar oscillation modes have taken many forms, but have been limited due to the difficulty of the problem.
For example, early attempts to understand magnetic effects on non-radial oscillations involved introducing a magnetic field as a small perturbation \citep[e.g.,][]{goossens1972perturbation,goossens1976stellar,goossens1976non,mathis2021probing}.
Some of these perturbative calculations have promisingly suggested that core magnetic fields may leave imprints on the mixed-mode period spacing \citep{prat2019period,prat2020period,bugnet2021magnetic,bugnet2022magnetic}, in addition to their impact on dipole mode visibilities.
%---however, because mode splittings are second-order in the field,
However, magnetic mode splittings are often small except for fields large enough to strongly couple with Alfv\'en waves, where a perturbative treatment is largely inappropriate \citep{cantiello2016asteroseismic}.
While other analyses have assumed a purely horizontal field \citep{rogers2010interaction,mathis2011low,macgregor2011reflection,dhouib2022detecting}, such studies are not applicable to the general case where the radial component of the field dominates the interaction with the gravity waves.

%\citep[see, e.g.,][]{fuller2015asteroseismology}. The first study emphasizing the dominance of the radial component of the magnetic field over interactions with gravity waves was presented by 
\citet{fuller2015asteroseismology} used a Wentzel--Kramers--Brillouin (WKB) approximation in both components of the wavenumber to show that magnetogravity waves are forced to be evanescent when the mode frequency lies below a characteristic frequency given by
\begin{equation} \label{omegamg}
    \omega_B = \left(\frac{\ell(\ell+1)B_0^2N^2}{\pi\rho_0r^2}\right)^{1/4} \, .
\end{equation}

\noindent where $\ell$, $B_0$, $N$, $\rho_0$, and $r$ are the angular degree, radial magnetic field, Brunt--V\"ais\"al\"a frequency, density, and radius, respectively.
This result can also be recovered exactly when considering the coupling of gravity waves to an exactly uniform radial field geometry (see Section \ref{monopole}).
However, while setting a useful scale for strong coupling between gravity waves and the magnetic field, this analysis relies on the assumption that the radial magnetic field is uniform at a given radius (which is not physical).

Other studies have probed the behavior of magnetogravity waves under arbitrarily complicated magnetic field geometries using a flexible ray-tracing method \citep{loi2018effects,loi2020magneto,loi2020effect}.
However, crucially, this method relies heavily upon the (WKB) approximation that both the radial \textit{and} horizontal components of the wavenumber are large compared to the variation scales of the magnetic field and stellar structure.
In reality, the horizontal wavenumber  $k_h=\sqrt{\ell(\ell+1)}/r$ of the observable $\ell \! \lesssim \! 3$ modes likely has a comparable length scale to that of the magnetic field \nzraddone{gradient}.
It is clear that a fuller understanding of magnetogravity waves must account for a magnetic field which is allowed to vary with latitude and longitude, without assuming an unrealistically large horizontal wavenumber.

Some progress on this front was made by \citet{lecoanet2016conversion}, who solve for the eigenmodes of a two-dimensional Cartesian analogue of a multipole magnetic field geometry, demonstrating that modes in their model cannot propagate in regions whose magnetic field exceeds a critical strength (see Section \ref{lecoanet}) close to the estimate of Equation \ref{omegamg}.
However, since their analysis cannot capture modes which propagate horizontally relative to the field (i.e., non-axisymmetric modes), the possibility is left open that such non-axisymmetric modes may propagate deeper into a star.
Later, \citet{lecoanet2022asteroseismic} extended this analysis numerically to more general tesseral/sectoral ($m\neq0$) modes using the \textsc{dedalus} code in order to probe the interior field of a main sequence B-type star HD 43317.
However, explanations for many qualitative properties of the solution have heretofore remained elusive.

%\textcolor{red}{INSERT SUMMARY HERE - state the goal of the paper, and then list the sections one by one and say what's in those.}
%\textcolor{red}{after the rest of the paper is done: state the goal of the paper, and then list the sections one by one and describe them}
%In this work, we determine the behavior of the oscillation modes of a star (i.e., self-gravitating plasma sphere) which is stably stratified (i.e., radiative) and magnetized with a field which has a dipole angular dependence.
%\textcolor{red}{claim that this behavior is probably generic, mention other geometry}
%Under a Wentzel–Kramers–Brillouin (WKB) approximation in the radial direction only, we demonstrate that (1) finite eigenfunctions with $m\neq0$ become strictly equatorially confined to a band with width $\Delta(\cos\theta)=2\omega/\omega_A$, (2) there are no propagating waves above a certain critical field (both when $m=0$ and $m\neq0$), and (3) at large fields, the $m\neq0$ modes only become a function of $\ell-|m|$, rather than $\ell$ and $m$ separately.\textcolor{red}{look this over again}

\begin{figure}
    \centering
    \includegraphics[width=0.47\textwidth]{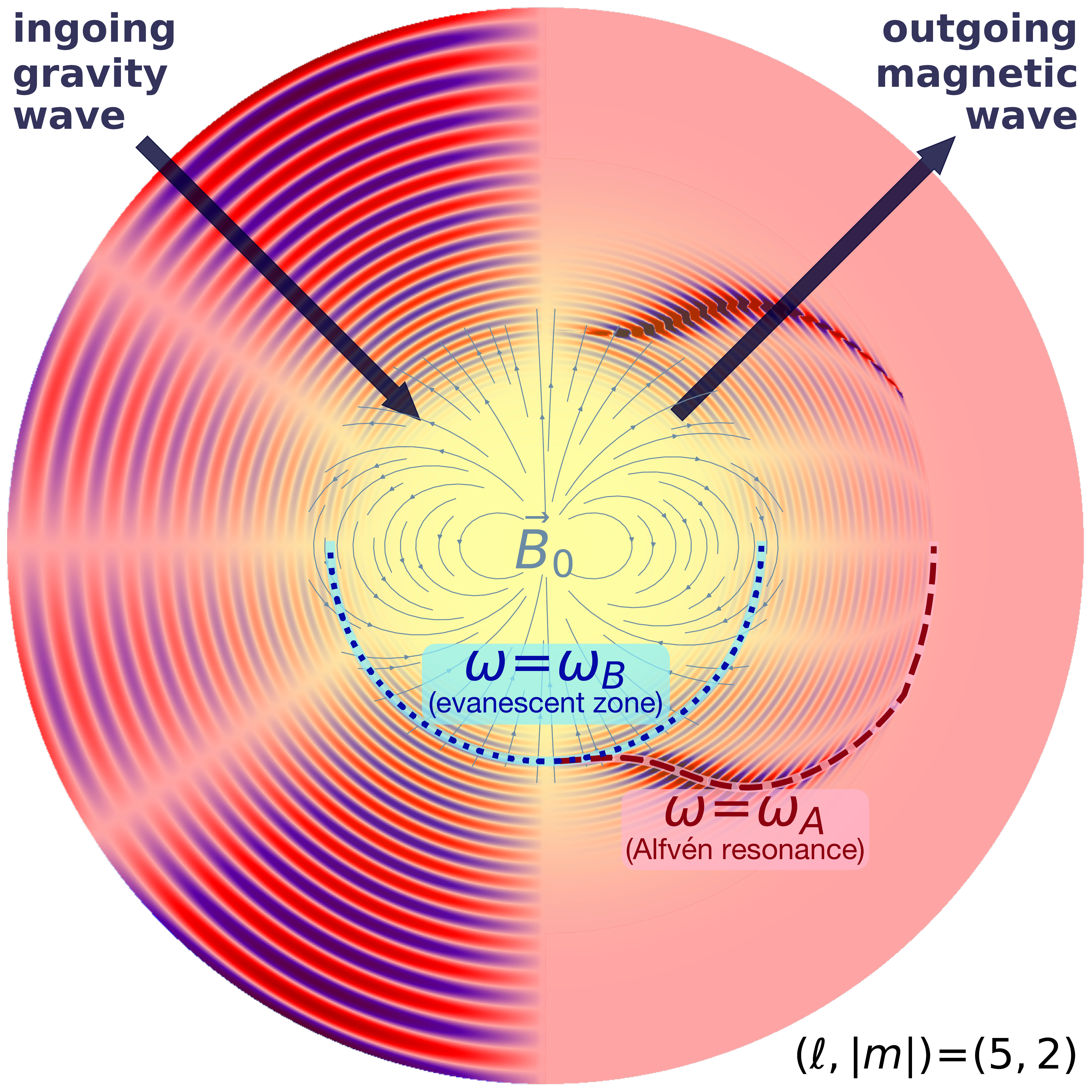}
    \caption{A meridional slice of the $\theta$ displacement $\xi_\theta$ for a magnetogravity mode with $(\ell,|m|)=(5,2)$, with the left half showing an ingoing gravity wave and the right half showing an outgoing \nzraddone{(slow) magnetic wave} (calculated in Section \ref{numerical2}), which approaches an infinite wavenumber at a cutoff radius (where it is dissipated).
    The eigenfunctions become large at the Alfven resonance $\omega=\omega_A$ ($b\cos\theta=\pm1$; red dashed line), and they become evanescent past the turning point $\omega=\omega_B$ ($a\sim1$; blue dotted line), where the solutions are given in Section \ref{numerical1}.
    This diagram is not to scale, as the stellar profile has been modified to better show the spatial structure of the magnetic wave.}
    \label{fig:thirdbranch_cartoon}
\end{figure}

% We use a WKB approximation in the radial direction but not in the latitudinal direction, where we numerically solve for the magnetogravity mode eigenfunctions.
%first characterizing their behavior under the assumption of formally zero dissipation and then numerically calculating them under a few cases of interest.
In this work, we analyze the horizontal structure of $g$ modes under a strong magnetic field.
We assume that the wavevector is primarily radial, and the radial wavelengths of the perturbations are much smaller than the stellar structure length scale (the radial WKB approximation), and numerically solve for the magnetogravity mode eigenfunctions.
We find that such $g$ modes contain sharp features in the fluid displacements at the locations of resonances with Alfv\'en waves (so-called ``critical latitudes''), and that the general structure of their branches and eigenfunctions are sensitive to even vanishingly small amounts of dissipation.
We also discuss the importance of the horizontal component of the field near these critical latitudes, as well as near the equator.
%We also discuss how the inclusion of additional non-WKB terms can smooth the eigenfunction near the critical latitude.
Nevertheless, %even when including oscillations which propagate in a direction perpendicular both to the magnetic field and its gradient (an extension of previous efforts),
we still find that $g$ modes cannot propagate arbitrarily deep in sufficiently magnetized stars, and are likely converted into outgoing slow \nzraddone{magnetic}\nzrremoveone{(Alfv\'en-like)} waves that dissipate inside of the star.
An outline of the solution described in this work is shown in Figure \ref{fig:thirdbranch_cartoon}.

We organize this paper as follows.
In Section \ref{problemstatement}, we describe the problem setup: a stably stratified, magnetized star obeying the %magneto-Boussinesq equations 
incompressible MHD equations
(Section \ref{lfe}), whose essential physics are governed by the relationship between the mode, Alfv\'en, and magnetogravity frequencies $\omega$, $\omega_A$, and $\omega_B$ (Section \ref{nondim}). For the majority of this work, we specialize to a dipole magnetic field (Section \ref{dipole}).
In the WKB limit, the resulting differential eigenproblem contains singularities at critical latitudes corresponding to resonances with the Alfv\'en spectrum.
%---we
%We therefore discuss in Section \ref{three} a number of a priori considerations of such problems,
We point out a close analogy with the rotational problem (Section \ref{tar}), then comment on previous work on internally singular eigenproblems (Section \ref{sl}), and lastly investigate the behavior of eigenfunctions around those critical latitudes (Section \ref{frob}).
In Section \ref{sols}, we present zonal ($m=0$; Section \ref{mequalszero}) and sectoral/tesseral ($m\neq0$; Section \ref{mnotequalszero}) solutions to the problem.
%where dissipation has been formally set to zero---
We then comment on the origin and behavior of the continuous Alfv\'en wave spectrum (Section \ref{alfven}).
However, since vanishingly small dissipation can qualitatively affect the mode spectrum, we present numerical solutions of dissipative solutions in Section \ref{viscous}, first allowing for evanescent solutions (Section \ref{numerical1}) and then constraining the \nzrremoveone{Alfv\'en frequency to be real}\nzraddone{radial phase velocity} (Section \ref{numerical2}).
Finally, in Section \ref{fr}, we discuss the importance of horizontal field terms near the critical latitudes and equator (Section \ref{wkb}), nonharmonic solutions of singular differential equations (Section \ref{nonharmonic}), the effects of more general magnetic field geometries (Section \ref{general}),
%describe the effect of a magnetic tangling shear modulus (Section \ref{tangle}),
and the possibility of magnetically stabilized $g$ modes in convective zones (Section \ref{g}).
Section \ref{conc} concludes.

\section{Problem statement} \label{problemstatement}

In this work, we consider a spherically symmetric star in hydrostatic equilibrium, with a possibly large equilibrium magnetic field (which is not spherically symmetric).
It is assumed that the magnetic field does not act on the background structure, i.e., it is not strong enough to introduce substantial departures from a spherically symmetric stellar profile.
For simplicity, we ignore rotation and use the incompressible and Cowling approximations, such that buoyancy and magnetic forces dominate the dynamics.
%assume that the only restorative forces are buoyancy and magnetism (e.g., there are no pressure, shear, or inertial waves).
These forces are likely to dominate in, e.g., the slowly rotating radiative cores of red giants.
%However, while we use the terminology of ``star'' to describe the problem, the model is fairly generic (under the parameters outlined above).

\nzraddone{Throughout this work, we use the term ``magnetogravity wave'' to refer to the general phenomenon of a gravity wave propagating through a highly conductive, magnetized fluid.
In sufficiently magnetized stars, ingoing magnetogravity waves are refracted outwards, and (as we will show in Sections \ref{mequalszero}, \ref{mnotequalszero}, and \ref{numerical2}) approach infinite radial wavenumber at a finite height---we refer to such waves as ``slow magnetic waves.''
Such branches are ``slow'' in the sense that their phase and group velocities approach zero as waves propagate outwards.
This medium also sustains ``Alfv\'en waves,'' which are confined to
%(and quantized on)
magnetic field lines and appear
%in the dissipationless problem (Section \ref{mnotequalszero})
as highly localized, linearly independent toroidal solutions to the fluid equations (see Section \ref{alfven}).}
%, and they are generally smaller than the magnetogravity wave branch.

%We will demonstrate (Sections \ref{mnotequalszero}, \ref{numerical1}, and \ref{numerical2}) that, once a magnetogravity wave's frequency becomes somewhere resonant with the Alfv\'en frequency, even a small amount of dissipation will enforce that the magnetogravity waves appear in a particular linear combination with the Alfv\'en waves. We therefore term such waves ``Alfv\'enic waves,'' and usually ``slow Alfv\'enic waves'' since this behavior occurs soon before divergence to infinite wavenumber.

In this Section, we first introduce the linearized fluid equations (Section \ref{lfe}).
We then identify the most important dimensionless parameters governing the physics (Section \ref{nondim}).
Finally, we specialize to the case of a dipole magnetic field (Section \ref{dipole}), to which the majority of this work is dedicated.

\subsection{Linearized fluid equations} \label{lfe}

The linearized
%magneto-Boussinesq
incompressible MHD equations are
\begin{subequations} \label{magnetoboussinesq}
    \begin{gather}
        \nabla\cdot\vec{\xi} = 0 \label{mb_xi} \\
        \rho_0\partial_t^2\vec{\xi} = -\nabla\left(p' + \frac{1}{4\pi}\vec{B}_0\cdot\vec{B}'\right) - \rho'g\hat{r} + \frac{1}{4\pi}\left(\vec{B}_0\cdot\nabla\right)\vec{B}' \label{mb_mtm} \\
        \rho' = \frac{\rho_0N^2}{g}\xi_r \label{mb_rho} \\
        \vec{B}' = \left(\vec{B}_0\cdot\nabla\right)\vec{\xi} \label{mb_b}
    \end{gather}
\end{subequations}

\noindent where $\vec{\xi}$ is the perturbed fluid displacement, while $\rho'$, $p'$, and $\vec{B}'$ are the Eulerian density, pressure, and magnetic field perturbations, and $0$ subscripts indicate non-perturbed quantities \citep{proctor1982magnetoconvection}. Here, $N$ is the Brunt--V\"ais\"al\"a frequency, and $g=g(r)$ is the inward gravitational acceleration.
Here, we have assumed the WKB approximation in the radial direction \textit{only}, and have made the Cowling approximation ($g'\approx0$).
Additionally, as implied by Equation \ref{mb_rho}, we only consider adiabatic oscillations.
Equation \ref{mb_b} is simply the induction equation in ideal magnetohydrodynamics, written in the WKB limit (for $\vec{B}_0$ varying radially on a length scale $\sim r$).
Throughout this paper, we will focus on solving for oscillation modes with harmonic time dependence, i.e., those with $\propto e^{i\omega t}$ (although this assumption is discussed in Section \ref{nonharmonic}).

Describing an incompressible fluid under ideal magnetohydrodynamics, these equations admit modes which are restored by buoyancy and magnetism (i.e., there are no acoustic waves).
Gravity waves are expected to have large radial wavenumbers which are much larger than both their horizontal wavenumbers ($k_r/k_h\sim N/\omega\sim10^2$ in typical red giant cores)
\nzraddone{and the star's structural variation scale $1/H$. However, the horizontal wavenumber $k_h\simeq\sqrt{\ell(\ell+1)}/r$, so low-$\ell$ magnetogravity modes
vary horizontally on similar length scales to large-scale magnetic fields.}
%to the stellar structure as long as $\ell\simeq1$ (as is true of all currently observable oscillation modes in red giants).}
Therefore, we have adopted a WKB approximation in the radial direction only (i.e., $\partial/\partial r \approx -i k_r$) such that $k_r$ is assumed to be larger than any \nzrremoveone{horizontal}\nzraddone{structural} gradients.

We define the Alfv\'en frequency $\omega_A = \vec{k}\cdot\vec{v}_A$, where $\vec{v}_A = \vec{B}_0/\sqrt{4 \pi \rho_0}$ is the Alfv\'en \nzrremoveone{speed}\nzraddone{velocity}.
Then \nzrremoveone{a WKB approximation}\nzraddone{the assumption that $k_r\gg k_h$} entails \nzraddone{that} $\omega_A\propto\left(\vec{B}\cdot\vec{k}\right)=B_rk_r+B_hk_h \simeq B_r k_r$, such that the horizontal component of the magnetic field is unimportant as long as $B_r$ and $B_h$ are comparable.
This approximation is made by \citet{fuller2015asteroseismology}, and is very analogous to the ``traditional approximation of rotation'' (see Section \ref{tar}).
We discuss the importance of $B_h$ terms in Section \ref{wkb}.% and Appendix \ref{nonWKB}.

When a WKB approximation is made in both the vertical and horizontal directions (or if a monopolar field is considered; Appendix \ref{toys}), the dispersion relation is given by
%\begin{equation} \label{cartesian}
%    \omega^2 - \frac{k_h^2}{k_r^2}N^2 - k_r^2v_A^2 = 0
%\end{equation}
\begin{equation} \label{cartesian}
    \omega^2 - \frac{k_h^2}{k_r^2}N^2 - k_r^2v_A^2 = 0
\end{equation}
%\noindent where $v_A$ is the Alfv\'en speed \citep{unno1989nonradial}.
%where $\omega_{A,0} = |v_{A,r}|/r$ is an Alfv\'en travel frequency.
\noindent where \nzrremoveone{$v_A$}\nzraddone{$v_A=|v_{A,r}|$} is the \nzraddone{radial component of the Alfv\'en velocity} \citep{unno1989nonradial}.
If both buoyancy and magnetism are important, all three terms in Equation \ref{cartesian} are of the same order.
This defines a hierarchy of variables: letting $\epsilon$ be a small quantity around which we implicitly expand, we see that, if $\omega,k_h\sim\mathcal{O}(1)$, then $N,k_r\sim\mathcal{O}(\epsilon^{-1})$ are ``large'' and $v_A\sim\mathcal{O}(\epsilon)$ is ``small.''
Hereafter, we only retain terms leading-order in $\epsilon$, which is realistic as long as $k_r\gg k_h$.

\subsection{Important frequency scales} \label{nondim}

To understand the nature of this magnetogravity problem, we can non-dimensionalize the relevant physics equations.
All formulations of the magnetogravity problem (see, e.g., Appendix \ref{toys}) that make similar assumptions to ours can be formulated as the following \nzraddone{horizontal} eigenproblem \nzraddone{at a given radius (see Section \ref{general})}:
\begin{equation} \label{genev}
    \mathcal{L}^{k_rv_A/\omega}p' + \left(\frac{\omega^2}{N^2}r^2k_r^2\right)p' = 0
\end{equation}

\noindent where $\mathcal{L}^{k_rv_A/\omega}$ is some geometry-dependent differential operator that depends on the ratio of the Alfv\'en frequency $\omega_A\sim k_rv_A$ to the mode frequency $\omega$.
\nzraddone{In Equation \ref{genev}, $v_A$ is a measure of the the Alfv\'en velocity at a given radius. Although the magnetic field strength clearly varies as a function of $\theta$ and $\phi$, hereafter we use $v_A$ to denote its \textit{maximum} value at a given radius.}

The Buckingham $\pi$ theorem \citep{vaschy1892lois,federman1911some,riabouchinsky1911methode,buckingham1914physically} states that, for some equations depending on $p$ dimensionful quantities in $q$ independent dimensions, those equations can be written in terms of $p-q$ dimensionless quantities which completely determine their behavior.
In this particular problem, Equation \ref{genev} depends on the $p=5$ dimensionful quantities $\omega$, $N$, $k_r$, and $r$, and $v_A$ over the $q=2$ independent dimensions, length and time.
Therefore, the essential behavior of the magnetogravity problem can be understood by understanding the interaction of $p-q=3$ dimensionless quantities.

One natural dimensionless quantity to construct is $r k_r$, the radial wavenumber rescaled to the characteristic length scale of the star.
Fortuitously, because $rk_r\gg1$ according to the radial WKB approximation, the non-dimensionalized version of Equation \ref{genev} will not actually depend on this quantity.
Next, because $\mathcal{L}^{k_rv_A/\omega}$ depends only on the combination $k_rv_A/\omega$ (which describes the presence/location of resonances between modes and Alfv\'en waves), it is natural to choose this to be another dimensionless quantity:
\begin{equation} \label{b}
    b = \frac{k_rv_A}{\omega} \simeq \frac{\omega_A}{\omega}
\end{equation}

Finally, if one seeks to non-dimensionalize Equation \ref{genev} using a third quantity which does not depend on the spatial structure of the mode itself (i.e., independent of $k_r$), the remaining dimensionless quantity must depend solely on some ``depth parameter'' $a$, given by
\begin{equation} \label{a}
    a = \left(\frac{N}{\omega}\right)\left(\frac{v_A/r}{\omega}\right)
\end{equation}
We refer to $a$ as a depth parameter because $N$ and $v_A$ often increase with depth in stars such as red giants, so we expect $a$ to increase with depth.
\nzraddone{It is possible that $a$ could reach a maximum at some finite radius which would admit a weakly magnetized inner region. In practice,
%unless the maximum value of $a$ is very fine-tuned for a given frequency,
this inner region will be nearly decoupled from the rest of the star by an evanescent region and will be effectively unobservable, except for finely tuned frequencies.
In a red giant, $N^2$ peaks near the H-burning shell, where the value of $a$ will likely peak as well.}
%and, in order to couple to the inner region of the star, one requires that the peak $a$ barely graze the ``strong'' (non-perturbative) regime, i.e., $\omega\sim2\pi\nu_{\mathrm{max}}\simeq\omega_B$, where $\nu_{\mathrm{max}}$ is the frequency of maximum power \citep[e.g.,][]{chaplin2013asteroseismology}. Hence, we hereafter assume the generic case that the inner propagating region is decoupled from the surface, so that it can be ignored.

In the terminology of \citet{fuller2015asteroseismology}, $a\sim\omega_B^2/\omega^2$ where
\begin{equation}
   \omega_B\sim\sqrt{Nv_A/r}\label{omegab} 
\end{equation}
is the magnetogravity frequency, below which modes cannot be spatially propagating.
We thus argue that the frequency scale $\omega_B$ defining strong magnetogravity waves (identified by \citealt{fuller2015asteroseismology} under some specific assumptions) arises as the most natural mode-independent frequency scale in the problem.

Adopting $b$ and $a$ as our dimensionless parameters, Equation \ref{genev} can be rewritten as
\begin{equation}
    \mathcal{L}^bp' + \frac{b^2}{a^2}p' = 0
\end{equation}
For the hierarchy of variables adopted in Section \ref{lfe}, we see that both $a$ and $b$ are of order unity within the domain of interest, where magnetic forces and buoyancy forces are comparable.
Consequently, when non-dimensionalizing the fluid equations, specifying $a$ (which is independent of $k_r$) determines the spectrum of allowed $b$.
For a fixed mode frequency $\omega$, the resulting dispersion relation will therefore relate $\omega_B$ to the allowed $\omega_A\propto k_r$.

For the remainder of this, we will study the magnetogravity problem in terms of these two dimensionless quantities, which relate the mode ($\omega$), Alfv\'en ($\omega_A$), and magnetogravity ($\omega_B$) frequencies to each other.

%This formalizes the notion of some frequency scale $\sim\omega_B$ as determining qualitative changes in behavior of the spectrum of $k_r$ (e.g., determining when radially propagating solutions exist), a result noted by \citet{fuller2015asteroseismology} and shown to be applicable to more realistic geometries by \citet{lecoanet2016conversion}.

%We show in Appendix \ref{toys} that the geometries considered by \citet{fuller2015asteroseismology} and \citet{lecoanet2016conversion} can both be non-dimensionalized in a similar way, and can be posed as similar eigenproblems as considered in this work.

\subsection{Dipole geometry} \label{dipole}

We give special attention to the case of a magnetic field whose radial component is dipolar,
\begin{equation} \label{dipolegeo}
    \vec{B}_0 = B_0(r)\cos\theta\,\hat{r} + B_\theta(r,\theta,\phi)\,\hat{\theta} + B_\phi(r,\theta,\phi)\,\hat{\phi} \sim B_0(r)\cos\theta\,\hat{r}
\end{equation}
Because the wavenumbers of gravity waves are predominantly radial, the radial component of the field couples most efficiently to them \citep{fuller2015asteroseismology}, and the horizontal field components can be neglected at lowest order.
%\noindent where we have ignored the horizontal components (which, while not small in an absolute sense, couple negligibly to gravity waves).
This generic dipole angular dependence encompasses as special cases the force-free dipole ($B_0(r)\propto r^{-3}$) and uniform $B_0\hat{z}$ ($B_0(r)=\mathrm{const.}$) field geometries, as well as the mixed poloidal--toroidal field solution of \citet{prendergast1956equilibrium}.

For this special case\nzraddone{, and adopting a radial WKB approximation}, Equations \ref{magnetoboussinesq} can be written in spherical polar coordinates as
\begin{subequations} \label{subeq}
    \begin{gather}
        ik_r\xi_r + \frac{1}{r}\frac{\mathrm{d}}{\mathrm{d}\mu}\left(\xi_\theta\sqrt{1-\mu^2}\right) - \frac{im}{r\sqrt{1-\mu^2}}\xi_\phi = 0 \label{continuitydipole} \\
        \rho_0N^2\xi_r = ik_rp' \\
        \rho_0\omega^2\xi_\theta = -\frac{\sqrt{1-\mu^2}}{r}\frac{\mathrm{d}p'}{\mathrm{d}\mu} + \frac{1}{4\pi}k_r^2B_0^2\mu^2\xi_\theta \label{thetamtmdipole} \\
        \rho_0\omega^2\xi_\phi = \frac{im}{r\sqrt{1-\mu^2}}p' + \frac{1}{4\pi}k_r^2B_0^2\mu^2\xi_\phi
        \label{phidipole}
    \end{gather}
\end{subequations}

\noindent where we have substituted Equation \ref{mb_rho} into the radial component of Equation \ref{mb_xi}, Equation \ref{mb_b} into the horizontal components of Equation \ref{mb_xi}, and kept only leading-order terms.
Here, $\mu\equiv\cos\theta$, and the axisymmetry of this geometry entails eigenfunctions with $\partial/\partial\phi\rightarrow im$ for an integer $m$.

In terms of the pressure perturbation $p'$, the other perturbations become
\begin{subequations} \label{perts}
    \begin{gather}
        \xi_r = \frac{ik_r}{\rho_0N^2}p' \label{xir} = \frac{i}{\rho\omega^2r}\left(\frac{\omega}{N}\right)\frac{b}{a}p' \\
        \xi_\theta = -\frac{\sqrt{1-\mu^2}}{\rho_0\omega^2r\left(1-b^2\mu^2\right)}\frac{\mathrm{d}p'}{\mathrm{d}\mu} \label{xitheta} \\
        \xi_\phi = \frac{im}{\rho_0\omega^2r\sqrt{1-\mu^2}\left(1-b^2\mu^2\right)}p' \label{xiphi} \\
        \rho' = \frac{ik_r}{g}p'
    \end{gather}
\end{subequations}

\noindent \nzraddone{and $\vec{B}'=-iB_0 \mu k_r\vec{\xi}$.}

%\textcolor{red}{old equations below}
%\begin{equation} \label{firstorder2}
%    \begin{split}
        %\frac{\mathrm{d}\mathcal{P}}{\mathrm{d}\mu} &= -\frac{1-b^2\mu^2-if(a,b,\mu)}{1-\mu^2}\mathcal{Z} \\
        %\frac{\mathrm{d}\mathcal{Z}}{\mathrm{d}\mu} &= \left(\frac{b^2}{a^2} - \frac{m^2}{(1-\mu^2)(1-b^2\mu^2-if(a,b,\mu))}\right)\mathcal{P} \\
%    \end{split}
%\end{equation}

%where in the dipole geometry $v_A=B_0/\sqrt{4\pi\rho_0}$.
When Equations \ref{xir}, \ref{xitheta}, and \ref{xiphi} for the displacements are substituted into the continuity equation (Equation \ref{continuitydipole}), we obtain
\begin{equation} \label{mte}
    \mathcal{L}_{\mathrm{mag}}^{m,b}p'(\mu) + \frac{b^2}{a^2}p'(\mu) = 0
\end{equation}

\noindent where
\begin{equation} \label{mto}
     \mathcal{L}^{m,b}_{\mathrm{mag}}p'(\mu) = \frac{\mathrm{d}}{\mathrm{d}\mu}\left(\frac{1-\mu^2}{1-b^2\mu^2}\frac{\mathrm{d}p'(\mu)}{\mathrm{d}\mu}\right) - \frac{m^2}{\left(1-\mu^2\right)\left(1-b^2\mu^2\right)}p'(\mu)
\end{equation}

Equation \ref{mte} can be viewed as an eigenvalue equation for the unusual operator $\mathcal{L}^{m,b}_{\mathrm{mag}}$.
Letting $\lambda$ be the (conventionally negative) eigenvalues of $\mathcal{L}_{\mathrm{mag}}^{m,b}$, Equation \ref{mte} is
\begin{equation}
    \mathcal{L}_{\mathrm{mag}}^{m,b}p' + \lambda p' = 0
\end{equation}
\noindent with
\begin{equation} \label{dispersionrelation}
    \lambda = b^2/a^2
\end{equation}

\noindent constitutes the dispersion relation for magnetogravity waves.

In the limit of zero magnetic field, $\mathcal{L}^{m,b}_{\mathrm{mag}}$ approaches the usual generalized Legendre operator (whose eigenfunctions are associated Legendre polynomials).
\nzrremoveone{In the zero-field limit, w}\nzraddone{Here, w}hile $a$ and $b$ individually approach zero, the combination $\lambda=b^2/a^2$ ($=r^2k_r^2\omega^2/N^2$) approaches $\ell(\ell+1)$, matching the zero-field result that $k_h=\sqrt{\ell(\ell+1)}/r$).
In this case, Equation \ref{dispersionrelation} approaches the unusual internal gravity wave dispersion relation $\omega/N=k_h/k_r$.

In this work, we index mode branches using $\ell$ and $|m|$, corresponding to the angular degree and order of the branch at zero field (note that modes of $+m$ and $-m$ have identical spectra).
Hereafter, we refer to mode branches as an ordered pair $(\ell,|m|)$, e.g., the $(2,1)$ branch corresponds to the branch which, at zero field, has a horizontal dependence of a spherical harmonic with $\ell=2$ and $m=\pm1$.
However, note that the eigenvalue of $\mathcal{L}_{\mathrm{mag}}^{m,b}$ does not equal $\lambda=\ell(\ell+1)$ except precisely in the $b=0$ (zero-field) case, and the index $\ell$ is just used for indexing purposes.

\section{Important features of the magnetogravity eigenproblem} \label{three}

\subsection{Close analogy to the rotational problem} \label{tar}

In the study of nonradial pulsations under uniform rotation, it is common to consider only the influence of the Coriolis force, which dominates the rotational effect for small $\Omega$.
Specializing further to the case where $k_r\gg k_h$, it is common also to ignore the horizontal component of the rotational vector $\vec{\Omega}$, since the product $\vec{k}\cdot\vec{\Omega}=k_r\Omega_r+k_h\Omega_h\approx k_r\Omega_r$ will be dominated by the radial term \citep[see, e.g.,][]{lee1997low,chen2009calculation,wang2016computation}.
Under this approximation (the ``traditional approximation of rotation''), the radial and horizontal fluid equations become separable, and the following eigenproblem appears:
\begin{equation}
    \mathcal{L}_{\mathrm{rot}}^{m,\nu}p'(\mu) + \lambda p'(\mu) = 0
\end{equation}

\noindent where $\mathcal{L}_{\mathrm{rot}}^{m,\nu}$ (called the ``Laplace tidal operator'') is given by
\begin{equation} \label{lto}
    \begin{split}
        \mathcal{L}^{m,\nu}_{\mathrm{rot}}p'(\mu) = \frac{\mathrm{d}}{\mathrm{d}\mu}&\left(\frac{1-\mu^2}{1-\mu^2\nu^2}\frac{\mathrm{d}p'(\mu)}{\mathrm{d}\mu}\right) - \frac{m^2}{\left(1-\mu^2\right)\left(1-\mu^2\nu^2\right)}p'(\mu) \\
        &- \frac{m\nu\left(1+\mu^2\nu^2\right)}{\left(1-\mu^2\nu^2\right)^2}p'(\mu) 
    \end{split}
\end{equation}

\noindent where $\nu=2\Omega/\omega$ describes the influence of rotation.

Comparing $\mathcal{L}_{\mathrm{mag}}^{m,b}$ and $\mathcal{L}^{m,\nu}_{\mathrm{rot}}$ suggests a close analogy---the latter is identical to the former (with $\nu$ playing the role of $b$) except for the presence of an extra term (the second term in Equation \ref{lto}) which distinguishes prograde ($m\nu<0$) and retrograde ($m\nu>0$) modes \citep{lee1997low}.
Because a dipole magnetic field does not privilege either clockwise or counterclockwise-propagating oscillations, the symmetries of the problem do not permit this term to exist in the magnetogravity problem.

The eigenfunctions of $\mathcal{L}^{m,\nu}_{\mathrm{rot}}$ \nzradd{(whose eigenvalues we denote by $\lambda^\nu_{\ell m}$)} are called Hough functions \citep{hough1898application,hough1898v}, and \nzrremoveone{its}\nzraddone{their} properties have been widely studied, both analytically \citep{homer1990boundary,townsend2003asymptotic,townsend2020improved} and numerically \citep{bildsten1996ocean,lee1997low,chen2009calculation,fuller2014dynamical,wang2016computation}.
In Section \ref{mequalszero}, we show that the exact correspondence between $\mathcal{L}_{\mathrm{mag}}^{m,b}$ and $\mathcal{L}^{m,\nu}_{\mathrm{rot}}$ in the zonal ($m=0$) case allows us to identify Hough functions as eigensolutions of the magnetogravity problem.

We note that, for $|\nu|>1$, the coefficients in the Laplace tidal operator $\mathcal{L}_{\mathrm{rot}}^{m,\nu}$ (Equation \ref{lto}) switch \nzrremove{\nzraddone{relative} }signs on the domain, and Sturm--Liouville theory no longer guarantees that its eigenvalues are positive-definite (see Section \ref{sl}), and indeed there are an infinite number of \nzrremove{$\lambda^\nu_{\ell0}<0$}\nzradd{$\lambda^\nu_{\ell m}<0$} branches occupying the range $|\nu|>1$ which diverge to negative infinity as $|\nu|=1$ is approached \citep[e.g.,][]{lee1997low}.
In the rotation problem, these negative \nzrremove{$\lambda^\nu_{\ell0}$}\nzradd{$\lambda^\nu_{\ell m}$} branches correspond physically to \nzraddone{oscillatory convective modes (e.g., Section \ref{g}).
Notably, in the retrograde case for $|m|\neq0$, some of these branches of eigenvalues actually rise above $0$ and physically correspond to Rossby waves \citep{lee1997low}.}\nzrremoveone{Rossby waves.}
In the magnetogravity problem, these negative eigenvalue branches are not directly relevant in radiative regions (see Section \ref{mequalszero} for a discussion of this), although \nzrremove{it}\nzradd{their existence} may imply magnetically stabilized $g$ modes in convective regions (see Section \ref{g}).

In the general $m$ case (Section \ref{mnotequalszero}), $\mathcal{L}^{m,b}_{\mathrm{mag}}$ and $\mathcal{L}^{m,\nu}_{\mathrm{rot}}$ no longer coincide.
However, the Laplace tidal equation can at least provide some basic expectations about the behavior of the magnetogravity eigenfunctions, although \nzrremoveone{they}\nzraddone{the latter} are significantly more pathological.

\subsection{Sturm--Liouville problems with internal singularities} \label{sl}

The magnetogravity problem is dependent on the behavior of the eigenvalue problem stated in Equation \ref{mte}, which contains a differential operator whose coefficients have singularities on the interior of the domain\nzraddone{, at least, when $\omega$ and $k_r$ are real} (at $\mu=\pm1/b$).
To inform our \nzrremoveone{particular problem}\nzraddone{procedure}, we summarize in this Section the previous body of work on such Sturm--Liouville problems with internal singularities.

Consider the following general eigenvalue problem
\begin{equation} \label{selfadjointform}
    \left(P(x)y'(x)\right)' - Q(x)y(x) + \lambda y(x) \equiv \mathcal{L}y(x) + \lambda y(x) = 0
\end{equation}

\noindent where $P(x)$ and $Q(x)$ are real functions of $x$ on the open range $x\in(a,b)$, and primes denote derivatives in $x$.
If the value of $f(x)^*P(x)g'(x)$ matches at the endpoints $x=a$ and $x=b$ for any two functions $f(x)$ and $g(x)$ satisfying some boundary conditions, then the operator $\mathcal{L}$ is Hermitian with respect to the inner product
\begin{equation}
    \langle f, g\rangle = \int^b_af(x)^*\mathcal{L}g(x)\,\mathrm{d}x
\end{equation}

\noindent for those boundary conditions.
Standard Sturm--Liouville theory then implies that $\mathcal{L}$ has a large number of ``nice'' properties such as an orthonormal basis of eigenfunctions with real eigenvalues \citep[e.g.,][]{al2008sturm}.
Specific properties held by $P(x)$ and $Q(x)$ often imply bounds on those eigenvalues.
An important example is that, if $P(x),Q(x)>0$ on $(a,b)$, then all of the eigenvalues $\lambda$ must be positive.
This can be seen by multiplying Equation \ref{selfadjointform} by $y(x)^*$, integrating over the domain, and solving for $\lambda$
\begin{equation} \label{rayleigh}
    \begin{split}
        \lambda &= \frac{-\int^b_ay(x)^*\left(P(x)y'(x)\right)'\,\mathrm{d}x+\int^b_aQ(x)y(x)^*y(x)\,\mathrm{d}x}{\int^b_ay(x)^*y(x)\,\mathrm{d}x} \\
        &= \frac{\int^b_aP(x)|y'(x)|^2\,\mathrm{d}x+\int^b_aQ(x)|y(x)|^2\,\mathrm{d}x}{\int^b_a|y(x)|^2\,\mathrm{d}x} \\
    \end{split}
\end{equation}

\noindent where in the second equality we have integrated by parts, applying our boundary condition to discard the boundary term.
Equation \ref{rayleigh} is called the Rayleigh quotient, and the fact that all of the integrands that appear are positive-definite implies that $\lambda$ must be positive.
We will apply this result in later sections.

While the differential operator $\mathcal{L}^{m,b}_{\mathrm{mag}}$ (for real $b$) appears superficially similar to $\mathcal{L}$ as written in Equation \ref{selfadjointform}, the comparison is thwarted by the interior singularities which appear in $P$ and $Q$ at $\mu=\pm1/b$ (for $|b|\leq1$, Sturm--Liouville theory indeed applies).
Although we show in Section \ref{mequalszero} that solutions in the $m=0$ case are Hough functions which are second-differentiable everywhere, solutions with $m\neq0$ do not generally have this property, and have a number of unusual attributes (physically reflecting resonant interaction of gravity modes with Alfv\'en waves).

Motivated by problems in atmospheric physics \citep{boyd1976thesis,boyd1982influence}, \citet{boyd1981sturm} wrote down \nzrremoveone{the}\nzraddone{a} prototypical eigenvalue problem with an interior singularity,
\begin{equation} \label{boyd}
    \frac{\mathrm{d}^2y(x)}{\mathrm{d}x^2} - \frac{1}{x}y(x) + \lambda y(x) = 0
\end{equation}
Equation \ref{boyd} is called the Boyd problem, and its interesting mathematical properties have been the subject of some study \citep{boyd1981sturm,everitt1987some,gunson1987perturbation,atkinson1988regularization}.
The most interesting case is when it is considered over the domain $x\in(a,b)$ where $a<0<b$, so that there is an interior, non-integrable singularity at $x=0$.
It is common to consider this problem over the direct sum domain $x\in(a,0)\cup(0,b)$, over which \citet{everitt1987some} show that Equation \ref{boyd} possesses an orthonormal basis of discrete eigenfunctions with real $\lambda$.
These eigenfunctions are continuous over the entire range $x\in(a,b)$ (including over the singularity), but not necessarily differentiable.

\nzraddone{\citet{boyd1981sturm} and \citet{everitt1987some} note that, for a given real $\lambda$, $y(x)$ has two linearly independent solutions defined in terms of the Whittaker functions, $M_{-\kappa,1/2}(-x/\kappa)$ and $W_{-\kappa,1/2}(-x/\kappa)$ (with $1/\kappa\equiv2\sqrt{\lambda}$), themselves defined via confluent hypergeometric functions \citep{whittaker1903expression}.
While the former is analytic, the latter has a logarithmic divergence whose coefficient is proportional to $M_{-\kappa,1/2}(-x/\kappa)$.
As we will show, these properties are shared by the magnetogravity wave (analogous to $M_{-\kappa,1/2}(-x/\kappa)$; Section \ref{mnotequalszero}) and Alfv\'en wave (analogous to $W_{-\kappa,1/2}(-x/\kappa)$; Section \ref{alfven}) parts of the eigenfunctions of $\mathcal{L}^{m,b}_{\mathrm{mag}}$.
Notably, the former solution $M_{-\kappa,1/2}(-x/\kappa)$ \textit{vanishes} at $x=0$.}
\nzrremoveone{Moreover, the requirement that $\lambda\in\textbf{R}$ implies that the \textit{value} of these eigenfunctions must vanish at $x=0$ (e.g., Boyd 1981).}

The Boyd problem shares many properties with the magnetogravity problem (Equation \ref{mte}).
In particular, the singularity in the Boyd problem appears in $Q$, and the singularity in $Q$ in the magnetogravity is responsible for the unusual behavior of its eigenfunctions (as shown in Section \ref{mequalszero}, the magnetogravity problem is numerically well-behaved when $Q=0$).
We will see in Section \ref{mnotequalszero} that $p'$ eigenfunctions of the $m\neq0$ eigenproblem also vanish at the critical latitudes.
However, we shall also see that the displacements $\vec{\xi}$ are discontinuous for $m\neq0$, even though $p'$ is continuous, making the solutions unphysical.

\subsection{Power series expansion around singularity} \label{frob}

%When $|b|<1$ (where $b$ is real), Equation \ref{mte} with $m\neq0$ is a standard Sturm--Liouville problem which can easily be solved numerically in the same manner as the $m=0$ solutions as in Section \ref{mequalszero}.
When $|b|>1$ for real $b$, Equation \ref{mte} develops a singularity at the critical latitudes $\mu=\pm1/b$ where the mode frequency exactly matches the Alfv\'en frequency, and in this case na\"ively trying to numerically solve for these modes produces erratic behavior.

In order to characterize the behavior of Equation \ref{mte} in the $|b|>1$ case, we can perform a Frobenius power series expansion of the form
\begin{equation}
    p'(\mu) = (\mu-1/b)^\alpha\sum^\infty_{n=0}c_n(\mu-1/b)^n
\end{equation}
The leading-order term is the indicial equation, and can be solved to yield \nzraddone{$\alpha=0$}\nzrremoveone{$r=0$} and \nzraddone{$\alpha=2$}\nzrremoveone{$r=2$}, implying either that the leading-order dependence of the eigenfunctions around the singularity must either be constant or quadratic.
Enforcing equality at the next two lowest orders for \nzraddone{$\alpha=0$}\nzrremoveone{$r=0$} (the constant case) yields
\begin{subequations}
    \begin{gather}
        0 = \frac{(b^2-1)^3}{b^5}c_1 \\
        0 = \frac{b^2-1}{b^4}\left[(b^4 - 6b^2 + 5)c_1 + b^3m^2c_0\right]
    \end{gather}
\end{subequations}
% \begin{equation}
%     \begin{split}
%         0 &= \frac{(b^2-1)^3}{b^5}c_1 \\
%         0 &= \frac{b^2-1}{b^4}\left[(b^4 - 6b^2 + 5)c_1 + b^3m^2c_0\right] \\
%     \end{split}
% \end{equation}

\noindent indicating that $c_1=0$ (the first derivative vanishes) and also $m^2c_0=0$ (the value of the function also vanishes when $m\neq0$).
Therefore, the pressure perturbation of eigenfunctions which can be expanded in this way must vanish at the critical latitudes, as must their first derivatives.
Note that, while the first derivative at $\mu=\pm1/b$ must also vanish in the $m=0$ case (consistent with numerical solutions in Section \ref{mequalszero}), the value of the pressure perturbation need not vanish.

This result may also be seen in a more straightforward fashion from Equation \ref{mte} by multiplying the singular factor to the numerator.
One thereby obtains
\begin{equation} \label{multipliedout}
    \begin{split}
        \left(1-\mu^2\right)&\left(1-b^2\mu^2\right)\frac{\mathrm{d}^2p'(\mu)}{\mathrm{d}\mu^2} + 2\mu\left(b^2-1\right)\frac{\mathrm{d}p'(\mu)}{\mathrm{d}\mu} \\
        &+ \left(\frac{b^2}{a^2}\left(1-b^2\mu^2\right) - \frac{m^2}{1-\mu^2} \right)\left(1-b^2\mu^2\right)p'(\mu) = 0 \\
    \end{split}
\end{equation}
If the pressure perturbation $p'$ is everywhere finite, then Equation \ref{multipliedout} implies that $\mathrm{d}p'/\mathrm{d}\mu=0$ when $\mu=\pm 1/b$ (for any value of $m$).

To show that the value of the function must also vanish for $m=0$, we require not just that the horizontal gradient of $p'$ vanish in the direction across the critical latitude but the more general result that it vanish in all directions on this curve, i.e., that $p'$ must be a constant on connected curves of $|b|=1$.
We will show this in Section \ref{general} for magnetic fields which are more general functions of $\theta$ and $\phi$).
Then the only way to enforce both that $p'\propto e^{im\phi}$ and $p'=\mathrm{const.}$ on a critical latitude is for $p'$ itself to vanish.
This result can be compared to the vanishing of the finite eigenfunctions of the Boyd equation around $x=0$ (Section \ref{sl}).
In Section \ref{mnotequalszero}, we will demonstrate that this fact requires that the $m\neq0$ solutions must be exactly confined to an equatorial band with width $\Delta\mu=2/|b|$, in the sense of having exactly zero amplitude outside of it.

\section{Oscillation modes without dissipation} \label{sols}

\begin{figure}
    \centering
    \includegraphics[width=0.47\textwidth]{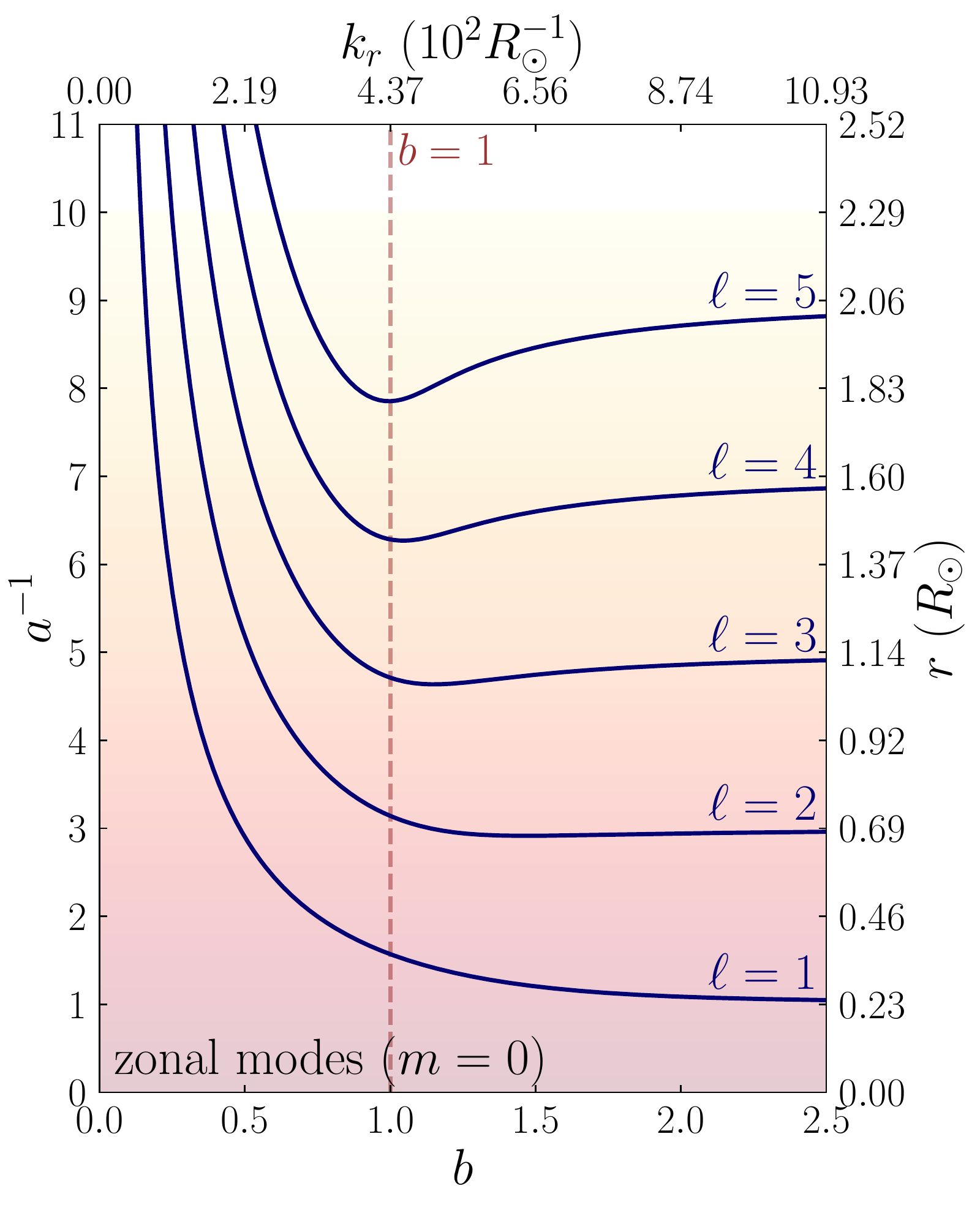}
    \caption{The inverse depth parameter \nzrremoveone{$1/a$}\nzraddone{$a^{-1}$} plotted against $b$ for zonal ($m=0$) modes (Section \ref{mequalszero}).
    The quantities \nzrremoveone{$1/a$}\nzraddone{$a^{-1}$} and $b$ have been roughly translated to $r$ and $k_r$ using \nzraddone{constant values} $\omega=2\pi\times10^2\,\mu\mathrm{Hz}$, $N=10^2\omega$, $v_A=0.1\,\mathrm{km}\,\mathrm{s}^{-1}$, and $R=10R_\odot$, reasonable parameters near the hydrogen burning shell in a first-ascent red giant.
    Ingoing gravity waves of different $\ell$ follow the tracks to the right, such that they never propagate back towards the surface of the star, and are converted to \nzraddone{slow magnetic}\nzrremoveone{Alfv\'en-like} waves with high radial wavenumber.}
    \label{fig:fig_evals_0}
\end{figure}

\begin{figure*}
    \centering
    \includegraphics[width=\textwidth]{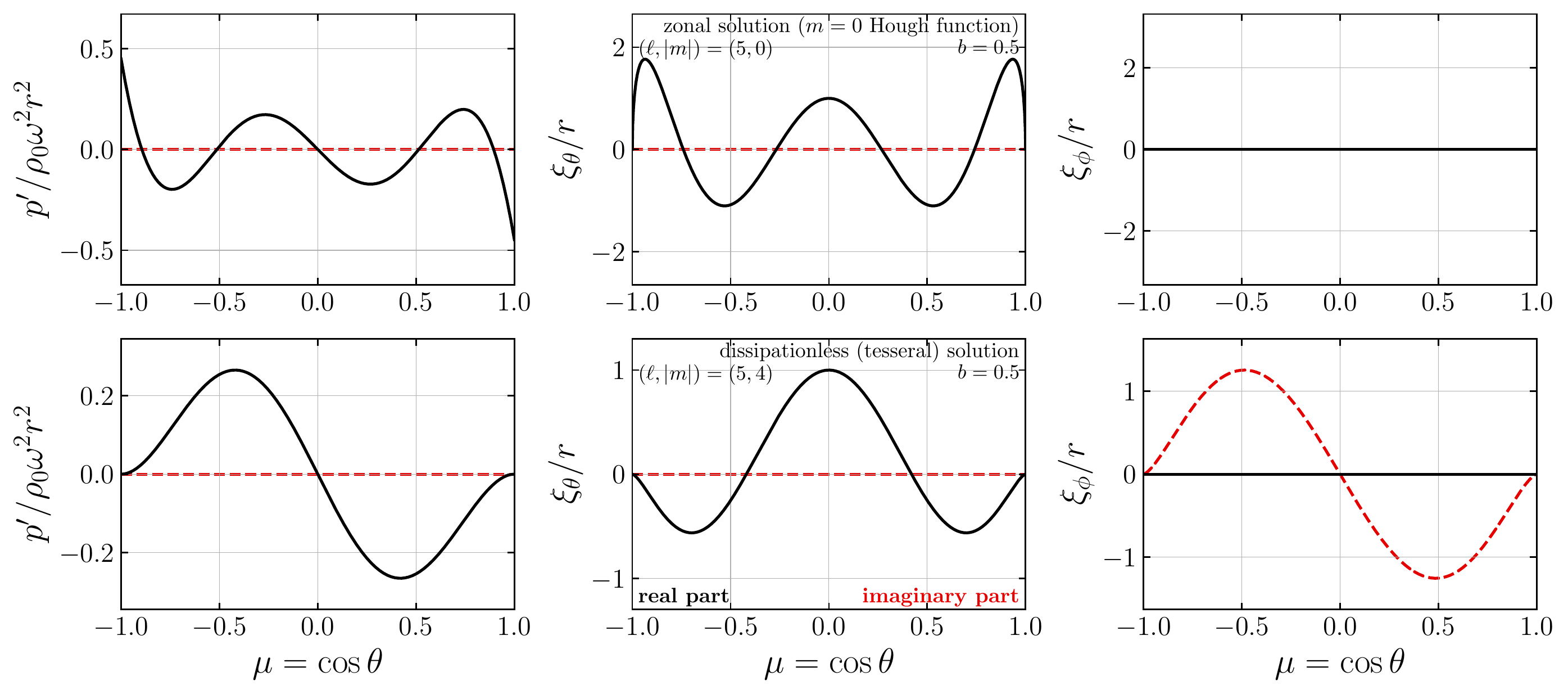}
    \caption{Fluid perturbations for the zonal $(\ell,|m|)=(5,0)$ (\textit{top}; Section \ref{mequalszero}) and tesseral $(5,4)$ modes (\textit{bottom:}; Section \ref{mnotequalszero}) as a function of the latitude $\mu=\cos\theta$, for $b=k_rv_A/\omega=0.5$.
    The \textit{left}, \textit{center}, and \textit{right} columns are the non-dimensionalized $p'$, $\xi_\theta$, and $\xi_\phi$ perturbations, respectively, with black solid lines representing the real part and red dashed lines representing the imaginary part.
    For low $b$, the eigenfunctions are close to spherical harmonics.}
    \label{fig:eigenfunctions_1}
\end{figure*}

\begin{figure*}
    \centering
    \includegraphics[width=\textwidth]{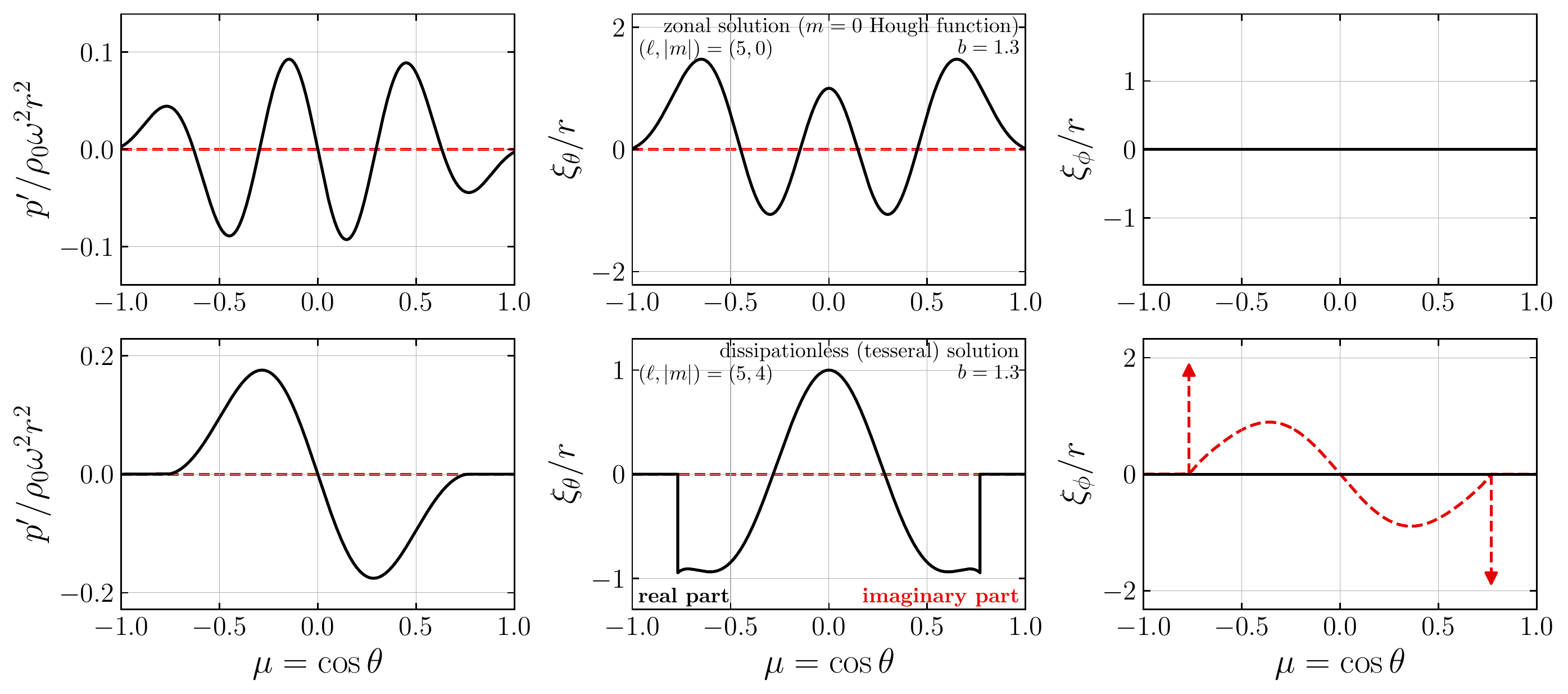}
    \caption{Fluid perturbations for the same mode branches as in Figure \ref{fig:eigenfunctions_1}, but for $b=k_rv_A/\omega=1.3$.
    The vertical arrows on the bottom right panel indicate the locations and phases of delta functions.
    For $b>1$ (when parts of the mode are resonant with Alfv\'en waves), both the $m=0$ and $m\neq0$ modes become localized to the equator, but only the $m\neq0$ modes gain sharp latitudinal features in $\xi_\theta$ and $\xi_\phi$ (owing to their vanishing outside of the critical latitudes).}
    \label{fig:eigenfunctions_0}
\end{figure*}

In this Section, we give solutions for the zonal ($m=0$; Section \ref{mequalszero}), tesseral/sectoral ($m\neq0$; Section \ref{mnotequalszero}), and Alfv\'en continuum (Section \ref{alfven}) modes for the singular eigenvalue problem discussed in Section \ref{three}.
The inclusion of viscous terms neutralizes the singularity and is discussed in Section \ref{viscous}.
%while specifically setting dissipation precisely to zero.
This is similar to the treatment given by authors such as \citet{boyd1981sturm} and similar authors investigating internally singular eigenvalue problems (Section \ref{sl}).
We refer to the solutions obtained in this way as \textit{dissipationless solutions}, and caution that this is distinct from the \textit{limit} as the dissipation is taken to zero (\textit{dissipative solutions}; Section \ref{viscous}).
The $m=0$ modes in the dissipationless solutions do not contain any discontinuous behavior at the critical latitude, and are exactly approached in the low-dissipation limit.
However, as we show in Section \ref{viscous}, any nonzero dissipation implies important qualitative differences in the $m\neq0$ modes, even in the very high Reynolds number, near-ideal magnetohydrodynamic flows in real stars.

\subsection{Zonal ($m=0$) solutions} \label{mequalszero}

In Section \ref{tar}, we noted the correspondence between \nzraddone{$\mathcal{L}^{m,b}_{\mathrm{mag}}$}\nzrremoveone{$\mathcal{L}^{m,b}_{\mathrm{rot}}$} and the $\mathcal{L}^{m,\nu}_{\mathrm{rot}}$ operator which appears in the rotational problem.
The latter's eigenfunctions are the Hough functions $\Theta^\nu_{\ell m}(\mu)$ with eigenvalues $\lambda^\nu_{\ell m}$, where $\ell$ denotes the degree of associated Legendre polynomial obtained by following a given Hough function branch to $\nu=0$.
When $m=0$, the correspondence becomes exact, and
\begin{equation}
    \mathcal{L}^{m,b}_{\mathrm{mag}}p'(\mu) = \mathcal{L}^{m,b}_{\mathrm{rot}}p'(\mu) = \frac{\mathrm{d}}{\mathrm{d}\mu}\left(\frac{1-\mu^2}{1-b^2\mu^2}\frac{\mathrm{d}p'(\mu)}{\mathrm{d}\mu}\right)
\end{equation}
It can therefore be seen that the Hough functions $\Theta^b_{\ell m}(\mu)$ are also horizontal pressure \nzrremoveone{(}$p'(\mu)$ eigenfunctions of the $m=0$ case of the magnetic problem.
Known properties of Hough functions thus greatly inform the behavior of these eigenfunctions.
In particular, because (a real value of) $b$ sets $1/|b|$ as a length scale \nzraddone{with respect to $\mu$}\nzrremoveone{ for $\mu$}\nzraddone{in $\mathcal{L}^{m,b}_{\mathrm{mag}}$}, Hough functions become approximately confined to an equatorial band of width $\Delta\mu\approx2/|b|$.

Additionally, it is known that the Hough function eigenvalue $\lambda^b_{\ell0}\approx(2\ell-1)^2b^2$ when $|b|$ is large, where the degree $\ell$ is equal to the number of latitudinal nodes for the $m=0$ case.
A heuristic argument for this behavior was given by \citet{bildsten1996ocean}, who argue that the quadratic scaling with $b$ arises from requiring that the eigenfunctions' zero crossings be localized to the aforementioned equatorial band.
The asymptotic behavior of the eigenvalues of the Hough functions was later derived more rigorously by \citet{townsend2003asymptotic} \citep[and more recently, to higher orders, by][]{townsend2020improved}.

By setting $\lambda^b_{\ell0}$ equal to $b^2/a^2$ (as required by the dispersion relation, Equation \ref{dispersionrelation}), one obtains for the zonal modes that $b$ \nzrremoveone{will }diverge\nzraddone{s} to infinity at some finite cutoff height \nzrremoveone{$a=a_{\ell0}^{(c)}$}\nzraddone{$a=a_\infty^{\ell0}$} defined by
\begin{equation} \label{cutoffa}
    a_\infty^{\ell0} = \frac{1}{2\ell-1}
\end{equation}
In other words, the ``cutoff height'' for these modes occurs at a radial magnetic field strength
\begin{equation}
    B_0^\infty = \frac{\sqrt{4\pi\rho_0}\omega^2r}{(2\ell-1)N}
\end{equation}
This is approximately equal to the critical magnetic field strength derived in \cite{fuller2015asteroseismology}\nzraddone{, although conceptually different}.
For $\ell=1$, \nzraddone{we find numerically that} the incoming wave approaches the cutoff height from above, and approaches infinite wavenumber before reaching a turning point (as can be seen in Figure \ref{fig:fig_evals_0}).
However, for all other values of $\ell$, \nzraddone{we find that} the incoming wave first refracts outwards before approaching the cutoff height from below.

\nzrremoveone{For stronger magnetic fields with $B_0 > B_0^{(c)}$}\nzraddone{In addition, for each mode, there is some critical field $B_c$ such that, for $B_0>B_c$} (\nzraddone{or $a > a_c$)}, there is no solution for a real value of $b$.
Only complex values of $b$ allow for solutions, implying \nzraddone{(for real $\omega$)} complex wavenumbers $k_r$ and evanescent waves similar to those discussed in \cite{fuller2015asteroseismology} and \cite{lecoanet2016conversion}.
Physically, this means that $m=0$ modes will refract off of strong magnetic fields as discussed in the works above.
This is different from the rotation problem where gravito-inertial waves can propagate at all radii where $N>\omega$, regardless of the rotation rate.

Using a relaxation method (see Appendix \ref{numa}), we solve for the $m=0$ eigenvalues 
and shown in Figure \ref{fig:fig_evals_0}, and the eigenfunctions shown in the top panels of Figures \ref{fig:eigenfunctions_1} and \ref{fig:eigenfunctions_0}.
Because $\lambda^b_{\ell0}$ approaches a constant $\ell(\ell+1)$ when $b$ approaches zero, $a^{-1}=\sqrt{\lambda^b_{\ell0}}/b$ diverges as $b$ vanishes.
%(note that the $\ell=0$ eigenfunction is unphysical, since purely radial low-frequency waves cannot propagate in stably stratified regions).
In most cases, an internal gravity wave branch increases in $|b|$ ($\propto|k_r|$) as it is followed to higher $a$ ($\propto Nv_A/r$), until it connects to a slow \nzraddone{magnetic} \nzrremoveone{Alfv\'enic }branch.
The wave then reaches a turning point at a maximum value of \nzrremoveone{$a$}\nzraddone{$a=a_c$} \nzraddone{(the ``critical depth'')}, and it is then forced to propagate back out to smaller values of $a$ (i.e., larger radii within a star) although $|b|$ continues to increase.
The value of $|b|$ and the radial wavenumber then diverge at the cutoff height defined in Equation \ref{cutoffa}.
This behavior is consistent with \citet{lecoanet2016conversion} (see Appendix \ref{lecoanet}) who discovered the same behavior in Cartesian geometry.

\begin{table}
\begin{tabular}{l@{\hspace{8pt}}l@{\hspace{8pt}}l@{\hspace{8pt}}l@{\hspace{8pt}}c}
    \hline
    Section & $(\ell,|m|)$ & $b_c$ & $|a_c|^{-1}$ & \multicolumn{1}{l}{$|a_\infty|^{-1}$} \\
    \hline\hline
    \multirow{5}{*}{\makecell{dissipationless\\$m=0$\\(Section \ref{mequalszero})}} & $(1,0)$ & $\infty$ & $a_\infty^{-1}$ & \multirow{5}{*}{\makecell{$a_\infty^{-1}=2\ell-1$\\(Equation \ref{cutoffa})}} \\
    & $(2,0)$ & $1.46$ & $2.92$ & \\
    & $(3,0)$ & $1.15$ & $4.64$ & \\
    & $(4,0)$ & $1.04$ & $6.27$ & \\
    & $(5,0)$ & $0.99$ & $7.85$ & \\
    \hline
    \multirow{15}{*}{\makecell{dissipationless\\$m\neq0$\\(Section \ref{mnotequalszero})}} & $(1,1)$ & $0.94$ & $1.99$ & \multirow{15}{*}{\makecell{$a_\infty^{-1}\approx\begin{cases}2.26 & \ell-|m|=0\\4.29 & \ell-|m|=1\\6.30 & \ell-|m|=2\\8.30 & \ell-|m|=3\\10.31 & \ell-|m|=4\\\vdots\end{cases}$\\(eigenvalues of Equation \ref{restricted2})}} \\
    & $(2,1)$ & $0.93$ & $3.45$ & \\
    & $(3,1)$ & $0.92$ & $4.93$ & \\
    & $(4,1)$ & $0.92$ & $6.42$ & \\
    & $(5,1)$ & $0.92$ & $7.91$ & \\
    & $(2,2)$ & $\infty$ & $a_\infty^{-1}$ & \\
    & $(3,2)$ & $1.66$ & $4.28$ & \\
    & $(4,2)$ & $1.02$ & $5.90$ & \\
    & $(5,2)$ & $0.99$ & $7.44$ & \\
    & $(3,3)$ & $\infty$ & $a_\infty^{-1}$ & \\
    & $(4,3)$ & $\infty$ & $a_\infty^{-1}$ & \\
    & $(5,3)$ & $\infty$ & $a_\infty^{-1}$ & \\
    & $(4,4)$ & $\infty$ & $a_\infty^{-1}$ & \\
    & $(5,4)$ & $\infty$ & $a_\infty^{-1}$ & \\
    & $(5,5)$ & $\infty$ & $a_\infty^{-1}$ & \\
    \hline
    \multirow{15}{*}{\makecell{dissipative\\real-$v_{{\rm p},r}$\\(Section \ref{numerical2})}} & $(1,1)$ & $0.94$ & $1.99$ & \multirow{15}{*}{\makecell{$|a_\infty|^{-1}\approx2(\ell-|m|)+3$\\(Equation \ref{cutoff2})}}\\
    & $(2,1)$ & $0.93$ & $3.45$ & \\
    & $(3,1)$ & $0.92$ & $4.93$ & \\
    & $(4,1)$ & $0.92$ & $6.42$ & \\
    & $(5,1)$ & $0.92$ & $7.91$ & \\
    & $(2,2)$ & $1.31$ & $2.66$ & \\
    & $(3,2)$ & $1.09$ & $4.33$ & \\
    & $(4,2)$ & $1.02$ & $5.90$ & \\
    & $(5,2)$ & $0.99$ & $7.44$ & \\
    & $(3,3)$ & $1.73$ & $2.95$ & \\
    & $(4,3)$ & $1.38$ & $4.81$ & \\
    & $(5,3)$ & $1.21$ & $6.53$ & \\
    & $(4,4)$ & $2.24$ & $3.09$ & \\
    & $(5,4)$ & $1.74$ & $5.07$ & \\
    & $(5,5)$ & $\infty$ & $|a_\infty|^{-1}$ & \\
    \hline
\end{tabular}
\caption{\nzraddone{For the mode branches computed in Sections \ref{mequalszero}, \ref{mnotequalszero}, and \ref{numerical2}, values of $b=b_c$ and $|a_c|^{-1}$ at the critical depth (the wave turning point), as well as values of the cutoff height $|a_\infty|^{-1}$.
Rows with $b_c=\infty$ and $|a_c|^{-1}=|a_\infty|^{-1}$ denote cases where the mode branch approaches $|a_\infty|^{-1}$ from above. Because our calculations only extend to $b=2.5$, it is possible that some branches reported as having $b_c = \infty$ have turning points at $b_c>2.5$.
%, \textit{without} turning around from a deeper depth $|a|^{-1}$.
}} \label{tablecutoffs}
\end{table}

The one exception is the $\ell=1$ case, where the wavenumber of the internal gravity wave branch directly diverges when approaching $a$ from below---there is no turning point, and no distinct slow \nzrremoveone{Alfv\'enic}\nzraddone{magnetic} branch.
In both cases there is a maximum $a$ (minimum radius) to which the wave can propagate, and the wavenumber $k_r$ diverges at a cutoff height within the star.
We thus find that the conclusions of \citet{fuller2015asteroseismology} and \citet{lecoanet2016conversion} that zonal modes cannot propagate arbitrarily deep in a sufficiently magnetized star to be robust for a dipole field geometry.
\nzraddone{In Table \ref{tablecutoffs}, we report values of the critical depth $a_c^{-1}$ and cutoff depths $a_\infty^{-1}$ for these mode branches.}

\begin{figure}
    \centering
    \includegraphics[width=0.5\textwidth]{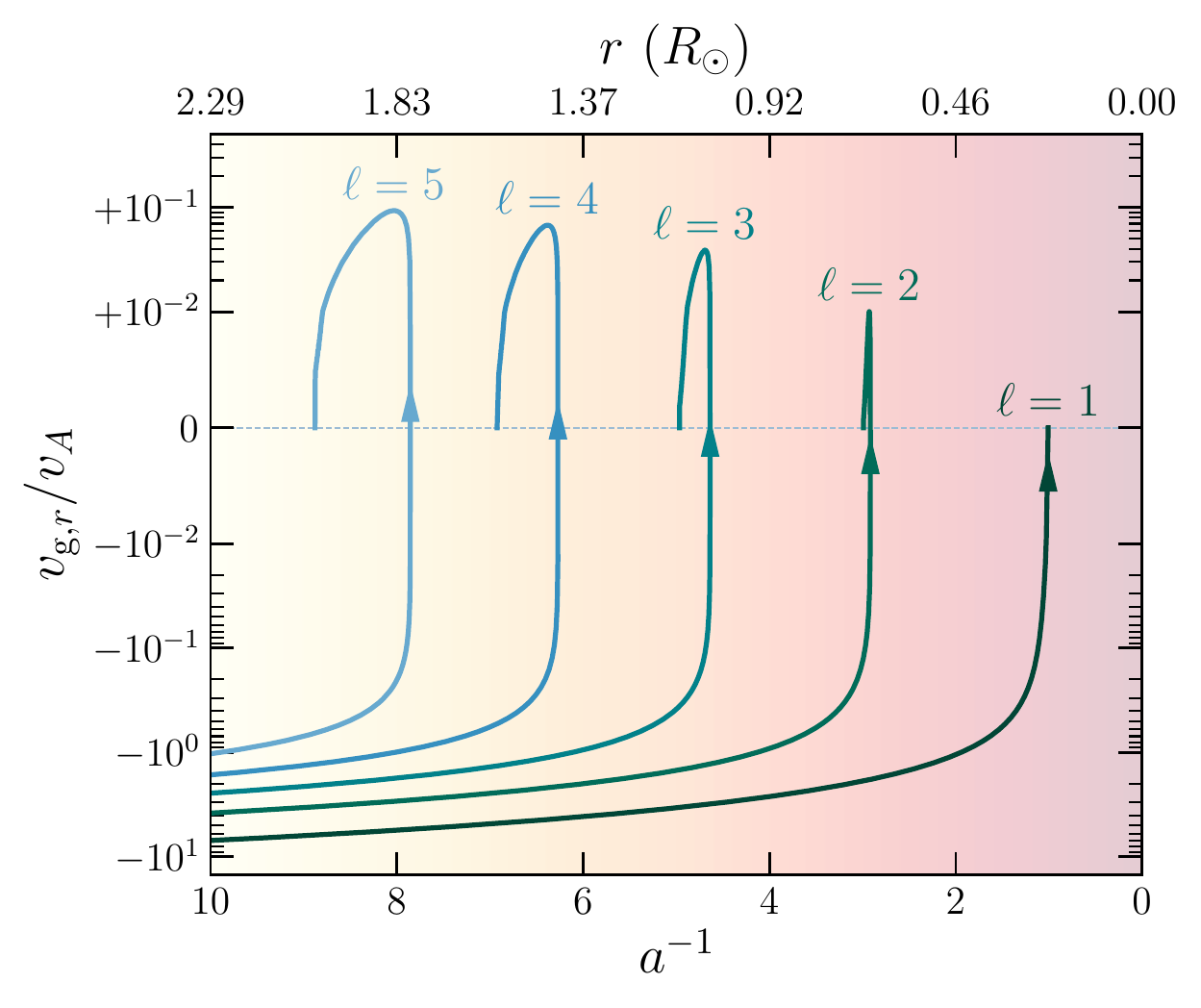}
    \caption{\nzraddone{The group velocities $v_{{\rm g},r}$ for $1\leq\ell\leq5$ zonal ($m=0$) magnetogravity wave branches.
    For most branches, ingoing magnetogravity waves refract back to larger radii at a critical depth $a_c$ before approaching infinite radial wavenumber (as slow magnetic waves) at some cutoff depth $a_\infty$.
    For the $\ell=1$ branch, the ingoing magnetogravity wave approaches the cutoff depth from above, without refracting outwards.
    The inverse depth parameter $a^{-1}$ has been roughly translated to a physical radius $r$ in the same manner as in Figure \ref{fig:fig_evals_0}.}}
    \label{fig:vgr}
\end{figure}

\nzradd{Assuming that $|\vec{k}|\approx k_r$,}\nzraddone{\nzrremove{T}he radial components of the phase and group velocities $v_{{\rm p},r}$ and $v_{{\rm g},r}$ can be specified in terms of $a$ and $b$ as}

\begin{subequations}
    \begin{gather}
        \frac{v_{{\rm p},r}}{v_A} = \frac{1}{v_A}\frac{\omega}{k_r} = \frac{1}{b} \\
        \frac{v_{{\rm g},r}}{v_A} = \frac{1}{v_A}\frac{\partial\omega}{\partial k_r} = \frac{1}{v_A}\left(\frac{\mathrm{d}\omega}{\mathrm{d}b}\right)\left(\frac{\mathrm{d}k_r}{\mathrm{d}b}\right)^{-1} = -\frac{\mathrm{d}a/\mathrm{d}b}{2a-b\,\mathrm{d}a/\mathrm{d}b}
    \end{gather}
\end{subequations}
\noindent \nzraddone{where we have used $\omega=\sqrt{Nv_A/r}a^{-1/2}$ and $k_r=\sqrt{N/v_Ar}ba^{-1/2}$ (from Equations \ref{b} and \ref{a}).}

\nzraddone{While $v_{{\rm p}, r}$ follows the motion of the wave pattern, $v_{{\rm g},r}$ tracks the transport of wave energy.
Figure \ref{fig:vgr} shows $v_{{\rm g},r}$ as a function of $a^{-1}$.
Ingoing gravity waves (whose $v_{{\rm p},r}$ and $v_{{\rm g},r}$ are in opposite directions) refract at the critical depth $a_c$ where $v_{{\rm g},r}=0$.
They then propagate outwards as slow magnetic waves with $v_{{\rm p},r}$ and $v_{{\rm g},r}$ in the same direction, with progressively slower group velocities as they approach the cutoff height.
The group velocities for $m\neq0$ modes (Sections \ref{mnotequalszero} and \ref{numerical2}) have similar behavior.}

In Section \ref{tar}, we pointed out that the Laplace tidal operator $\mathcal{L}^{m,\nu}_{\mathrm{rot}}$ (defined in Equation \ref{lto}) has \nzraddone{branches of mostly} negative eigenvalues for $\nu>1$\nzraddone{, which manifest as Rossby waves on the segments of the branches which are positive}\nzrremoveone{, corresponding to Rossby waves}.
However, in the magnetic problem, these branches are irrelevant when $\omega$ and $N$ are real\nzraddone{, since the eigenvalues on these branches are always negative}.
When $\lambda^b_{\ell0}<0$, this implies that $b$ ($\propto k_r$) is imaginary (i.e., that the wave is evanescent).
However, if $b$ is imaginary, then $b^2=-|b|^2$, and $\mathcal{L}^{m,b}_{\mathrm{mag}}$ becomes
\begin{equation} \label{mto2}
    \mathcal{L}^{m,b}_{\mathrm{mag}}p'(\mu) = \frac{\mathrm{d}}{\mathrm{d}\mu}\left(\frac{1-\mu^2}{1+|b|^2\mu^2}\frac{\mathrm{d}p'(\mu)}{\mathrm{d}\mu}\right) - \frac{m^2}{\left(1-\mu^2\right)\left(1+|b|^2\mu^2\right)}p'(\mu)
\end{equation}
Equation \ref{mto2} clearly has a positive $P,Q$ on the domain of the eigenproblem, with no internal singularities at all.
Sturm--Liouville theory thus implies (contrary to our initial assumption) that $\lambda^b_{\ell0}$ must be positive (see Section \ref{sl}).
This contradiction implies not only that these $\lambda^b_{\ell0}<0$ branches are irrelevant to the magnetogravity problem but also that the magnetogravity problem does not admit purely spatially evanescent solutions (for real $\omega$).

\subsection{Tesseral and sectoral ($m\neq0$) solutions} \label{mnotequalszero}

When $|b|<1$, the $m\neq0$ horizontal eigenfunctions (representing tesseral and sectoral modes) are simply solutions of a standard Sturm--Liouville problem with no internal singularities, and can be solved numerically \nzrremoveone{in the standard fashion}\nzraddone{using standard techniques}.
However, in the $|b|>1$ case, the mode and Alfv\'en frequencies are resonant at a critical latitude, where Equation \ref{mte} develops an internal singularity (Section \ref{sl}).
We discuss the implications of this critical latitude in the succeeding paragraphs.

In Section \ref{frob}, it is argued (vis-\`a-vis power series expansion) that both the pressure perturbation $p'$ and its first derivative $\mathrm{d}p'/\mathrm{d}\mu$ must vanish in the vicinity of the critical latitudes $\mu=\pm1/b$.
We first consider an eigenfunction with eigenvalue $\lambda$, and form a ``Rayleigh quotient'' (\nzrremoveone{compare}\nzraddone{cf.} Equation \ref{rayleigh}), but only over the portion of the domain bounded between $\mu\in(-1/b,+1/b)$ with $b>0$:

\begin{equation} \label{rayleigh2}
    \lambda = \frac{\int^{+1/b}_{-1/b}\frac{1-\mu^2}{1-b^2\mu^2}\left|\frac{\mathrm{d}p'(\mu)}{\mathrm{d}\mu}\right|^2\,\mathrm{d}x+\int^{+1/b}_{-1/b}\frac{m^2}{\left(1-\mu^2\right)\left(1-b^2\mu^2\right)}|p'(\mu)|^2\,\mathrm{d}x}{\int^{+1/b}_{-1/b}|p'(\mu)|^2\,\mathrm{d}\mu}
\end{equation}

\noindent where the vanishing pressure perturbation and gradient justify discarding the boundary term.
It is easily seen that each of the integrands above is positive-definite over the entire subdomain, and therefore $\lambda>0$.

However, one may write a similar Rayleigh quotient over the range $\mu\in(1/b,1)$,
\begin{equation} \label{rayleigh3}
    \lambda = \frac{\int^1_{1/b}\frac{1-\mu^2}{1-b^2\mu^2}\left|\frac{\mathrm{d}p'(\mu)}{\mathrm{d}\mu}\right|^2\,\mathrm{d}x+\int^1_{1/b}\frac{m^2}{\left(1-\mu^2\right)\left(1-b^2\mu^2\right)}|p'(\mu)|^2\,\mathrm{d}x}{\int^1_{1/b}|p'(\mu)|^2\,\mathrm{d}\mu}
\end{equation}

\noindent where it can be verified that the integrands in the numerator are now \textit{negative}-definite.
In Equation \ref{rayleigh3}, we have similarly discarded the boundary terms---this can be done at the outer boundary $\mu=1$ so long as $p'$ and its derivative are finite there.
This, in turn, implies that $\lambda<0$.

Of course, by definition, an eigenfunction must have just a single eigenvalue across the entire domain.
There are two ways to rectify these apparently contradictory conclusions.
One possibility is that the $\lambda>0$ eigenfunctions vanish outside of the critical latitudes, i.e., they are localized to a band of width $\Delta\mu=2/|b|$, bounded by the critical latitudes on each side (as demonstrated in Section \ref{mequalszero}, the $\lambda<0$ eigenvalues are not physical in this problem).
A second possibility is that only complex values of $b$ (and hence evanescent waves) exist when the real part of $b$ is greater than unity.

\begin{figure}
    \centering
    \includegraphics[width=0.47\textwidth]{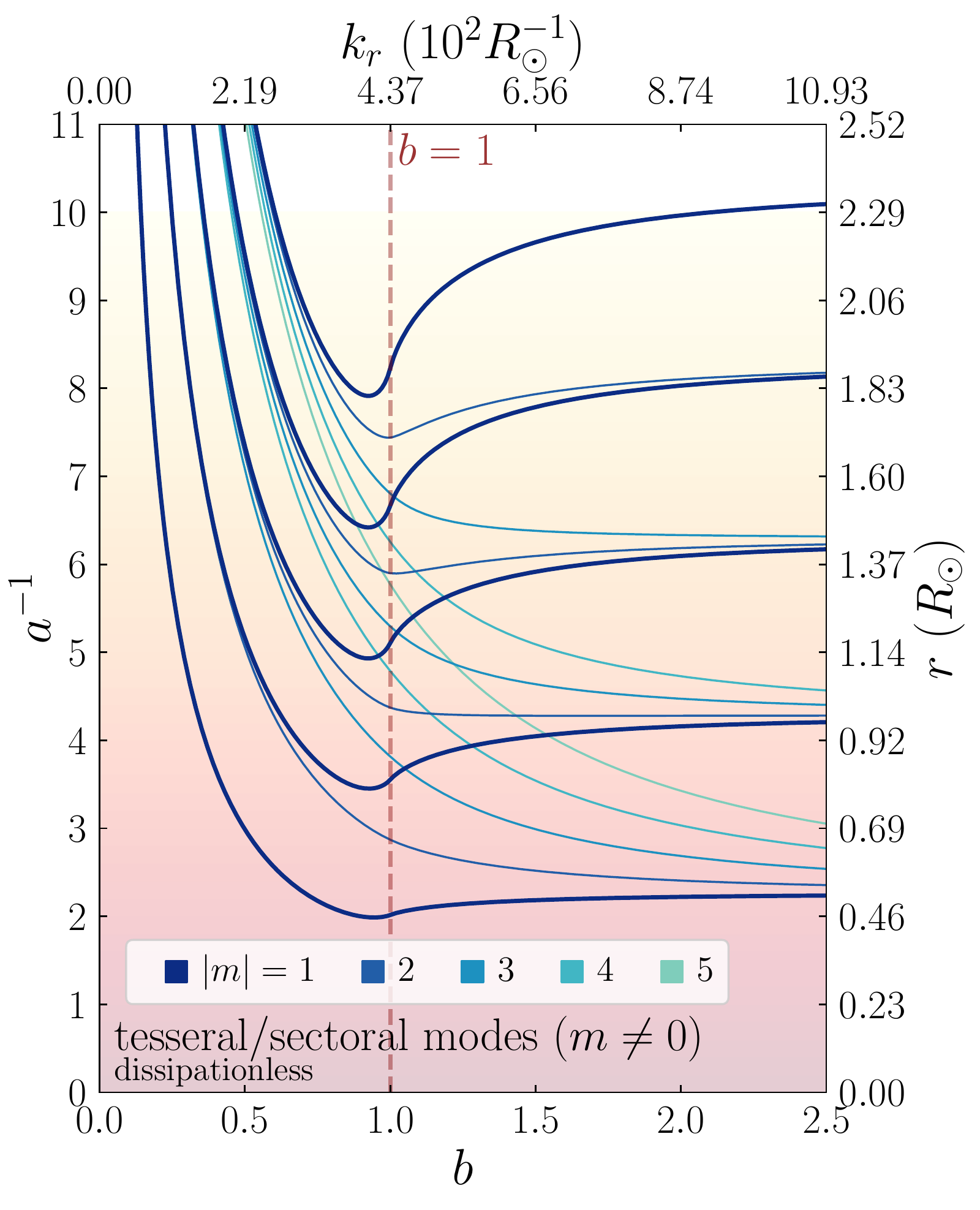}
    \caption{The inverse depth parameter \nzrremoveone{$1/a$}\nzraddone{$a^{-1}$} plotted against $b$ for tesseral/sectoral ($m\neq0$) modes for the singular eigenvalue problem described in Section \ref{mnotequalszero}.
    Both \nzrremoveone{$1/a$}\nzraddone{$a^{-1}$} and $b$ have been translated to $r$ and $k_r$ as in Figure \ref{fig:fig_evals_0}. As in Figure \ref{fig:fig_evals_0}, ingoing gravity waves follow the lines to the right, and are converted to \nzrremoveone{Alfv\'en}\nzraddone{slow magnetic} waves that do not propagate back to the surface.}
    \label{fig:fig_evals_1}
\end{figure}

In the first case, because the eigenfunction is confined to the range $\mu\in(-1/b,+1/b)$, we can restate the problem as a standard Sturm--Liouville problem (with no internal singularities) over this subinterval.
In particular, Equation \ref{mte} can be rewritten using $x=b\mu$ as
\begin{equation} \label{restricted}
    \frac{\mathrm{d}}{\mathrm{d}x}\left(\frac{b^2-x^2}{b^2\left(1-x^2\right)}\frac{\mathrm{d}p'(x)}{\mathrm{d}x}\right) - \frac{m^2}{\left(1-x^2\right)\left(1-b^2x^2\right)}p'(x) + \frac{1}{a^2}p'(x) = 0
\end{equation}

\noindent over the range $x\in(-1,+1)$.
We solve for both the eigenvalues and eigenfunctions by solving Equation \ref{mte} when $b<1$ and Equation \ref{restricted} when $b>1$, again using the relaxation method (Appendix \ref{numa}).
The eigenvalues for \nzrremoveone{$\ell,m\leq5$}\nzraddone{$\ell,|m|\leq5$} are shown in Figure \ref{fig:fig_evals_1}, and example eigenfunctions are shown in the bottom panels of Figures \ref{fig:eigenfunctions_1} (for $b<1$) and \ref{fig:eigenfunctions_0} (for $b>1$), respectively.
While the eigenfunctions are close to spherical harmonics for low $b$ (Figure \ref{fig:eigenfunctions_1}), they become formally confined between the critical latitudes when $b>1$, corresponding to resonances with Alfv\'en waves.
This is in contrast to the $m=0$ solutions which, although also experiencing some degree of equatorial confinement, are not forced to vanish outside of the resonant latitudes.

When $b$ is large (compared to $|m|$), Equation \ref{restricted} approaches
\begin{equation} \label{restricted2}
    \frac{\mathrm{d}}{\mathrm{d}x}\left(\frac{1}{1-x^2}\frac{\mathrm{d}p'(x)}{\mathrm{d}x}\right) + \frac{1}{a^2}p'(x) = 0
\end{equation}

Equation \ref{restricted2} is a generalized eigenvalue problem with eigenvalues $1/a^2$.
Therefore, we see that $a$ approaches a constant cutoff value \nzrremoveone{$a_{\ell m}^{(c)}$}\nzraddone{$a_\infty^{\ell m}$} in the large $b$ limit---in other words, when approaching some \nzrremoveone{critical} \nzraddone{cutoff} value \nzrremoveone{$a=a_{\ell m}^{(c)}$}\nzraddone{$a=a_\infty^{\ell m}$} from either above or below, $b$ \nzrremoveone{will diverge}\nzraddone{diverges}.
Moreover, since Equation \ref{restricted2} does not depend on $m$, \nzrremoveone{$a_{\ell m}^{(c)}$}\nzraddone{$a_\infty^{\ell m}$} only depends on the specific solution of Equation \ref{restricted2} which is approached by a given branch.
Therefore, \nzrremoveone{$a=a_{\ell m}^{(c)}$}\nzraddone{$a=a_\infty^{\ell m}$} is a function of $\ell-|m|$, which defines the number of nodes possessed by the generalized Legendre operator.
The cutoff values roughly lie between the $m=0$ cutoff values \nzrremoveone{$a_{\ell0}^{(c)}$}\nzraddone{$a_\infty^{\ell0}$} (defined in Equation \ref{cutoffa}), which do not follow the same pattern (see Figure \ref{fig:fig_evals_1}).
\nzraddone{Table \ref{tablecutoffs} reports the eigenvalues of Equation \ref{restricted2}, which give the cutoff depths $a_\infty^{-1}$ for these $m\neq0$ mode branches (as well as the critical depths $a_c^{-1}$).}

Another very important implication of Equation \ref{restricted} is that the $m\neq0$ branches cannot extend to arbitrarily large $a$, i.e., \textit{in a sufficiently magnetized star, propagating modes cannot extend arbitrarily deeply}.
When compared to Equation \ref{selfadjointform}, the differential operator which appears in Equation \ref{restricted} has $P,Q>0$ everywhere on the domain, implying that $1/a^2>0$, i.e., $a$ cannot \nzrremoveone{approach}\nzraddone{be} infinity for any finite $b$.
Furthermore, because the differential operator in Equation \ref{restricted2} (the large-$b$ limit of Equation \ref{restricted}) \nzrremoveone{also }has \nzraddone{$P>0$ and $Q=0$}\nzrremoveone{$P,Q>0$} everywhere on the domain\nzraddone{, the Rayleigh quotient (Equation \ref{rayleigh}) still implies that $1/a^2>0$ (in the large-$b$ limit) strictly, so long as $\mathrm{d}p'(x)/\mathrm{d}x\neq0$ somewhere on the domain.
As this is guaranteed to be the case for any perturbation for which $p'(x)\neq0$ (since it must vary from its boundary values $p'(\pm1)=0$)},\nzrremoveone{ its lowest eigenvalue ($=1/a^2$) cannot be zero either, and therefore} $a$ may not approach infinity \nzraddone{even in the limit that} $b$ does.
If we consider the second possibility discussed above, that $b$ becomes complex, the waves become evanescent at large values of $a$, meaning they no longer propagate.
This extends the conclusions of \citet{fuller2015asteroseismology} and \citet{lecoanet2016conversion} to the general $m\neq0$ case that propagating magnetogravity waves cannot exist arbitrarily deeply in a \nzrremoveone{highly }magnetized\nzraddone{-enough} star.

However, the localized nature of the pressure perturbations of the $m\neq0$ modes has important implications for the other perturbations (which also vanish outside of the critical latitudes, by Equations \ref{perts}). 
For example, since the leading-order dependence of the $p'$ eigenfunction near the singularity is quadratic \nzrremoveone{in general }(Section \ref{frob}), the discontinuity of $\mathrm{d}^2p'/\mathrm{d}\mu^2$ across the critical latitudes implies via Equation \ref{xitheta} that the \textit{value} of $\xi_\theta$ is discontinuous.
The fact that $\xi_\theta$ behaves as a step function near the singularity further implies (by the continuity equation) that $\xi_\phi$ contains a delta function at the critical latitude. This behavior is discussed in depth in \cite{goedbloed2004principles}, and
we comment further on this behavior in Section \ref{alfven}.

Because of the singular denominator factors in $\mathcal{L}_{\mathrm{mag}}^{m,b}$ and implied discontinuous eigenfunctions, it is important to consider that even \nzrremoveone{miniscule}\nzraddone{infinitesimally little} viscosity/\nzraddone{Ohmic} diffusivity can induce finite damping as well as global changes to the eigenfunctions.
We further discuss these effects in Section \ref{viscous}.
Nevertheless, the dissipationless solutions provide some analytic insight to qualitative features that they share with dissipative solutions to the magnetogravity wave problem.

\subsection{Alfv\'en wave solutions} \label{alfven}

In Sections \ref{frob} and \ref{mnotequalszero}, we performed a power series expansion to probe the behavior of the perturbations around the critical latitude $\mu=1/b$ and solved for the  $m\neq0$ solutions.
However, as a second-order differential equation, one na\"ively expects there to be \textit{two} linearly independent solutions.
More formally, when performing a Frobenius expansion, one obtains an indicial equation which can be solved to yield two solutions for the power law dependence of the solution very near the singularity (as in Section \ref{frob}).
When these two values are not separated by an integer, one immediately obtains these two linearly independent solutions.

However, the values of \nzraddone{the indicial root $\alpha$}\nzrremoveone{$r$} found in Section \ref{frob} \textit{are} separated by an integer, so a Frobenius expansion in $p'$ is not particularly helpful in the search for the other solution.
Instead, by substituting Equations \ref{xir} and \ref{xiphi} into the continuity equation (Equation \ref{continuitydipole}), solving for $p'$ in terms of $\xi_\theta$, and then substituting the result into the $\theta$ momentum equation (Equation \ref{thetamtmdipole}), one obtains
\begin{equation}
    \frac{\mathrm{d}}{\mathrm{d}\mu}\left[\left(\frac{b^2}{a^2}-\frac{m^2}{\left(1-\mu^2\right)\left(1-b^2\mu^2\right)}\right)^{-1}\frac{\mathrm{d}\mathcal{Z}_\theta(\mu)}{\mathrm{d}\mu}\right] + \frac{1-b^2\mu^2}{1-\mu^2}\mathcal{Z}_\theta(\mu) = 0
\end{equation}

\noindent where
\begin{equation}
    \mathcal{Z}_\theta(\mu) \equiv \sqrt{1-\mu^2}\frac{\xi_\theta(\mu)}{r}
\end{equation}

A power series expansion of the form
\begin{equation}
    \mathcal{Z}_\theta(\mu) = (\mu-1/b)^\alpha\sum^\infty_{n=0}c_n(\mu-1/b)^n
\end{equation}

\noindent gives an indicial equation which has a double root at \nzraddone{$\alpha=0$}\nzrremoveone{$r=0$}, consistent with the results of \citealt{goedbloed2004principles} on a similar magnetohydrodynamic problem (see their Section 7.4).

Hereafter, for illustrative purposes, we focus on the restricted problem over the interval $\mu\in(0,1)$ in order to focus on the critical latitude at $\mu=1/b$ (this is justified in Section \ref{numa}).
The choice of \nzraddone{$\alpha=0$}\nzrremoveone{$r=0$} gives a single everywhere-finite solution which can be called $\mathcal{Z}_\theta(\mu)=\mathcal{Z}_1(\mu)=u(\mu)$.
In this case, a second linearly independent solution is given by
\begin{equation}
    \mathcal{Z}_2(\mu) = u(\mu)\ln\left\lvert\mu-1/b\right\rvert + v(\mu)
\end{equation}

\noindent which contains a logarithmic divergence at the critical latitude.
\citealt{goedbloed2004principles} show that, while the coefficient $\mathcal{Z}_1$ may differ on either side of the singularity, the coefficient in front of $\mathcal{Z}_2$ may not.
The general solution for $\mathcal{Z}_\theta$ is thus given by
\begin{equation}
    \mathcal{Z}_\theta(\mu) = \left[A_1\Theta(\mu-1/b) + A_2\Theta(1/b-\mu)\right]\mathcal{Z}_1(\mu) + A_3\mathcal{Z}_2(\mu)
\end{equation}

\noindent where $\Theta$ is the Heaviside step function.
Note that that the presence of \textit{three} undetermined coefficients $A_1$, $A_2$, and $A_3$ constrained by only two boundary conditions implies a continuous spectrum of modes.
This is a well-established consequence of singularities in differential equations, especially those corresponding to Alfv\'en resonances in plasma physics \citep{appert1974continuous,poedts1985continuous,rauf1995alfven,appert1998continuous,widdowson1998continuum,rincon2003oscillations,goedbloed2004principles,reese2004oscillations,pinter2007global,loi2017torsional}.
Physically, the continuous Alfv\'en spectrum arises out of a lack of discretization in the $\theta$ direction, associated with mode localization in \nzraddone{geometries with field/plasma inhomogeneity}\nzrremoveone{inhomogeneous field/plasma structures}.

In the treatment in this work, we do not explicitly impose boundary conditions in the radial direction.
However, doing so would discretize the allowed values of $k_r$ both for the global modes and the \nzrremoveone{torsional }Alfv\'en \nzrremoveone{resonances}\nzraddone{waves} \citep[see, e.g.,][]{loi2017torsional}.
Alfv\'en resonances can exist whenever $\omega=k_rv_A|\mu|$, i.e., $b = 1/|\mu|$.
The continuum Alfv\'en spectrum therefore occupies all frequencies $\omega$ with $|b|\geq1$ (i.e., every point to the right of $b=1$ in Figure \ref{fig:fig_evals_1}).
In practice, because each field line has a discrete spectrum of Alfv\'en \nzrremoveone{resonances}\nzraddone{waves} (which are analogous to oscillations on a closed loop), a real global mode resonates with the Alfv\'en spectrum at only a finite (but large) number of locations \citep{loi2017torsional}.
%Though $k_r$ is actually discretized along a closed field loop, $|\mu|$ runs continuously from $0$ to $1$, and thus each $k_r$ is compatible with an infinite number of mode frequencies $\omega$ (resonating at different latitudes).
%The continuum Alfv\'en spectrum therefore occupies all frequencies $\omega$ with $|b|\geq1$ (i.e., every point to the right of $b=1$ in Figure \ref{fig:fig_evals_1}).

In problems possessing even vanishingly small amounts of dissipation, the Alfv\'en continuum has important implications both for the global forms of the eigenfunctions and wave damping.
\citet{hoven2011magnetar} note that any dissipation couples fluid displacements across flux surfaces, destroying the continuum nature of the Alfv\'en spectrum (see Section \ref{wkb}).
In Section \ref{viscous}, we find that including dissipation produces discrete spectra for which only a specific linear combination of $u(\mu)$ and $v(\mu)$ are truly eigenfunctions.

\nzraddone{Because Alfv\'en waves are \nzrremove{incompressible}\nzradd{not associated with a pressure perturbation}, \nzradd{the Lagrangian temperature perturbation vanishes and therefore does not produce bulk brightness fluctuations which would be asteroseismically detectable in the light curve \citep{houdek2015interaction}.}\nzrremove{they likely cannot directly be detected using standard asteroseismic methods, which rely on changes in brightness due to surface temperature perturbations (Houdek \& Dupret 2015).}It may be possible to observe their signature in surface velocity fluctuations, if the waves do not damp before reaching the surface.}
%However, Alfv\'en waves may still leave an indirect impact on the light curve by contributing to nonradial mode damping through processes such as phase mixing \citep{loi2017torsional}. Together with refraction of magnetogravity waves to large wavenumbers \citep{lecoanet2016conversion}, phase mixing may contribute to the highly suppressed nonradial mode amplitudes observed in the power spectrum of many red giants \citep{garcia2014study,stello2016prevalence}.}

\section{Oscillation modes with dissipation} \label{viscous}

So far, we have considered the mathematical problem where we have formally set all dissipation to zero.
In this Section, we consider the important role played by even small amounts of dissipation in shaping the horizontal structure of magnetogravity modes.

As discussed in Section \ref{alfven}, the magnetogravity problem possesses a continuum of Alfv\'en modes, each localized to \nzraddone{a} magnetic field line\nzrremoveone{s}.
Adjacent Alfv\'en modes will oscillate at slightly different frequencies, corresponding to the slightly different Alfv\'en frequencies of their field lines.
This quickly leads to a dephasing process called ``phase mixing,'' a kind of quasi-damping which, while formally reversible in ideal magnetohydrodynamics, leads to \nzrremoveone{real}\nzraddone{finite} energy damping under any \nzraddone{(arbitrarily small)} amount of dissipation.
Interestingly, this energy damping approaches a finite value in the limit of even a vanishingly small dissipation, meaning that its role cannot be ignored even in stars where dissipative processes are usually considered to be negligible.
For further discussion of phase mixing and its associated energy dissipation, see \citet{goedbloed2004principles}.

If dissipation, in the form of fluid viscosity and Ohmic diffusion, are included, \nzraddone{the linearized horizontal momentum and induction equations are modified to}
\begin{subequations} \label{mb2}
    \begin{gather}
        -\rho_0\omega^2\vec{\xi}_h = -\nabla p' + \frac{1}{4\pi}\left(\vec{B}_0\cdot\nabla\right)\vec{B}' + i\omega\rho_0\nu \nabla^2 \vec{\xi}_h \label{mb_mtm2} \\
        \vec{B}' = \left(\vec{B}_0\cdot\nabla\right)\vec{\xi} + \eta\nabla^2\vec{B}' \label{mb_b2}
    \end{gather}
\end{subequations}

\noindent \nzraddone{where we continue to assume the hierarchy of variables described in Section \ref{lfe} (including taking $\nabla^2\approx-k_r^2$).
In Equations \ref{mb2}, $\nu$ and $\eta$ denote the kinematic viscosity and magnetic diffusivity, respectively.
We note in passing that the latter is expected to dominate the overall dissipation, but that both terms have a similar impact on the solutions.}

\nzraddone{Equations \ref{mb_mtm2} and \ref{mb_b2} can be combined to obtain}
\begin{equation} \label{newp}
    \nabla_hp' = \rho_0\omega^2\left(1 - b^2\mu^2 - ic\right)\vec{\xi}_h
\end{equation}

\noindent \nzraddone{where $c$ is given by}
\begin{equation} \label{cccc}
    c = c_\nu b^2 + c_\eta b^4\mu^2
\end{equation}

\noindent \nzraddone{where}
\begin{subequations}
    \begin{gather}
        c_\nu = \frac{\nu\omega}{v_A^2} \\
        c_\eta = \frac{\eta\omega}{v_A^2}
    \end{gather}
\end{subequations}
% \begin{equation}
%     \begin{split}
%         c_{\mathrm{kin}} &= \left(\frac{\nu}{r^2\omega}\right)\left(\frac{N}{\omega}\right)^2 \\
%         c_{\mathrm{mag}} &= \left(\frac{\eta}{r^2\omega}\right)\left(\frac{N}{\omega}\right)^2 \\
%     \end{split}
% \end{equation}
\nzraddone{In deriving Equations \ref{newp} and \ref{cccc}, we have assumed that $k_r^2\eta/\omega\ll1$.
Note that, because the effect of $c\ll1$ is to shift the poles slightly off of the real line into the complex plane, the exact form of $c$ does not matter, and it suffices to take it to be a small, real constant.
Moreover, since dissipation is most important near the critical latitudes $\mu=\pm1/b$, both terms scale roughly as $\propto b^2$ in the most affected regions.}

\nzraddone{Overall, }the operator $\mathcal{L}^{m,b}_{\mathrm{mag}}$\nzraddone{then takes the new form}\nzrremoveone{ gets modified to}
\begin{equation} \label{vmto}
    \begin{split}
        \mathcal{L}^{m,b}_{\mathrm{mag}}p'(\mu) = \frac{\mathrm{d}}{\mathrm{d}\mu}&\left(\frac{1-\mu^2}{1-b^2\mu^2-ic}\frac{\mathrm{d}p'(\mu)}{\mathrm{d}\mu}\right) \\
        &- \frac{m^2}{\left(1-\mu^2\right)\left(1-b^2\mu^2-ic\right)}p'(\mu)
    \end{split}
\end{equation}

\noindent where $c$ encodes the dissipative processes in the problem, and ``softens'' the singularity.
\nzrremoveone{For a kinematic viscosity $\nu$ and a magnetic diffusivity $\eta$ (expected to dominate), $c$ is given by}
% \begin{equation}
%     c = c_{\mathrm{kin}}\left(b^2/a^2\right) + c_{\mathrm{mag}}\left(b^4/a^2\right)\mu^2
% \end{equation}

% \noindent where
% \begin{equation}
%     \begin{split}
%         c_{\mathrm{kin}} &= \left(\frac{\nu}{r^2\omega}\right)\left(\frac{N}{\omega}\right)^2 \\
%         c_{\mathrm{mag}} &= \left(\frac{\eta}{r^2\omega}\right)\left(\frac{N}{\omega}\right)^2 \\
%     \end{split}
% \end{equation}

\nzrremoveone{where we have assumed that $k_r^2\eta/\omega$ is small, and that the radial wavenumber dominates the dissipation.
However, because the effect of $c\ll1$ is to shift the poles slightly off of the real line into the complex plane, the exact form of $c$ does not matter, and it suffices to take it to be a small, real constant.}

We note in passing that terms dependent on the horizontal field $B_h$ may be significant at the critical latitudes where dissipation is expected to be most important.
The inclusion of such terms introduces higher-order horizontal derivatives to the linearized equations and greatly increases their complexity.
Nevertheless, we expect that the parameterization above in terms of $c$ will still physically select the right branch of solutions, in the limit of small dissipation.
In Section \ref{wkb}, we comment further on the importance of such terms near the critical latitudes.

\begin{figure}
    \centering
    \includegraphics[width=0.47\textwidth]{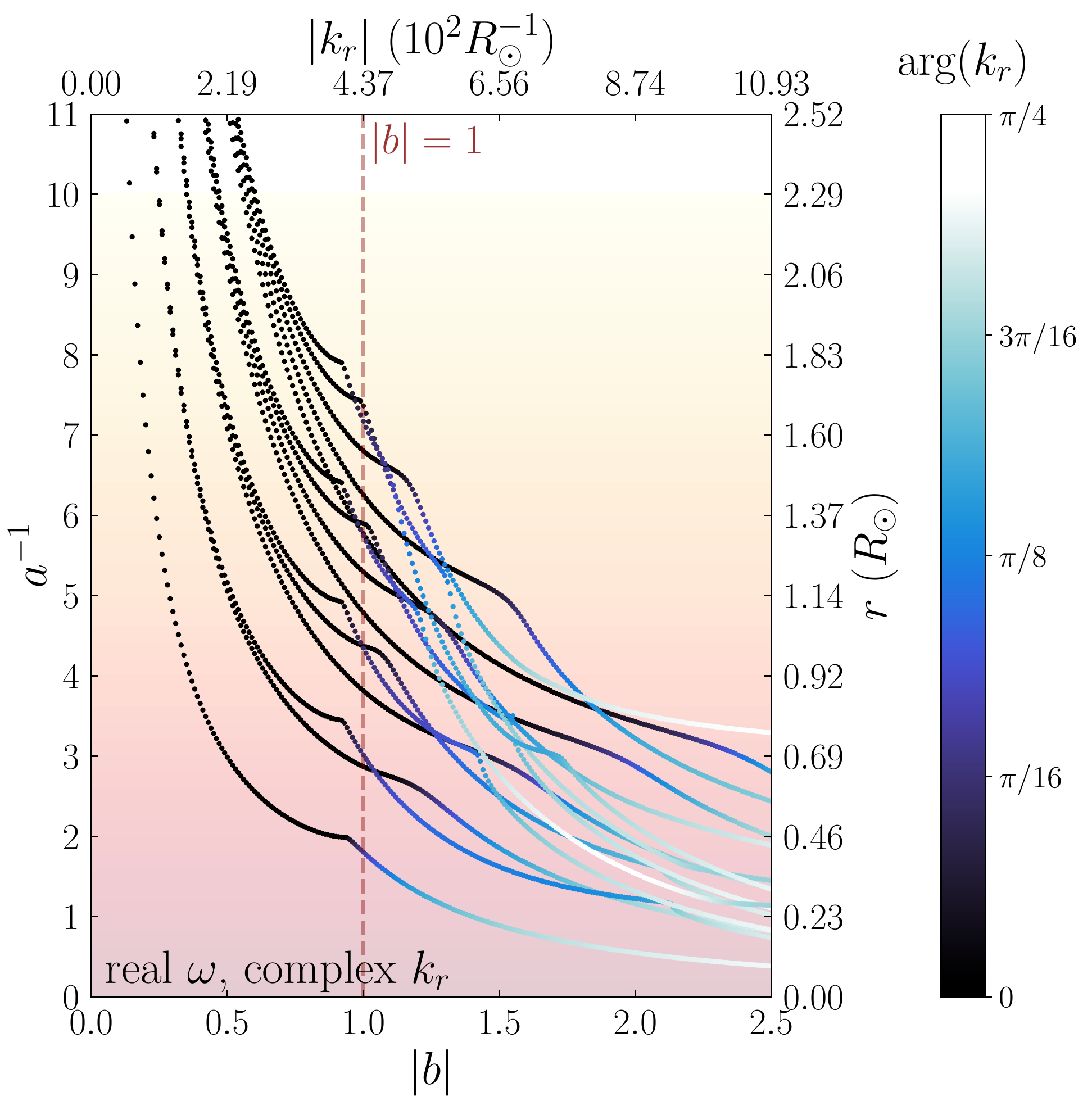}
    \caption{The inverse depth parameter \nzrremoveone{$1/a$}\nzraddone{$a^{-1}$} plotted against $b$ for non-axisymmetric $m\neq0$ modes, with finite dissipation and real $\omega$ (Section \ref{numerical1}).
    The color represents the complex argument of $k_r$, with the lower branches representing spatially evanescent solutions.
    Both \nzrremoveone{$1/a$}\nzraddone{$a^{-1}$} and $b$ have been translated to $r$ and $k_r$ as in Figure \ref{fig:fig_evals_0}. Unlike previous figures, gravity waves do not propagate into the colored portions of the lines, where they become strongly evanescent. Instead, they are refracted upwards onto a\nzrremoveone{n} \nzraddone{slow magnetic} \nzrremoveone{Alfv\'enic} wave branch not shown here (see Figure \ref{fig:fig_evals_3}).}
    \label{fig:fig_evals_2}
\end{figure}

In the following subsections, we present numerical solutions for the dissipative magnetogravity eigenproblem (details in Appendix \ref{numb}).
Section \ref{numerical1} considers modes with real $\omega$ but complex $k_r$, i.e., possibly spatially evanescent modes, and Section \ref{numerical2} considers modes with \nzraddone{real radial phase velocity $v_{{\rm p},r}=\omega/k_r$}\nzrremoveone{ possibly complex $k_r$ and $\omega$, but subject to the constraint that their ratio $k_r/\omega$ be real} (approximating the case of propagating waves).
We will show that, while the analysis of Section \ref{mnotequalszero} provides insights into realistic modes, the presence of dissipation introduces notable deviations from the idealized behavior.

\begin{figure*}
    \centering
    \includegraphics[width=\textwidth]{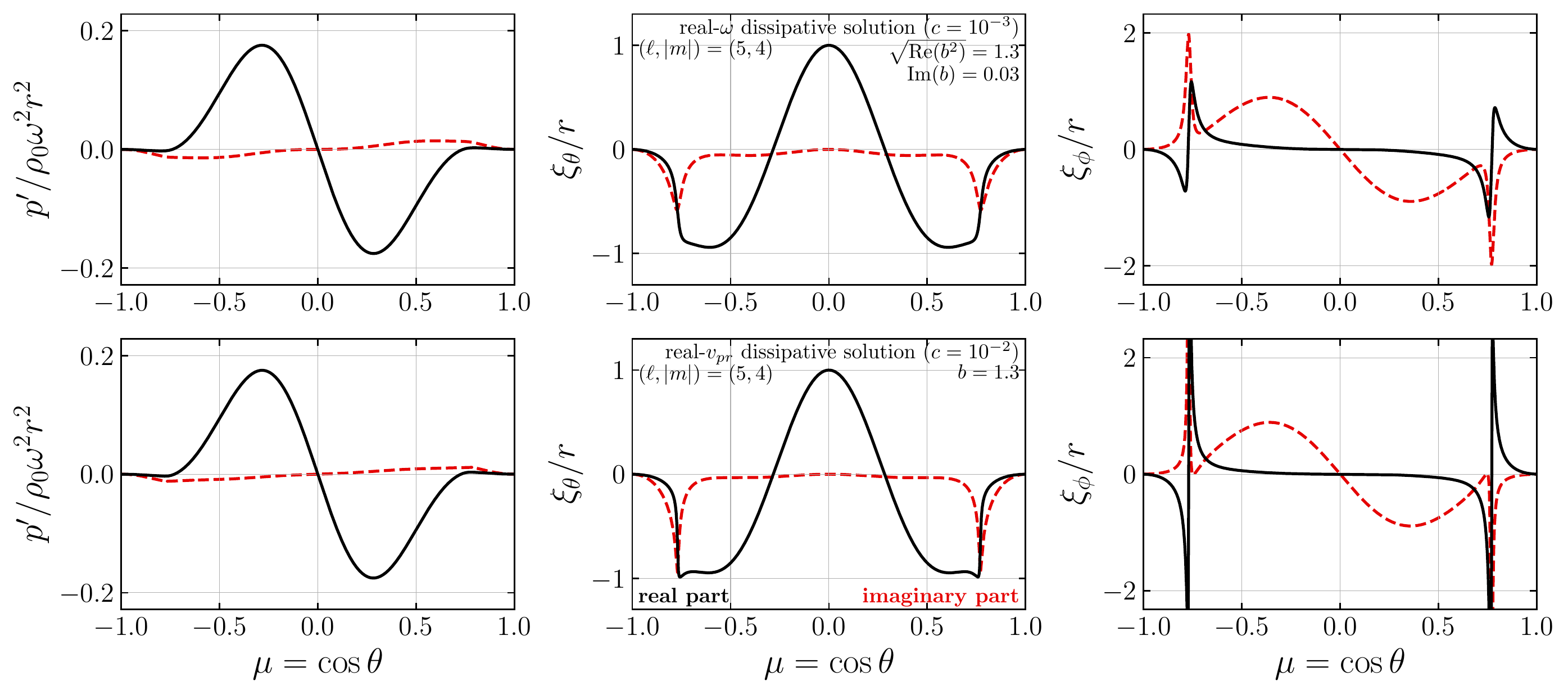}
    \caption{Fluid perturbations for the tesseral $(\ell,|m|)=(5,4)$ modes for $\sqrt{\mathrm{Re}(b^2)}=1.3$\, calculated numerically in the dissipative case where either $\omega$ is real (\textit{top}; Section \ref{numerical1}), or \nzrremoveone{$k_r/\omega$}\nzraddone{$v_{{\rm p},r}=\omega/k_r$} is real (\textit{bottom}; Section \ref{numerical2}).
    These eigenfunctions should be compared to those shown in the bottom panels of Figure \ref{fig:eigenfunctions_1} for the discontinuous case.}
    \label{fig:eigenfunctions_2}
\end{figure*}

%\subsection{Real $\omega$, complex $k_r$}
\subsection{Numerical solutions of the evanescent branch}\label{numerical1}

We first consider the case where $\omega$ is real but $k_r$ is allowed to be complex (i.e., allowing solutions to be spatially evanescent).
This corresponds to fixing $a\propto\omega^{-2}$ to be real but allowing $b\propto k_r/\omega$ to be complex.
As described in Appendix \ref{numb}, we solve for the eigenfunctions of the operator in Equation \ref{vmto} up to $\ell,|m|=5$ (using $c=10^{-3}$) while allowing the complex argument of $b$ to vary.
For consistency, we search for only solutions with $\mathrm{Im}(b)\geq0$, although each such evanescent branch is accompanied by a conjugate branch of solutions.

The eigenvalues are shown in Figure \ref{fig:fig_evals_2} as a function of $|b|$.
%For concreteness, we have adopted for this problem a dissipation value $c=10^{-3}$.
When $|b|\lesssim1$ (i.e., weak magnetic fields), the singularity does not lie on the domain and dissipation does not play a major role. For decreasing values of $c$, $b$ approaches a real number, as expected, and the solutions are nearly identical to those discussed in Section \ref{mnotequalszero}.

However, for $|b|>1$, there are significant qualitative differences between the discontinous solutions of Section \ref{mnotequalszero} and the dissipative solutions.
Even in the limit of \nzrremoveone{small $c$}$c\rightarrow0$, the imaginary part of $b$ \nzrremoveone{never becomes exactly zero}\nzraddone{does not correspondingly vanish}, although (as we discuss below) \nzrremoveone{it sometimes approaches a small and constant limit}\nzraddone{its limiting value is sometimes quite small}.
This implies that the corresponding eigenfunctions are still ``smoothed'' with respect to the discontinuous solutions even in the $c\rightarrow0$ limit.

For some branches of modes\nzrremoveone{ (usually those with larger $\ell$ and $|m|$)}, there is a range extending from $|b|=1$ to some intermediate value of $|b|$ where $\mathrm{Im}(b)$ is small when $c\approx0$.
In these intermediate ranges, the real parts of $p'$, $\xi_\theta$, and $\xi_\phi$ strongly resemble smoothed versions of the discontinuous solutions described in Section \ref{mnotequalszero} (e.g., the top panel of Figure \ref{fig:eigenfunctions_2}).
In particular, $\xi_\theta$ has a smoothed step-like jump across the singularity, and $\xi_\phi$ retains a smoothed, but narrow, peak there.
Interestingly, the imaginary part of $\xi_\theta$ approaches the logarithmic Alfv\'en ``spike'' solutions described in Section \ref{alfven}---the numerical solution is thus a close approximation of a superposition of these two solutions predicted in Section \ref{alfven}.
These solutions can be visualized as equatorially focused magnetogravity modes which oscillate $\pi/2$ out of phase with an Alfv\'en mode.
This closely resembles the example shown in Figure 11.2 by \citet{goedbloed2004principles} (in a similar magnetohydrodynamic problem), as well as the numerical results of \citet{lecoanet2022asteroseismic}.
We emphasize that, because the imaginary part of $b$ does not approach zero in the $c\rightarrow0$ limit, the ``smoothing'' \textit{does not} go away even in this limit.
\nzraddone{It appears that the size of the intermediate range of $|b|$ for which $\mathrm{Im}(b)$ is small appears to increase with $|m|$ for fixed $\ell$.
%, and decrease slightly with $\ell$ for fixed $|m|$.
However,
%the dependence of this width on $\ell$ and $|m|$ do not seem to trade off in any obvious algebraic combination, and
the origin of this trend is so far unclear.}

% Branches with larger $\ell$ and higher $m$ (i.e., those with high degree but low number of nodes in the polar direction, or those with more sectoral character) are more likely to have such a range, and furthermore tend to have larger ranges.
% For example, while we find the $(\ell,|m|)=(1,1)$ and $(5,1)$ branches do not have any such range (the solver jumps onto the evanescent branch even before $|b|=1$), the $(5,5)$ sectoral mode retains this character over the whole range of $|b|$ shown in Figure \ref{fig:fig_evals_2}, and its eigenvalues are always close to the FDS expectation.

\begin{figure*}
    \centering
    \includegraphics[width=\textwidth]{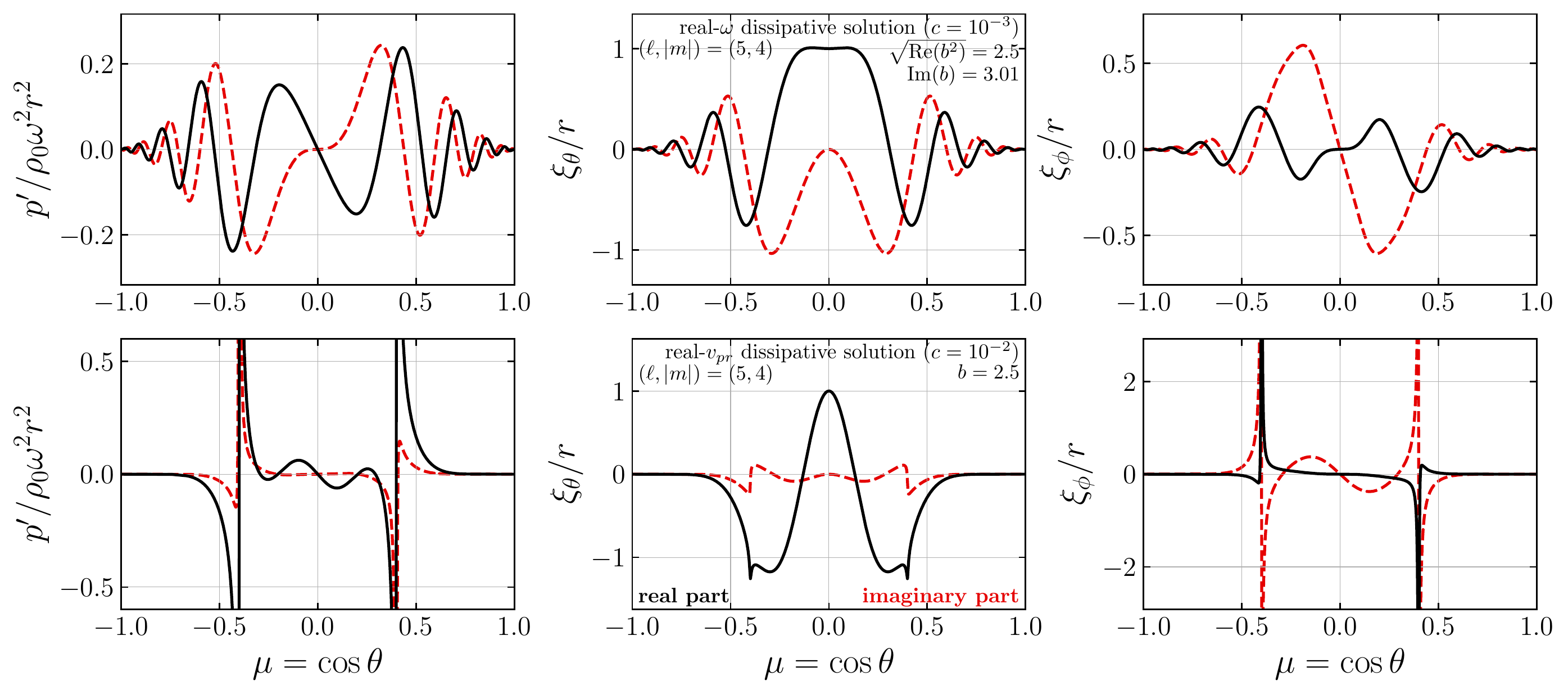}
    \caption{Fluid perturbations for the same mode branches as in Figure \ref{fig:eigenfunctions_2}, but for $\sqrt{\mathrm{Re}(b^2)}=2.5$.}
    \label{fig:eigenfunctions_3}
\end{figure*}

In all branches, for large enough \nzrremoveone{$|b|\simeq\mathrm{few}$}\nzraddone{$|b|\gtrsim\mathrm{few}$}, the imaginary part of $b$ found by the solver becomes large, and $a^{-1}$ dips as the solver follows an evanescent branch deeper into the star.
%Most of the time, $a$ increases monotonically as $|b|$ is increased, although, in very few cases (e.g., $(3,1)$), the solver actually follows the dissipationless solutions back to lower $a$ for a short range of $|b|$ before ``snapping'' to an evanescent branch.
At large $b$, all of the evanescent mode branches we solve for approach $\mathrm{Im}(k_r)/\mathrm{Re}(k_r)=1$ (i.e., $\mathrm{arg}(k_r)\rightarrow+\pi/4$) such that waves radially decay in the same direction as they travel.
%Almost all of the evanescent mode branches we solve for approach $\mathrm{Im}(k_r)/\mathrm{Re}(k_r)=1$ (i.e., $\mathrm{arg}(k_r)\rightarrow+\pi/4$; waves radially decay in the same direction as travel) with the exception of the $(3,1)$ branch, which approaches $\mathrm{Im}(k_r)/\mathrm{Re}(k_r)=-1$ (i.e., $\mathrm{arg}(k_r)\rightarrow-\pi/4$; waves radially decay in the opposite direction of travel) after undergoing the aforementioned ``snapping'' behavior.
In this regime, the eigenfunctions approach horizontally traveling waves which propagate away from the equator (e.g., top panel of Figure \ref{fig:eigenfunctions_3}), \nzraddone{as shown by the relative phases of the real and imaginary eigenfunctions.}
The conjugate branches are expected to have the opposite behavior, with the eigenfunctions approaching horizontally traveling waves which propagate toward the equator.
%, except in the $(3,1)$ case where they travel toward the equator.
%We note that the presence (or absence) of such numerical ``snapping'' behavior and the sign of the imaginary part of $k_r$ in a given branch depends on the value of $c$ adopted---tests at lower $c$ reveal that the presence of snapping and negative $\mathrm{Im}(k_r)$ are not specific to the $(3,1)$ branch.
\nzraddone{Note that, because the values of $b$ (and therefore $k_r$) for these equator-ward and pole-ward traveling solutions are complex conjugates of each other, they exponentially decay in radius in opposite directions, and it is generally not possible to superpose them to form a wavefunction which is a horizontal standing wave at all radii.}

Overall, the behavior at $|b|>1$ is very complex and difficult to characterize from first principles.
Branches often have multiple ``kinks'' in addition to the initial one at $|b|=1$ characterizing the transition from propagation to evanescence.
We suspect these kinks are related to avoided crossings between different evanescent branches of magnetogravity waves.

However, these branches represent modes that are evanescent on short length scales, implying very little wave energy propagates to larger depths. 
Hence, it seems clear that in the dissipative case, there are no propagating mode branches which extend arbitrarily deep into the star.
This extends the two-dimensional results of \citet{lecoanet2016conversion} to non-axisymmetric modes.
\nzraddone{Physically, evanescent waves indicate the presence of either total internal reflection or (in this case) refraction.
\nzrremove{Such waves cannot transport wave energy arbitrarily deeply, and conservation of energy therefore enforces that, in the limit of small dissipation, all of the wave energy must be converted into an outgoing propagating wave.}}
\nzrremove{\nzrremoveone{U}\nzraddone{Thus, u}pon evanescence, the wave energy is refracted onto propagating branches that extend back out towards the surface of the star, discussed in the following section.}
\nzradd{Unless the radial extent of the core is $\lesssim1/\mathrm{Im}(k_r)$, the wave power transmitted by these evanescent waves through the core is vanishingly small, and conservation of energy thereby enforces that the rest of the energy (which is the vast majority) be converted into some kind of outgoing propagating wave (Section \ref{numerical2}).}
%Some branches also increase in $a$ only very slowly, and it is not clear whether they actually approach $a\rightarrow\infty$.

%It is natural to suspect that the high $k_r$ ``slow magnetosonic'' waves found in Section \ref{mnotequalszero} may persist in the dissipative case, where it might exist as a branch of nearly propagating waves that may merge with the fast magnetogravity branch and connect with the fully evanescent mode branches.However, despite searches for eigenfunctions with the expected structure and eigenvalues, we are unable to find these modes.

%\subsection{Real $k_r/\omega$} 
\subsection{Numerical solutions of the propagating branch}\label{numerical2}

\begin{figure}
    \centering
    \includegraphics[width=0.47\textwidth]{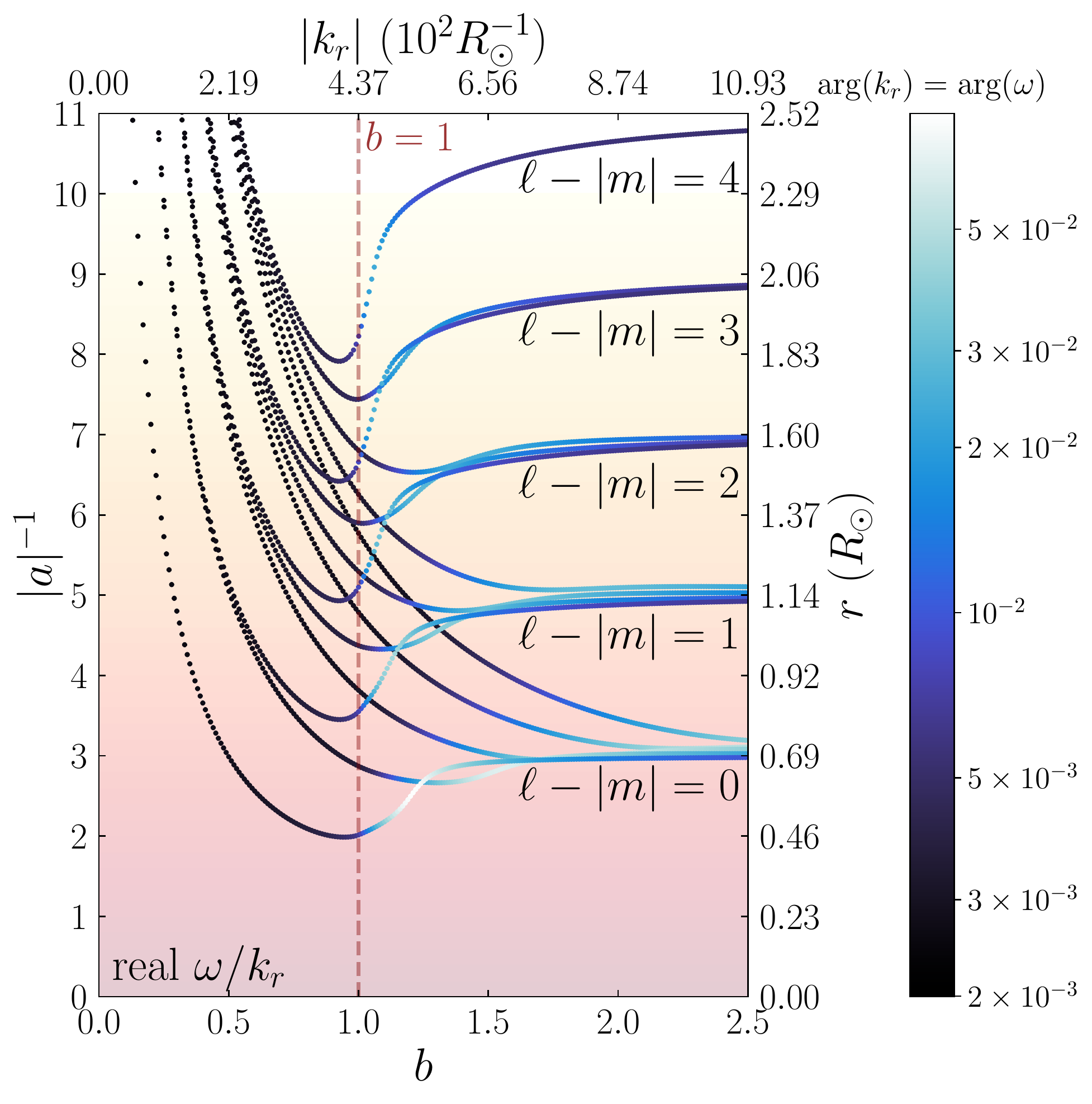}
    \caption{The inverse depth parameter \nzrremoveone{$1/a$}\nzraddone{$a^{-1}$} plotted against $b$ for tesseral/sectoral ($m\neq0$) modes, with finite dissipation and real \nzrremoveone{$k_r/\omega$}\nzraddone{$v_{{\rm p},r}=\omega/k_r$} (Section \ref{numerical2}).
    The color represents the complex argument of $\omega$, which is enforced to be equal to the complex argument of $k_r$.
    Both \nzrremoveone{$1/a$}\nzraddone{$a^{-1}$} and $b$ have been translated to $r$ and $k_r$ as in Figure \ref{fig:fig_evals_0}. As in previous figures, waves follow these tracks to the right as they are converted from gravity waves to \nzraddone{slow magnetic}\nzrremoveone{ Alfv\'enic} waves.}
    \label{fig:fig_evals_3}
\end{figure}

%In the context of physics which is dissipative in nature, it is natural to consider not only spatially evanescent waves but also those which may decay with time. At first glance, 
It is natural to search for solutions where waves are purely propagating (real $k_r$) but $\omega$ is complex (corresponding to decay).
However, we find that our relaxation approach is unable to solve this particular problem formulation.
Instead, we consider the case \nzraddone{where the radial phase velocity $v_{{\rm p},r}=\omega/k_r$ is real, which is equivalent to taking}\nzrremoveone{where} \nzraddone{$b=v_A/v_{{\rm p},r}$} \nzrremoveone{is}\nzraddone{to be} real (placing the singularity as close to the real line as possible), but allowing $a\propto\omega^{-1}$ to be complex (i.e., so that $\mathrm{arg}(k_r)=\mathrm{arg}(\omega)\neq0$).
The eigenvalues for these calculations are shown in Figure \ref{fig:fig_evals_3}.

Interestingly, in this formulation, the eigenvalues have similar qualitative behavior to discontin\nzraddone{u}ous case in Figure \ref{fig:fig_evals_1}.
They reach some maximum $|a|$ at $b\sim1$, at which point the waves turn and propagate outwards onto a slow \nzrremoveone{Alfv\'enic}\nzraddone{magnetic} branch which asymptotes to a finite cutoff height at infinite wavenumber.
This corroborates the basic picture that propagating modes with real (or nearly real) $k_r$ and $\omega$ cannot exist in a strongly magnetized star, and that gravity waves are converted to \nzraddone{slow} \nzrremoveone{Alfv\'enic}\nzraddone{magnetic} waves by strong magnetic fields.
\nzraddone{Table \ref{tablecutoffs} reports the critical and cutoff depths $|a_c|^{-1}$ and $|a_\infty|^{-1}$ for these solutions.}

However, there are some interesting features unique to this problem, which were unanticipated by the discontinuous solutions.
First, the ``cutoff'' values of $|a|$ where the wavenumbers diverge are approximately 
\begin{equation} \label{cutoff2}
    %a_{\mathrm{diss},\ell m}^{(c)} \approx \frac{1}{2(\ell-|m|)+3}
    |a_\infty| \approx \frac{1}{2(\ell-|m|)+3} \, .
\end{equation}
%\noindent although there appears to be some branch-to-branch deviations in $a_{\mathrm{diss},\ell m}^{(c)}$ which prevent this formula from being exact.
This deviates from the expected cutoff heights for $m\neq0$ modes, which are the solutions to Equation \ref{restricted2} and lie close to even numbers rather than odd numbers.
The $m=0$ modes have \nzrremoveone{$a_c^{-1}=2\ell-1$}\nzraddone{$a_\infty^{-1}=2\ell-1$} (Figure \ref{fig:fig_evals_0}), offset by $4$ \nzraddone{(in inverse depth)} for the same values of $\ell-|m|$.
%offset in $\ell-|m|$ by $2$. 
This should not be too surprising\nzrremoveone{,} because\nzraddone{,} at $b\gtrsim1$, the mode eigenfunctions are very different in each case.
The modes described here gain substantial complex parts (unlike the $m=0$ modes), and logarithmic ``spike'' features appear in the real part of $p'$, as shown in the bottom panel of Figure \ref{fig:eigenfunctions_3}.

Figure \ref{fig:fig_evals_3} also shows that the imaginary components of $k_r$ and $\omega$ are largest for $b$ slightly larger than unity, reaching up to $\approx\mathrm{Re}(\omega)/12$ in the $(1,1)$ case.
For larger values of $b$, \nzraddone{the complex arguments of $k_r$ and $\omega$ appear to decrease to roughly constant values of $\mathrm{arg}(\omega)=\mathrm{arg}(k_r)\sim10^{-2}$.
However, due to numerical difficulty, we are unable to confirm this behavior for $b\gtrsim2.5$ or much lower values of $c$.}
\nzrremoveone{the imaginary components of $k_r$ and $\omega$ appear to decrease to a roughly constant value of ${\rm arg}(\omega) \! \sim \! 10^{-1}$ that is nearly independent of viscosity.
This reflects the finite energy dissipation rate even for vanishing viscosity, as discussed above.}

At values of $b$ just above unity, the eigenfunctions behave similarly to the discontinous solutions, with a sharp peak in $\xi_\phi$ and a discontinuity in $\xi_\theta$ at the critical latitude (Figure \ref{fig:eigenfunctions_2}). %features oscillating in phase with a nonzero value at the equator (e.g., bottom panel of Figure \ref{fig:eigenfunctions_2}).
In the dissipationless solutions (Section \ref{mnotequalszero}), we assumed that a given mode oscillates entirely in phase (i.e., each perturbation was either totally real or totally imaginary).
For dissipative modes with $b$ only slightly larger than $1$, this is still true---for example, for the $(5,4)$ mode at $b=1.3$ (lower panels of Figure \ref{fig:eigenfunctions_2}), the delta function feature in $\xi_\phi$ oscillates in phase with the bulk oscillation between the critical latitudes (both are imaginary).
However, at higher values of $b$ (lower panels of Figure \ref{fig:eigenfunctions_3}), the sharp/discontinuous features oscillate $\pi/2$ out of phase with the bulk oscillation (e.g., the delta function in $\xi_\phi$ becomes real).
The spike in $\mathrm{Im}(\omega)$ \nzraddone{(which occurs on the slow magnetic branch)} coincides with a transition between these two regimes.
This latter behavior is not captured by the non-dissipative solution, which assumes that $\xi_\phi$ and $\xi_\theta$ are purely real.
It is thus unsurprising that the cutoff depths \nzrremoveone{$a_c^{-1}$}\nzraddone{$a_\infty^{-1}$} predicted by the non-dissipative solution (Section \ref{mnotequalszero}) do not coincide with those predicted by Equation \ref{cutoff2}.

%Unlike in the real-$\omega$ case (Section \ref{numerical1}), in the real-$(k_r/\omega)$ case problem described above, the singularity is forced to always remain near the real axis since $b$ is forced to be real.
For increasingly small values of $c$, the spike features of the eigenfunctions at the critical latitudes become increasingly sharp and narrow.
This makes calculating the eigenfunctions increasingly numerically challenging for smaller viscosities (we have chosen $c=10^{-2}$ here).
Decreasing $c$ from this value appears to steadily decrease $\mathrm{Im}(\omega)$ for small values of $b$, but only marginally for large values of $b$.
\nzrremoveone{Once again w}\nzraddone{W}e suspect this is due to the finite damping rates that persist at vanishing viscosities\nzraddone{/diffusivities} for waves with these sorts of internal singularities, as discussed \nzrremoveone{above}\nzraddone{earlier}.
%is not entirely clear numerically whether it vanishes entirely in the limit as $c\rightarrow0$.
\nzradd{If true, the upward-propagating branch would also be radially evanescent: this complicates the energetic argument that all initially ingoing wave power must be carried by out by the upward-propagating branch rather than the ingoing evanescent one described in Section \ref{numerical1}.
However, since in this case the damping rate $\mathrm{Im}(\omega)$ remains finite, we believe it is most likely that the wave energy be dissipated on the upward-propagating branch, rather than transmitting through the core.
Moreover, as $k_r\rightarrow0$ on this branch, upward-propagating waves will eventually attain high enough wavenumbers that they should be efficiently damped by even arbitrarily small dissipation $c$: the argument that the wave energy is dissipated in the upward-propagating branch would then be the same as previous.}

%Additionally, we note that the dissipative solutions calculated in Section \ref{numerical1} and this Section (which assumed $\omega\in\textbf{R}$ and $k_r/\omega\in\textbf{R}$, respectively) only enforce two of the infinite number of possible constraints on the complex parts of $k_r$ and $\omega$. The actual magnetogravity oscillations in a real star may lie outside of the two narrow mode subspaces that we have examined, and may have interesting features in their own right. However, we consider exhaustively exploring this space numerically beyond the scope of this particular work.

\section{Further remarks} \label{fr}

\subsection{Behavior of the wavefunctions near the equator and critical latitudes} \label{wkb}

A primary assumption of our analysis is that perturbations vary much faster in the radial direction than the horizontal direction.
This allowed us to effectively decouple the radial dependence of the mode from the horizontal dependence, and solve the latter independently as a two-dimensional problem over the sphere.
\nzraddone{The problem then reduces to a more tractable one-dimensional eigenproblem by making an assumption that the equilibrium field is axisymmetric (although some analytical insight is still available if this assumption is relaxed; see Section \ref{general}).}
For gravity modes at zero field, the ratio of $k_r/k_h=N/\omega$ is large, and this assumption is very reasonable.
This assumption has also been instrumental in defining a hierarchy of variables whereby buoyancy and magnetism contribute at similar strengths to mode restoration (via Equation \ref{cartesian}), and that $k_r$ dominates the magnetic interaction.
However, this hierarchy can be subverted in a few ways.
% \begin{equation}
%     \omega \sim \frac{k_h}{k_r}N \sim k_rv_{Ar}
% \end{equation}

% This hierarchy implies that the mode frequency $\omega$ is much smaller than the buoyancy frequency $N$ (which is typical for gravity waves), but that it is similar to the \textit{radial} contribution to the Alfv\'en frequency.
%However, this hierarchy can be subverted in a few ways.

First, in regions where the magnetic field is nearly horizontal, $v_{Ar}\approx0$, and the magnetic interaction $(\Vec{k}\cdot\Vec{v}_A)^2=k_r^2v_{Ar}^2 + 2k_rk_hv_{Ar}v_{Ah} + k_h^2v_{Ah}^2$ is no longer dominated by the radial part.
The other magnetic terms become comparable when $k_rv_{Ar}\lesssim k_hv_{Ah}$, which is when $v_{Ar}/v_{Ah}\lesssim k_h/k_r\sim\omega/N$.
For a dipole field, this occurs in a very narrow band around the equator with angular extent $\delta\theta\sim\omega/N\ll1$.
It is possible that mode confinement between the critical latitudes found in our work may ``funnel'' refracted magnetogravity waves into radially propagating solutions which may produce detectable surface power in outgoing magnetogravity waves.
We further investigate such equatorially confined magnetogravity waves in Appendix \ref{mgextra}.
\nzraddone{While such waves may exist, they have large horizontal wavenumbers and very large radial wavenumbers, so they may be difficult to observe.}

% In this regime, assuming a large wavenumbers in all direction, the magnetogravity waves follow
% \begin{equation}
%     \omega^2 - \frac{k_h^2}{k_r^2}N^2 - k_h^2v_{Ah}^2 = 0
% \end{equation}

% Then
% \begin{equation}
%     k_r^2 \approx \frac{k_h^2N^2}{\omega^2-k_h^2v_{Ah}^2}
% \end{equation}

% \noindent where, if $k_r\gg k_h$, then $k_r$ is still approximately constant at a given radius.
% It can be seen that, when the mode frequency $|\omega|>|k_hv_{Ah}|$, radially propagating waves confined to an extremely narrow equatorial band could potentially bring some fraction of wave energy past the cutoff height to the surface of the star (where it could potentially be observed).
% Because this wave behavior subverts the variable hierarchy assumed in this paper, the solutions described throughout this work is not expected to capture it accurately.
% However, simulations of magnetogravity waves suggest that subversion of the usual hierarchy can manifest in highly unusual behavior very near the equator, as described above \citep[see Figure 6 in][]{lecoanet2016conversion}.
% In our solutions, outgoing magnetogravity waves are brought to higher and higher wavenumbers until reaching a finite cutoff radius where they are confined to an arbitrarily small equatorial band.
% We conjecture that mode confinement between the critical latitudes may ``funnel'' refracted magnetogravity waves into radially propagating solutions, that may produce detectable surface power in outgoing magnetogravity waves.
% We further explore this idea in Appendix \ref{mgextra}.

The usual hierarchy can also be subverted very near the critical latitudes, where the solutions described in this work attain very sharp horizontal features.
More specifically, our solutions predict that $\xi_\theta$ has a discontinuity and $\xi_\phi$ is a delta function according to the solutions in Section \ref{mnotequalszero}.
While the presence of dissipation (Section \ref{viscous}) may smooth these sharp features somewhat, the sharpness of the features is still cause for concern in realistic stars where these effects are small.
As in the example above, the $k_h v_{Ah}$ terms will become important near the critical latitude and can regula\nzraddone{te}\nzrremoveone{rize} the singularity in our equations.
%In the strong magnetic field problem, the horizontal gradient of the perturbations may be extremely large at the critical latitudes, where $\xi_\theta$ has a discontinuity and $\xi_\phi$ is a delta function according to the solutions in Section \ref{mnotequalszero}.
%In the vicinity of these sharp features, the horizontal derivative terms that we have dropped are expected to become non-negligible.

% We reiterate that one of the most powerful implications of the radial WKB approximation (together with $k_r\gg k_h$) is that $k_r$ is only a function of $r$ (and not $\theta$ and $\phi$).
% Relaxing this approximation means not only restoring horizontal gradient terms containing $k_h$ but allowing all such terms to vary arbitrarily in space.
% However, we note that, when $\omega^2\approx\omega_A^2$, Equation \ref{cartesian} implies that $k_hN/k_r$ is small.
% Therefore, if it is still true that $k_r\gg k_h$ very near the critical latitudes, we can make some rough estimate for the horizontal length scale over which such non-WKB terms cannot be neglected.

Now assuming a WKB approximation in both the radial and horizontal directions and a purely poloidal field ($B_\phi=0$), the horizontal momentum equations become
\begin{align}
    \omega^2\Vec{\xi}_h &= \left(\Vec{k}\cdot\vec{v}_A\right)^2\Vec{\xi}_h \nonumber \\
    &= \left(k_r^2v_{Ar}^2 + 2k_rk_\theta v_{Ar}v_{A\theta} + k_\theta^2v_{A\theta}^2\right)\Vec{\xi}_h
\end{align}

\noindent where we have ignored the pres\nzraddone{s}ure term (note that the Alfv\'en \nzrremoveone{resonances}\nzraddone{waves} which cause the sharp features at the critical latitudes cannot be restored by pressure).

Keeping the dominant terms \nzraddone{(and still assuming $k_r\gg k_h$, $B_r\sim B_h$)},
\begin{equation} \label{vrvh}
    \omega^2 - k_r^2v_{Ar}^2 = 2k_rk_\theta v_{Ar}v_{A\theta}
\end{equation}
Because the left-hand side of Equation \ref{vrvh} is close to zero near the critical latitude, we can perform a Taylor expansion in the horizontal direction:
\begin{equation}
\label{eqn:critalf}
    -k_r^2\frac{\partial v_{Ar}^2}{\partial\theta}\delta\theta \approx 2k_rk_\theta v_{Ar}v_{A\theta} \approx \frac{2k_rv_{Ar}v_{A\theta}}{r\delta\theta}
\end{equation}

\noindent where $\delta\theta$ is the horizontal angular distance from the resonance point where $|\omega|=|k_rv_{Ar}|$.
Here we have assumed that the displacements vary on an angular length scale $\delta\theta$ such that \nzrremoveone{$k_\theta\sim1/r\delta\theta$}\nzraddone{$k_\theta\sim1/(r\delta\theta)$}.
Appendix \ref{nonWKB} solves for the \nzraddone{``}wavefunction\nzraddone{''} \nzraddone{$\Vec{\xi}_h$} more precisely. 

From Equation \ref{eqn:critalf}, we then see that the horizontal field terms terms become important when
\begin{equation}
    |\delta\theta| \lesssim \sqrt{\frac{v_{A\theta}}{k_rr}\left\lvert\frac{\partial v_{Ar}}{\partial\theta}\right\rvert^{-1}}
\end{equation}
However, since $v_{Ar}\sim v_{A\theta}$ typically,
\begin{equation}
    \frac{1}{v_{A\theta}}\frac{\partial v_{Ar}}{\partial\theta} \sim \frac{\partial\ln v_A}{\partial\theta} \sim 1
\end{equation}
\noindent for a large-scale (e.g., dipole) magnetic field.
Therefore, near the critical latitudes, we expect that the wavefunction $\Vec{\xi}_h$ will vary over an angular scale%the threshold at which non-WKB magnetic tension terms begin to matter is
\begin{equation} \label{wkblimit}
    |\delta\theta| \sim \frac{1}{\sqrt{k_rr}}
\end{equation}
This angular scale also naturally appears in Appendix \ref{nonWKB}, where it describes the angular wavenumber of Alfv\'en waves near the critical latitude.
Note that, because magnetogravity waves with $m=0$ have $\xi_\phi=0$, they cannot couple to $m=0$ Alfv\'en waves, which are purely toroidal \citep[i.e., $\xi_\theta=0$;][]{loi2017torsional}.
This \nzrremoveone{heuristically}\nzraddone{physically} explains why sharp fluid features near critical latitudes do not appear in our $m=0$ solutions (Section \ref{mequalszero}), or earlier two-dimensional solutions \citep{lecoanet2016conversion}.
%This coupling cannot occur in two dimensions (or in our $m=0$ solutions), which is why it is not seen in \cite{lecoanet2016conversion}, and why there is no critical latitude for $m=0$ waves.
%the horizontal angular scale at which $\Vec{B}_0$ varies is on the order of a radian.

Physically, the Alfv\'en and magnetogravity waves, which are decoupled in the dispersion relation of Equation \ref{cartesian}, \nzraddone{may} become strongly coupled in a narrow band \nzraddone{due to additional small terms left out of Equation \ref{cartesian}.}
\nzraddone{Because the Alfv\'en waves are expected to have angular scales}\nzrremoveone{of width} $\delta\theta \! \sim \! 1/\sqrt{k_rr}$ \nzraddone{due to the effect of the horizontal field, coupling between Alfv\'en and magnetogravity waves should also occur within $\sim \! \delta\theta$ of a critical latitude (due to geometric overlap)}.
%\nzraddone{In real stars, strong coupling between two modes occurs when their phase velocities overlap, and so resonances between Alfv\'en and magnetogravity modes should in principle occur at only (a large) countable number of discrete points due to mode discretization due to radial boundary conditions (for magnetogravity waves) and periodic boundary conditions along field lines (for Alfv\'en waves) viz. e.g., \citet{loi2017torsional}. This discretization of resonances is not captured in our analysis where we do not impose these boundary conditions.}
\nzrremoveone{This}\nzraddone{This coupling} may allow a small amount of gravity wave power to be converted into outgoing Alfv\'en waves.
\nzraddone{These Alfv\'en waves would then propagate along a closed field line, eventually curving back inwards to the critical latitude on the opposite hemisphere of the star. Here, they could be converted back into outgoing gravity waves, potentially allowing for some wave power to escape the core.}
This possibility could be investigated with numerical simulations.

Additionally, the presence of shear stress would also cause quantifiable departures from the horizontal mode structure derived in this work.
While plasmas do not generally have shear restorative forces, \citet{hoven2011magnetar} argue that tangling in the equilibrium magnetic field at small scales can produce a small effective shear modulus.
\nzraddone{We investigate this possibility further in Appendix \ref{tangling}, finding that it causes the wave function to have an Airy function horizontal dependence near the critical latitude.}

Out of these effects, it is most likely that the horizontal field terms have the largest impact on the mode structure (i.e., $\delta\theta$ as given by Equation \ref{wkblimit} most accurately characterizes when our solutions break down).
Both dissipation and shear stress (due to, e.g., tangling) are likely to be small in real stars, but any physical equilibrium fields must have horizontal magnetic fields $B_h\sim B_r$.

%In general, the breakdown of the radial WKB approximation near the critical latitudes strongly suggests that a search for global solutions with this assumption relaxed is the natural next step for accurately characterizing strong-field $g$ modes.
In general, the importance of horizontal-field terms near the equator and critical latitudes strongly suggests that a search for global solutions with those terms included is the natural next step for accurately characterizing strong-field $g$ modes.
However, the solutions become non-separable in the radial and horizontal directions, and a solution of the full, coupled partial differential equations would be necessary.
A global treatment of magnetogravity modes dramatically increases the complexity of any numerical mode calculations, but is likely to reveal important (and hard to predict) departures from a separable treatment \citep[as it has in eigenmode problems in differentially rotating planets, e.g.,][]{takata2013rosette,dewberry2021constraining}.
While we believe our solutions to capture the basic behavior of the waves, the effects of horizontal magnetic fields discussed here are likely to be the more important effect in real stars, and should be examined more thoroughly in future work.

\subsection{The continuum spectrum and nonharmonic solutions} \label{nonharmonic}

In this work, we have focused on harmonic solutions with time dependence $\propto e^{i\omega t}$, for some global oscillation frequency $\omega$.
However, the unusual nuances introduced by the internal singularity suggest more general approaches may be appropriate.
For example, standard Sturm--Liouville theory only ensures that the eigenfunctions of $\mathcal{L}^{m,b}_{\mathrm{mag}}$ form a basis for a real $b\propto k_r$ in the absence of internal singularities.
Thus, while we have mostly discussed the discrete spectrum of eigenfunctions of $\mathcal{L}^{m,b}_{\mathrm{mag}}$, it is not guaranteed that an arbitrary perturbation can be decomposed into them, both because $b$ is not necessarily real and because different modes at the same radius are eigenfunctions of different differential operators (i.e., $\mathcal{L}^{m,b}_{\mathrm{mag}}$ for different $b$).
In general, the continuous spectrum of Alfv\'en waves (i.e., Section \ref{alfven}) plays a major role.

Similar frequency-dependent internal singularities often appear in problems related to differentially-rotating fluids.
In such problems, authors such as \citet{burger1966instability} and \citet{balbinski1984continuous} apply more general Laplace transform \nzraddone{techniques}\nzrremoveone{methods} involving contour integrals to solve for the time dependence of possible solutions. 
%which are intractable in the general case.
Specifically, \citet{balbinski1984continuous} find that the continuum spectrum in a differentially-rotating cylinder corresponds to perturbations which oscillate periodically and also decay as a power law in time.
In those \nzrremoveone{``quasi''-modal}\nzraddone{``quasi-modal''} solutions, the oscillation frequency depends on position, and hence the solutions are not separable in space and time.

\citet{levin2007theory} and \citet{hoven2011magnetar} intuitively explain the origin of such non-exponential time dependence in the context of the coupling of a magnetar crust mode to an Alfv\'en continuum in the magnetar bulk.
In a toy model analogous to this problem, a ``large'' oscillator (the crust mode) couples to a dense collection of ``small'' oscillators (the Alfv\'en modes). In our case, the ``large" oscillator would be an ingoing gravity wave.
At early times, the large oscillator's amplitude exponentially decays as energy is distributed among the small oscillators.
However, the time dependence transitions to algebraic decay to a finite, nonzero amplitude driven by coherent driving from small oscillators at the edge\nzraddone{s} of the continuum.
In the presence of dissipation, such edge modes retain energy for much longer than modes in the interior of the continuum.
It is unclear how these edge modes manifest in the simplified model of magnetogravity waves presented in this work.

%The origin of this non-exponential time dependence has been intuitively explained in the context of basic toy models.
%In the study of magnetar quasi-periodic oscillations, \citet{levin2007theory} and \citet{hoven2011magnetar} model 

Interestingly, \citet{boyd1981sturm} note in their Appendix B that the decomposition of perturbations into either real-eigenvalue continuum modes or complex-eigenvalue discrete modes are equivalent and complementary approaches. The eigenfunctions of the modes may diverge at some points, similar to an Alfv\'en wave confined to a single field line. However, a superposition of a continuous spectrum of modes can produce a finite-valued function. Hence, examining single continuum modes can be misleading, but they can be superposed to produce unusual decay behavior as in \citet{balbinski1984continuous}. In future work, application of these insights to the magnetogravity wave problem may shed more light on what to expect in real stars, including the possibility of quasi-modes with non-harmonic time dependence.

\subsection{Magnetogravity waves in general geometries} \label{general}

In this work, we have focused on dipolar magnetic field configurations whose radial components have angular dependence $\propto\cos\theta$ (Equation \ref{dipolegeo}).
However, many real stars have more complex field morphologies \citep{maxted2000wd,tout2004magnetic,donati2009magnetic,kochukhov2010extraordinary,szary2013non,kochukhov2016magnetic}.
In this Section, we generalize some of the arguments made in Section \ref{sols} to more general magnetic fields of the form
%\begin{equation}
%    \vec{B} = B_0(r)\psi(\theta,\phi)\hat{r} + \vec{B}_{h0} \sim B_0(r)\psi(\theta,\phi)\hat{r}
%\end{equation}
\begin{equation}
    B_r = B_0(r)\psi(\theta,\phi)
\end{equation}
\noindent where $\psi$ is a dimensionless function describing the horizontal dependence of the field.
As in Section \ref{dipole}, we use a WKB approximation such that terms dependent on the horizontal component of the field are small and can be dropped. %we drop terms horizontal field, which is not expected to be important for gravity waves.
Without loss of generality, we can rescale $\psi$\nzraddone{ and $B_0$} so that the maximum of $|\psi|$ on the sphere is $1$.

The general problem can be non-dimensionalized in the same way as described in Section \ref{nondim}.
In particular, we still define $b$ and $a$ as in Equations \ref{b} and \ref{a}, but interpreting $v_A$ as the maximum Alfv\'en speed at a given radius (which no longer necessarily occurs at the poles).
Via Equations \ref{magnetoboussinesq}, the perturbations are given by
\begin{subequations} \label{pertgen}
    \begin{gather}
        \xi_r = \frac{ik_r}{\rho_0N^2}p' \\
        \vec{\xi}_h = \frac{1}{\rho_0\omega^2r(1-b^2\psi(\theta,\phi)^2)} {\bm \nabla}_hp' \label{xigen} \\
        \rho' = \frac{ik_r}{g}p'
    \end{gather}
\end{subequations}

\noindent where we have defined the horizontal gradient,
\begin{equation}
    {\bm \nabla}_h = \hat{\theta}\frac{\partial}{\partial\theta} + \hat{\phi}\frac{1}{\sin\theta}\frac{\partial}{\partial\phi}
\end{equation}
\noindent with the factor of $1/r$ excluded.

Substituting Equations \ref{pertgen} into the continuity equation (Equation \ref{mb_xi}), we obtain
\begin{equation} \label{diff}
    0 = \frac{b^2}{a^2}p' + {\bm \nabla}_h\cdot\left(\frac{1}{1-b^2\psi^2} {\bm \nabla}_h p'\right)
\end{equation}

We see that Equation \ref{diff} can be viewed as a partial differential equation to be solved over a sphere of radius $a^{-1}$ (i.e., Equation \ref{diff} can be rewritten without $a$ after defining some $\nabla_a=a\nabla$).
In other words, the depth parameter $a$ parameterizes the effective ``curvature'' of the spherical domain over which the horizontal equations are to be solved.
\nzraddone{Since we have not assumed a specific magnetic field geometry here, the form of the differential eigenproblem in Equation \ref{genev} is generic.
In the case of an axisymmetric field, the two-dimensional angular differential operator in Equation \ref{diff} can be reduced to a differential operator in $\theta$ only (recovering, e.g., Equation \ref{mte}, for a dipole field).}

Equation \ref{xigen} can be rearranged to
\begin{equation}
    {\bm \nabla}_hp' = \rho_0\omega^2r\left(1-b^2\psi^2\right)\vec{\xi}_h
\end{equation}
We therefore see that, so long as $\vec{\xi}_h$ is finite, ${\bm \nabla}_h p'=0$ along any critical surface ($|\psi|=\pm1/|b|$) as long as $b$ is real.

The vanishing directional derivative of $p'$ across the critical surface generalizes an analogous result in Section \ref{frob} for the dipole geometry. 
%However, this analysis also reveals that $p'$ must be a constant on a given critical surface (at a given radius),
Physically, this result simply reflects that, at the site of an Alfv\'en resonance, magnetic tension completely accounts for the (horizontal) restoring force of the mode, and the pressure perturbation makes no contribution.
This fact was also used in Section \ref{mnotequalszero} to show that dissipationless $m\neq0$ solutions must vanish outside of the critical latitudes.

We can perform a similar analysis as in Section \ref{mnotequalszero} by multiplying Equation \ref{diff} by $p'^*$ and integrating over the region of the sphere where \nzrremoveone{$b\psi<1$}\nzraddone{$|b\psi|<1$} (i.e., where $\omega<\omega_A$), denoted by $S_<$:
\begin{equation}
    0 = \frac{b^2}{a^2}\int_{S_<}|p'|^2\,\mathrm{d}\Omega + \int_{S_<}p'^*{\bm \nabla}_h\cdot\left(\frac{1}{1-b^2\psi^2}{\bm \nabla}_hp'\right)\mathrm{d}\Omega
\end{equation}

The second term becomes
\begin{equation} \label{aaa}
    \begin{split}
        &\int_{S_<}p'^*{\bm \nabla}_h\cdot\left(\frac{1}{1-b^2\psi^2}{\bm \nabla}_hp'\right)\mathrm{d}\Omega \\
        &= \int_{S_<}{\bm \nabla}_h\cdot\left(\frac{1}{1-b^2\psi^2}p'^*{\bm \nabla}_hp'\right)\mathrm{d}\Omega - \int_{S_<}\frac{1}{1-b^2\psi^2}|{\bm \nabla}_h p'|^2\mathrm{d}\Omega \\
        &= \int_{\partial S_<}\frac{1}{1-b^2\psi^2}p'^*{\bm \nabla}_hp'\cdot\hat{n}\,\mathrm{d}\chi - \int_{S_<}\frac{1}{1-b^2\psi^2}|{\bm \nabla}_h p'|^2\mathrm{d}\Omega \\
    \end{split}
\end{equation}

\noindent where we have first integrated by parts, and then applied the divergence theorem to the first term ($\partial S_<$ denotes the boundary of $S_<$, $\mathrm{d}\chi$ is an angular line element, and $\hat{n}$ points out of $S_<$).
If the first (boundary) term in Equation \ref{aaa} vanishes, then
\begin{equation}
    \frac{b^2}{a^2} = \frac{\int_{S_<}\frac{1}{1-b^2\psi^2}|{\bm \nabla}_hp'|^2\,\mathrm{d}\Omega}{\int_{S_<}|p'|^2\,\mathrm{d}\Omega} > 0
\end{equation}

\noindent generalizes Equation \ref{rayleigh2}.
However, this process can be repeated for $S_>$, the region where \nzrremoveone{$b\psi>1$}\nzraddone{$|b\psi|>1$}, to obtain
\begin{equation}
    \frac{b^2}{a^2} = \frac{\int_{S_>}\frac{1}{1-b^2\psi^2}|{\bm \nabla}_hp'|^2\,\mathrm{d}\Omega}{\int_{S_>}|p'|^2\,\mathrm{d}\Omega} < 0
\end{equation}

Since $b^2/a^2$ may only have one sign or another for a given global mode, we see that modes for which $k_r$ and $\omega$ are both real (i.e., propagating and non-decaying) will be localized to the region where $\omega<\omega_A$ in the case when the boundary term in Equation \ref{aaa} vanishes.
This condition will be satisfied when the complex winding number enclosed by $\partial S_<$ is nonzero, since $p'=\mathrm{const.}$ on $\partial S_<$.
This argument generalizes the result described in Section \ref{mnotequalszero} that propagating, non-decaying $m\neq0$ modes in the dipole geometry must be localized between the critical latitudes.

%This generalizes the result demonstrated in Section \ref{frob} for the dipole geometry, and we highlight that the gradient vanishes not only in parallel to $\nabla_hp'$ but in all directions, i.e., $p'=\mathrm{const.}$.
%This more general result was needed to show in Sections \ref{frob} and \ref{sl} that the dissipationless $m\neq0$ solutions must vanish outside of the critical latitudes in the dipole case.
%It seems natural to suspect that $p'$ may also be forced to vanish on curves with $|\psi|=1/|b|$ in the general problem for some subset of modes, perhaps those which enclose subdomains of the sphere with nonzero complex winding number.

%Of course, the results stated above are analogous to the dipole-case FDSs that we describe in Section \ref{sols}. As we showed in Section \ref{viscous}, the presence of even a small amount of dissipation introduces qualitative departures from the dissipationless case. Those departures are, no doubt, also relevant in the more general geometries discussed in this Section. Nevertheless, formally dissipationless results in the general field case are likely to be similarly informative to the dissipative problem as the dipole-case FDSs were to the dissipative dipole problem.

\subsection{Stable $g$ modes in convective regions} \label{g}

\nzraddone{Standard mixing-length theory assumes a slight superadiabatic temperature gradient such that $N^2<0$ in convective zones.}
While $g$ modes in non-rotating, non-magnetized stars are only present in stably stratified (radiative) regions, \citet{lee1997low} show that buoyancy-restored oscillatory modes (real $\omega$) can be stabilized even in convective regions ($N^2<0$) by sufficiently high rotation.
In particular, when $|\Omega|>|\omega|/2$, the coefficient functions which appear in the Laplace tidal operator $\mathcal{L}^{m,\nu}_{\mathrm{rot}}$ (Equation \ref{lto}) are no longer strictly positive, and it will possess negative eigenvalues $\lambda<0$.
In this case, $k_h=\sqrt{\lambda}/r$ becomes imaginary, and there exist solutions to $k_h/k_r=\omega/N$ when $N$ is also imaginary.
While standard $g$ modes under strong rotation tend to be localized to the equator, these rotationally stabilized convective modes are instead localized near the poles \citep{lee1997low}.

However, the same argument can be applied to the magnetogravity problem\nzraddone{, and gives meaning to the negative branches of eigenvalues $\lambda$ implied by Equation \ref{mte}}.
In particular, like $\mathcal{L}^{m,\nu}_{\mathrm{rot}}$, the operator $\mathcal{L}^{m,b}_{\mathrm{mag}}$ (Equation \ref{mto}) also contains coefficients which switch signs over the domain.
In this formalism, for oscillatory solutions with real $\omega$, $a$ becomes imaginary, and one instead must solve
\begin{equation} \label{mteim}
    \mathcal{L}^{m,b}_{\mathrm{mag}}p' - \frac{b^2}{|a|^2}p' = 0
\end{equation}

\noindent where now the (negative) eigenvalues $\lambda$ of $\mathcal{L}^{m,b}_{\mathrm{mag}}$ must satisfy
\begin{equation}
    \lambda = -b^2/|a|^2 \, .
\end{equation}

\nzraddone{In the case of no buoyancy ($N^2=0$) and relaxing the Boussinesq assumption, convective regions are expected to sustain standard magnetohydrodynamic waves \citep{shu1991physics}.}
\nzraddone{On top of these modes, the aforementioned negative eigenvalue branches}\nzrremoveone{These negative eigenvalues} hint at the existence of buoyancy-\nzrremoveone{driven}\nzraddone{restored} oscillations in convective regions which are stabilized by magnetic forces.
By a similar argument as made in Section \ref{mnotequalszero}, Equation \ref{mteim} implies that such $\lambda<0$ modes would be exactly localized \textit{outside} (rather than inside) the critical latitudes.
Moreover, while they require $|b|>1$, there is no formal upper limit on the magnetic fields at which they can exist, meaning they may exist in the convective cores of strongly magnetized stars.

While the analogy to rotationally stabilized convective modes seems obvious, we note the magnitude of the Brunt--V\"ais\"al\"a frequency $|N|$ is typically extremely close to $0$ in convective zones, owing to the extremely efficient mixing caused by the convective instability.
\nzraddone{Note that this feature is not unique to the magnetogravity problem, and would also be true for the rotational problem considered by \citet{lee1997low}.}
This appears to violate a fundamental assumption of our analysis that $k_r/k_h\sim N/\omega$ is large\nzraddone{, or at least implies that stable convective oscillations which can accurately be described by our formalism must be of very low frequency}.
Therefore, we strongly caution against using the formalism in this work to make quantitative (or even strong qualitative) predictions about the properties of these modes.
More detailed analyses relaxing this assumption are necessary to characterize these modes accurately (if indeed they exist).
%, which violates our radial WKB approximation as well as the hierarchy assumed at the end of Section \ref{lfe}.

\section{Summary} \label{conc}

In this work, we have characterized the pulsation modes 
%(with time dependence $\propto e^{i\omega t}$)
of a spherically symmetric, stratified stellar structure with a strong dipole magnetic field. We focus on radiative zones with large Brunt--V\"ais\"al\"a frequencies such that magnetogravity waves have short radial wavelengths.
We have assumed that

\begin{itemize}
    \item the radial wavelength is everywhere much smaller than both the stellar structure length (the radial WKB approximation) and the horizontal wavelength (i.e., the wavevector is primarily radial),
    %the horizontal wavelength and stellar structure length scale (the radial WKB approximation),
    
    \item oscillations are incompressible and adiabatic,
    
    \item perturbations to the gravitational potential can be ignored (Cowling), and

    \item dissipative processes are either formally absent (Section \ref{sols}) or small (Section \ref{viscous}).
\end{itemize}

Our chief conclusions are as follows:

\begin{enumerate}
\item Propagating zonal ($m=0$) magnetogravity modes merge at a finite field with a branch of slow \nzrremoveone{Alfv\'enic}\nzraddone{magnetic} waves whose wavenumbers diverge at a finite cutoff radius. Their horizontal eigenfunctions are Hough functions for a dipolar magnetic field. Hence, ingoing gravity waves are converted into \nzraddone{slow} \nzrremoveone{Alfv\'enic}\nzraddone{magnetic} waves at a critical magnetic field strength similar to that derived in \cite{fuller2015asteroseismology}. Above this field strength, the modes become evanescent and cannot propagate.
%$a^{(c)}_{\ell0}=1/(2\ell-1)$.
This is in agreement with the results of \citealt{lecoanet2016conversion} in a similar geometry.

\item Propagating sectoral and tesseral ($m\neq0$) modes also merge with branches of slow \nzrremoveone{Alfv\'enic}\nzraddone{magnetic} waves whose wavenumbers diverge at a cutoff radius within the star. Like $m=0$ modes, ingoing gravity waves cannot propagate above a critical magnetic field strength, and are instead converted to outgoing \nzraddone{slow} \nzrremoveone{Alfv\'enic}\nzraddone{magnetic} waves.
%some separate $a^{(c)}_{\ell m}$.
%Their eigenfunctions are only nonzero within critical latitudes defined by the locations of resonances with Alfv\'en waves.
For strong fields and large wavenumbers, the modes are closely confined to the equator, and are bounded by sharp features in the fluid displacement profile at critical latitudes where the wave frequency is resonant with Alfv\'en waves.

\item Even vanishingly small dissipation can cause qualitative deviations from the problem where dissipation is formally set to zero.
This can be heuristically understood because viscosity allows for interaction between magnetogravity waves and the continuous Alfv\'en wave spectrum.
However, even for finite dissipation, the conclusion that sufficiently high magnetic fields will destroy all propagating magnetogravity modes is robust.

\item Near the critical latitudes and equator, magnetic tension terms associated with the horizontal field are likely to affect the mode structure significantly.
Thus, a global solution which includes such terms is necessary to confidently characterize the mode structure very near these regions.
We speculate that such an analysis might reveal that an observable amount of wave power may be able to escape a strongly magnetized stellar core through \nzraddone{coupling with} Alfv\'en waves (at critical latitudes) or extremely localized magnetogravity waves (near the equator).% can bring an observable amount of wave power out of a strongly magnetized stellar core.
%It is possible that wave power may be transferred to Alfv\'en waves (at critical latitudes) or extremely localized magnetogravity waves (near the equator) which may lead to observable consequences.
%This should be investigated further.
%higher-order terms in the radial WKB expansion (associated with the horizontal field)
%(equatorially confined 
%This may allow for some wave power to be exchanged with Alfv\'en waves at critical latitudes.
%Accounting for lower order terms in our WKB analysis can also allow for weak coupling between magnetogravity waves and the Alfv\'en continuum.
%This likely discretizes the modes of the Alfv\'en continuum, and smooths the magnetogravity waves eigenfunctions around critical latitudes.

%\item Strong magnetic fields may be able to stabilize $g$-mode oscillations within convective zones, although it is unclear if such modes would be able to exist in real stars in the manner described by our formalism.
\end{enumerate}

%The analysis presented in this work suggests a few clear directions for further progress.
%As we have shown, the magnetogravity problem admits both scattering to high wavenumber slow magnetosonic waves and coupling to a rapidly dephasing Alfv\'en continuum.
Our analysis reinforces conclusions from earlier studies that strong magnetic fields should convert gravity waves into \nzraddone{slow magnetic}\nzrremoveone{Alfv\'enic} waves that damp within stellar interiors, causing magnetic fields to suppress the amplitudes of gravity modes in red giant stars \citep{fuller2015asteroseismology,stello2016prevalence}.
However, it may remain possible that higher-order WKB terms (neglected in our analysis) or modes with non-harmonic time dependence (Section \ref{nonharmonic}) could allow for some signatures of mixed modes in observed power spectra as claimed by \citet{mosser2017dipole}.
%More effort (in, e.g., translating this work to observable consequences in mode frequencies and amplitudes) will be required to determine how much each of these two processes contribute to the suppressed dipole modes observed in many red giants.
More effort accounting for these effects will be required to robustly predict the magnetogravity pulsation spectra of stars with strong magnetic fields.
%Additionally, while the assumptions that we have made in this text are generally applicable to realistic red giant cores, the sharp latitudinal features found in the eigenfunctions are in tension with the assumption that the radial wavenumber dominates over the horizontal wavenumber. Relaxing this assumption may be important in fully understanding these modes, but necessitates the development of dramatically different numerical/analytical approaches.

\section*{Acknowledgements}

We thank Daniel Lecoanet, Yuri Levin, Sterl Phinney, Janosz Dewberry\nzraddone{, Joel Ong}, and Saul Teukolsky for their helpful advice and comments.
N.Z.R. acknowledges support from the National Science Foundation Graduate Research Fellowship under Grant No. DGE‐1745301.
J.F. is thankful for support through an Innovator Grant from The Rose Hills Foundation, and the Sloan Foundation through grant FG-2018-10515.
\nzraddone{We thank
the anonymous referee for their thorough review and helpful suggestions, which greatly improved the work.}

\section*{Data Availability}

The output of the oscillation mode calculations described in this work will be shared upon reasonable request to the corresponding author.

\bibliographystyle{mnras}
\bibliography{mnras_template}

\begin{thebibliography}{}
\makeatletter
\relax
\def\mn@urlcharsother{\let\do\@makeother \do\$\do\&\do\#\do\^\do\_\do\%\do\~}
\def\mn@doi{\begingroup\mn@urlcharsother \@ifnextchar [ {\mn@doi@}
  {\mn@doi@[]}}
\def\mn@doi@[#1]#2{\def\@tempa{#1}\ifx\@tempa\@empty \href
  {http://dx.doi.org/#2} {doi:#2}\else \href {http://dx.doi.org/#2} {#1}\fi
  \endgroup}
\def\mn@eprint#1#2{\mn@eprint@#1:#2::\@nil}
\def\mn@eprint@arXiv#1{\href {http://arxiv.org/abs/#1} {{\tt arXiv:#1}}}
\def\mn@eprint@dblp#1{\href {http://dblp.uni-trier.de/rec/bibtex/#1.xml}
  {dblp:#1}}
\def\mn@eprint@#1:#2:#3:#4\@nil{\def\@tempa {#1}\def\@tempb {#2}\def\@tempc
  {#3}\ifx \@tempc \@empty \let \@tempc \@tempb \let \@tempb \@tempa \fi \ifx
  \@tempb \@empty \def\@tempb {arXiv}\fi \@ifundefined
  {mn@eprint@\@tempb}{\@tempb:\@tempc}{\expandafter \expandafter \csname
  mn@eprint@\@tempb\endcsname \expandafter{\@tempc}}}

\bibitem[\protect\citeauthoryear{Al-Gwaiz}{Al-Gwaiz}{2008}]{al2008sturm}
Al-Gwaiz M.~A.,  2008, Sturm-Liouville theory and its applications.
~ Vol. 264, Springer

\bibitem[\protect\citeauthoryear{Angel}{Angel}{1977}]{angel1977magnetism}
Angel J.,  1977, The Astrophysical Journal, 216, 1

\bibitem[\protect\citeauthoryear{{Appert}, {Gruber}  \& {Vaclavik}}{{Appert}
  et~al.}{1974}]{appert1974continuous}
{Appert} K.,  {Gruber} R.,   {Vaclavik} J.,  1974, \mn@doi [Physics of Fluids]
  {10.1063/1.1694918}, \href
  {https://ui.adsabs.harvard.edu/abs/1974PhFl...17.1471A} {17, 1471}

\bibitem[\protect\citeauthoryear{Appert, Gruber  \& Vaclavik}{Appert
  et~al.}{1998}]{appert1998continuous}
Appert K.,  Gruber R.,   Vaclavik J.,  1998, Technical report, Continuous
  spectra of a cylindrical magnetohydrodynamic equilibrium: the derivation

\bibitem[\protect\citeauthoryear{Atkinson, Everitt  \& Zettl}{Atkinson
  et~al.}{1988}]{atkinson1988regularization}
Atkinson F.,  Everitt W.,   Zettl A.,  1988, Differential and Integral
  Equations, 1, 213

\bibitem[\protect\citeauthoryear{Balbinski}{Balbinski}{1984}]{balbinski1984continuous}
Balbinski E.,  1984, Monthly Notices of the Royal Astronomical Society, 209,
  145

\bibitem[\protect\citeauthoryear{Baliunas, Nesme-Ribes, Sokoloff  \&
  Soon}{Baliunas et~al.}{1996}]{baliunas1996dynamo}
Baliunas S.,  Nesme-Ribes E.,  Sokoloff D.,   Soon W.,  1996, The Astrophysical
  Journal, 460, 848

\bibitem[\protect\citeauthoryear{{Bildsten}, {Ushomirsky}  \&
  {Cutler}}{{Bildsten} et~al.}{1996}]{bildsten1996ocean}
{Bildsten} L.,  {Ushomirsky} G.,   {Cutler} C.,  1996, \mn@doi [\apj]
  {10.1086/177012}, \href
  {https://ui.adsabs.harvard.edu/abs/1996ApJ...460..827B} {460, 827}

\bibitem[\protect\citeauthoryear{Boyd}{Boyd}{1976}]{boyd1976thesis}
Boyd J.~P.,  1976, Ph. D. Thesis

\bibitem[\protect\citeauthoryear{Boyd}{Boyd}{1981}]{boyd1981sturm}
Boyd J.~P.,  1981, Journal of Mathematical Physics, 22, 1575

\bibitem[\protect\citeauthoryear{Boyd}{Boyd}{1982}]{boyd1982influence}
Boyd J.~P.,  1982, Journal of Atmospheric Sciences, 39, 770

\bibitem[\protect\citeauthoryear{Braithwaite \& Spruit}{Braithwaite \&
  Spruit}{2004}]{braithwaite2004fossil}
Braithwaite J.,  Spruit H.~C.,  2004, nature, 431, 819

\bibitem[\protect\citeauthoryear{Brun \& Browning}{Brun \&
  Browning}{2017}]{brun2017magnetism}
Brun A.~S.,  Browning M.~K.,  2017, Living Reviews in Solar Physics, 14, 1

\bibitem[\protect\citeauthoryear{Buckingham}{Buckingham}{1914}]{buckingham1914physically}
Buckingham E.,  1914, Physical review, 4, 345

\bibitem[\protect\citeauthoryear{Bugnet}{Bugnet}{2022}]{bugnet2022magnetic}
Bugnet L.,  2022, Astronomy \& Astrophysics, 667, A68

\bibitem[\protect\citeauthoryear{Bugnet et~al.,}{Bugnet
  et~al.}{2021}]{bugnet2021magnetic}
Bugnet L.,  et~al., 2021, Astronomy \& Astrophysics, 650, A53

\bibitem[\protect\citeauthoryear{Burger}{Burger}{1966}]{burger1966instability}
Burger A.,  1966, Journal of Atmospheric Sciences, 23, 272

\bibitem[\protect\citeauthoryear{Cantiello, Fuller  \& Bildsten}{Cantiello
  et~al.}{2016}]{cantiello2016asteroseismic}
Cantiello M.,  Fuller J.,   Bildsten L.,  2016, The Astrophysical Journal, 824,
  14

\bibitem[\protect\citeauthoryear{Chen \& L{\"u}}{Chen \&
  L{\"u}}{2009}]{chen2009calculation}
Chen Z.-Y.,  L{\"u} D.-R.,  2009, Chinese Journal of Geophysics, 52, 547

\bibitem[\protect\citeauthoryear{Dewberry, Mankovich, Fuller, Lai  \&
  Xu}{Dewberry et~al.}{2021}]{dewberry2021constraining}
Dewberry J.~W.,  Mankovich C.~R.,  Fuller J.,  Lai D.,   Xu W.,  2021, The
  Planetary Science Journal, 2, 198

\bibitem[\protect\citeauthoryear{Dhouib, Mathis, Bugnet, Van~Reeth  \&
  Aerts}{Dhouib et~al.}{2022}]{dhouib2022detecting}
Dhouib H.,  Mathis S.,  Bugnet L.,  Van~Reeth T.,   Aerts C.,  2022, Astronomy
  \& Astrophysics, 661, A133

\bibitem[\protect\citeauthoryear{Donati \& Landstreet}{Donati \&
  Landstreet}{2009}]{donati2009magnetic}
Donati J.,  Landstreet J.,  2009, arXiv preprint arXiv:0904.1938

\bibitem[\protect\citeauthoryear{Dudorov \& Khaibrakhmanov}{Dudorov \&
  Khaibrakhmanov}{2015}]{dudorov2015theory}
Dudorov A.~E.,  Khaibrakhmanov S.~A.,  2015, Advances in Space Research, 55,
  843

\bibitem[\protect\citeauthoryear{Everitt, Gunson  \& Zettl}{Everitt
  et~al.}{1987}]{everitt1987some}
Everitt W.,  Gunson J.,   Zettl A.,  1987, Zeitschrift f{\"u}r angewandte
  Mathematik und Physik ZAMP, 38, 813

\bibitem[\protect\citeauthoryear{Federman}{Federman}{1911}]{federman1911some}
Federman A.,  1911, Proc. St.-Petersburg Polytech. Inst. Sect. Tech. Nat. Sci.
  Math, 16, 97

\bibitem[\protect\citeauthoryear{Ferrario, Pringle, Tout  \&
  Wickramasinghe}{Ferrario et~al.}{2009}]{ferrario2009origin}
Ferrario L.,  Pringle J.,  Tout C.~A.,   Wickramasinghe D.,  2009, Monthly
  Notices of the Royal Astronomical Society: Letters, 400, L71

\bibitem[\protect\citeauthoryear{Ferrario, Melatos  \& Zrake}{Ferrario
  et~al.}{2015}]{ferrario2015magnetic}
Ferrario L.,  Melatos A.,   Zrake J.,  2015, Space Science Reviews, 191, 77

\bibitem[\protect\citeauthoryear{Fuller \& Lai}{Fuller \&
  Lai}{2014}]{fuller2014dynamical}
Fuller J.,  Lai D.,  2014, Monthly Notices of the Royal Astronomical Society,
  444, 3488

\bibitem[\protect\citeauthoryear{Fuller, Cantiello, Stello, Garcia  \&
  Bildsten}{Fuller et~al.}{2015}]{fuller2015asteroseismology}
Fuller J.,  Cantiello M.,  Stello D.,  Garcia R.~A.,   Bildsten L.,  2015,
  Science, 350, 423

\bibitem[\protect\citeauthoryear{Garc{\'\i}a et~al.,}{Garc{\'\i}a
  et~al.}{2014}]{garcia2014study}
Garc{\'\i}a R.,  et~al., 2014, Astronomy \& Astrophysics, 563, A84

\bibitem[\protect\citeauthoryear{Goedbloed \& Poedts}{Goedbloed \&
  Poedts}{2004}]{goedbloed2004principles}
Goedbloed H.,  Poedts S.,  2004, Principles of magnetohydrodynamics: with
  applications to laboratory and astrophysical plasmas.
Cambridge university press

\bibitem[\protect\citeauthoryear{Goossens}{Goossens}{1972}]{goossens1972perturbation}
Goossens M.,  1972, Astrophysics and Space Science, 16, 386

\bibitem[\protect\citeauthoryear{Goossens}{Goossens}{1976}]{goossens1976non}
Goossens M.,  1976, Astrophysics and Space Science, 43, 9

\bibitem[\protect\citeauthoryear{Goossens, Smeyers  \& Denis}{Goossens
  et~al.}{1976}]{goossens1976stellar}
Goossens M.,  Smeyers P.,   Denis J.,  1976, Astrophysics and Space Science,
  39, 257

\bibitem[\protect\citeauthoryear{Gunson}{Gunson}{1987}]{gunson1987perturbation}
Gunson J.,  1987, Proceedings of the Royal Society of London. A. Mathematical
  and Physical Sciences, 414, 255

\bibitem[\protect\citeauthoryear{Homer}{Homer}{1990}]{homer1990boundary}
Homer M.,  1990, Proceedings of the Royal Society of London. A. Mathematical
  and Physical Sciences, 428, 157

\bibitem[\protect\citeauthoryear{Houdek \& Dupret}{Houdek \&
  Dupret}{2015}]{houdek2015interaction}
Houdek G.,  Dupret M.-A.,  2015, Living Reviews in Solar Physics, 12, 1

\bibitem[\protect\citeauthoryear{Hough}{Hough}{1898a}]{hough1898v}
Hough S.~S.,  1898a, Philosophical Transactions of the Royal Society of London.
  Series A, Containing Papers of a Mathematical or Physical Character, pp
  139--185

\bibitem[\protect\citeauthoryear{Hough}{Hough}{1898b}]{hough1898application}
Hough S.~S.,  1898b, Philosophical Transactions of the Royal Society of London.
  Series A, Containing Papers of a Mathematical or Physical Character, 191, 139

\bibitem[\protect\citeauthoryear{Hoven \& Levin}{Hoven \&
  Levin}{2011}]{hoven2011magnetar}
Hoven M.~v.,  Levin Y.,  2011, Monthly Notices of the Royal Astronomical
  Society, 410, 1036

\bibitem[\protect\citeauthoryear{Kochukhov \& Wade}{Kochukhov \&
  Wade}{2016}]{kochukhov2016magnetic}
Kochukhov O.,  Wade G.,  2016, Astronomy \& Astrophysics, 586, A30

\bibitem[\protect\citeauthoryear{Kochukhov, Lundin, Romanyuk  \&
  Kudryavtsev}{Kochukhov et~al.}{2010}]{kochukhov2010extraordinary}
Kochukhov O.,  Lundin A.,  Romanyuk I.,   Kudryavtsev D.,  2010, The
  Astrophysical Journal, 726, 24

\bibitem[\protect\citeauthoryear{Kulkarni \& Thompson}{Kulkarni \&
  Thompson}{1998}]{kulkarni1998star}
Kulkarni S.,  Thompson C.,  1998, Nature, 393, 215

\bibitem[\protect\citeauthoryear{Lecoanet, Vasil, Fuller, Cantiello  \&
  Burns}{Lecoanet et~al.}{2017}]{lecoanet2016conversion}
Lecoanet D.,  Vasil G.~M.,  Fuller J.,  Cantiello M.,   Burns K.~J.,  2017,
  \mn@doi [Monthly Notices of the Royal Astronomical Society]
  {10.1093/mnras/stw3273}, 466, 2181

\bibitem[\protect\citeauthoryear{Lecoanet, Bowman  \& Van~Reeth}{Lecoanet
  et~al.}{2022}]{lecoanet2022asteroseismic}
Lecoanet D.,  Bowman D.~M.,   Van~Reeth T.,  2022, Monthly Notices of the Royal
  Astronomical Society: Letters, 512, L16

\bibitem[\protect\citeauthoryear{Lee \& Saio}{Lee \& Saio}{1997}]{lee1997low}
Lee U.,  Saio H.,  1997, The Astrophysical Journal, 491, 839

\bibitem[\protect\citeauthoryear{Levin}{Levin}{2006}]{levin2006qpos}
Levin Y.,  2006, Monthly Notices of the Royal Astronomical Society: Letters,
  368, L35

\bibitem[\protect\citeauthoryear{Levin}{Levin}{2007}]{levin2007theory}
Levin Y.,  2007, Monthly Notices of the Royal Astronomical Society, 377, 159

\bibitem[\protect\citeauthoryear{Li, Deheuvels, Ballot  \& Lignières}{Li
  et~al.}{2022}]{gang2022topology}
Li G.,  Deheuvels S.,  Ballot J.,   Lignières F.,  2022, 30 to 100-kG magnetic
  fields in the cores of red giant stars, \mn@doi{10.48550/ARXIV.2208.09487},
  \url {https://arxiv.org/abs/2208.09487}

\bibitem[\protect\citeauthoryear{Liebert, Bergeron  \& Holberg}{Liebert
  et~al.}{2003}]{liebert2003true}
Liebert J.,  Bergeron P.,   Holberg J.,  2003, The astronomical journal, 125,
  348

\bibitem[\protect\citeauthoryear{Loi}{Loi}{2020a}]{loi2020magneto}
Loi S.~T.,  2020a, Monthly Notices of the Royal Astronomical Society, 493, 5726

\bibitem[\protect\citeauthoryear{Loi}{Loi}{2020b}]{loi2020effect}
Loi S.~T.,  2020b, Monthly Notices of the Royal Astronomical Society, 496, 3829

\bibitem[\protect\citeauthoryear{Loi \& Papaloizou}{Loi \&
  Papaloizou}{2017}]{loi2017torsional}
Loi S.~T.,  Papaloizou J.~C.,  2017, Monthly Notices of the Royal Astronomical
  Society, 467, 3212

\bibitem[\protect\citeauthoryear{Loi \& Papaloizou}{Loi \&
  Papaloizou}{2018}]{loi2018effects}
Loi S.~T.,  Papaloizou J.~C.,  2018, Monthly Notices of the Royal Astronomical
  Society, 477, 5338

\bibitem[\protect\citeauthoryear{MacGregor \& Rogers}{MacGregor \&
  Rogers}{2011}]{macgregor2011reflection}
MacGregor K.~B.,  Rogers T.,  2011, Solar Physics, 270, 417

\bibitem[\protect\citeauthoryear{Maeder \& Meynet}{Maeder \&
  Meynet}{2005}]{maeder2005stellar}
Maeder A.,  Meynet G.,  2005, Astronomy \& Astrophysics, 440, 1041

\bibitem[\protect\citeauthoryear{Mathieu}{Mathieu}{1868}]{mathieu1868memoire}
Mathieu {\'E}.,  1868, Journal de math{\'e}matiques pures et appliqu{\'e}es,
  13, 137

\bibitem[\protect\citeauthoryear{Mathis \& De~Brye}{Mathis \&
  De~Brye}{2011}]{mathis2011low}
Mathis S.,  De~Brye N.,  2011, Astronomy \& Astrophysics, 526, A65

\bibitem[\protect\citeauthoryear{Mathis, Bugnet, Prat, Augustson, Mathur  \&
  Garcia}{Mathis et~al.}{2021}]{mathis2021probing}
Mathis S.,  Bugnet L.,  Prat V.,  Augustson K.,  Mathur S.,   Garcia R.~A.,
  2021, Astronomy \& Astrophysics, 647, A122

\bibitem[\protect\citeauthoryear{Maxted, Ferrario, Marsh  \&
  Wickramasinghe}{Maxted et~al.}{2000}]{maxted2000wd}
Maxted P.,  Ferrario L.,  Marsh T.,   Wickramasinghe D.,  2000, Monthly Notices
  of the Royal Astronomical Society, 315, L41

\bibitem[\protect\citeauthoryear{Mosser et~al.,}{Mosser
  et~al.}{2012}]{mosser2012characterization}
Mosser B.,  et~al., 2012, Astronomy \& Astrophysics, 537, A30

\bibitem[\protect\citeauthoryear{Mosser et~al.,}{Mosser
  et~al.}{2017}]{mosser2017dipole}
Mosser B.,  et~al., 2017, Astronomy \& Astrophysics, 598, A62

\bibitem[\protect\citeauthoryear{Pint{\'e}r, Erd{\'e}lyi  \&
  Goossens}{Pint{\'e}r et~al.}{2007}]{pinter2007global}
Pint{\'e}r B.,  Erd{\'e}lyi R.,   Goossens M.,  2007, Astronomy \&
  Astrophysics, 466, 377

\bibitem[\protect\citeauthoryear{Poedts, Hermans  \& Goossens}{Poedts
  et~al.}{1985}]{poedts1985continuous}
Poedts S.,  Hermans D.,   Goossens M.,  1985, Astronomy and astrophysics, 151,
  16

\bibitem[\protect\citeauthoryear{Prat, Mathis, Buysschaert, Van~Beeck, Bowman,
  Aerts  \& Neiner}{Prat et~al.}{2019}]{prat2019period}
Prat V.,  Mathis S.,  Buysschaert B.,  Van~Beeck J.,  Bowman D.~M.,  Aerts C.,
   Neiner C.,  2019, Astronomy \& Astrophysics, 627, A64

\bibitem[\protect\citeauthoryear{Prat, Mathis, Neiner, Van~Beeck, Bowman  \&
  Aerts}{Prat et~al.}{2020}]{prat2020period}
Prat V.,  Mathis S.,  Neiner C.,  Van~Beeck J.,  Bowman D.~M.,   Aerts C.,
  2020, \mn@doi [A\&A] {10.1051/0004-6361/201937398}, 636, A100

\bibitem[\protect\citeauthoryear{Prendergast}{Prendergast}{1956}]{prendergast1956equilibrium}
Prendergast K.~H.,  1956, The Astrophysical Journal, 123, 498

\bibitem[\protect\citeauthoryear{Press, Teukolsky, Vetterling  \&
  Flannery}{Press et~al.}{2007}]{press2007numerical}
Press W.~H.,  Teukolsky S.~A.,  Vetterling W.~T.,   Flannery B.~P.,  2007,
  Numerical Recipes with Source Code CD-ROM 3rd Edition: The Art of Scientific
  Computing.
Cambridge University Press

\bibitem[\protect\citeauthoryear{Proctor \& Weiss}{Proctor \&
  Weiss}{1982}]{proctor1982magnetoconvection}
Proctor M.,  Weiss N.,  1982, Reports on Progress in Physics, 45, 1317

\bibitem[\protect\citeauthoryear{Rauf \& Tataronis}{Rauf \&
  Tataronis}{1995}]{rauf1995alfven}
Rauf S.,  Tataronis J.,  1995, Physics of Plasmas, 2, 340

\bibitem[\protect\citeauthoryear{Reese, Rincon  \& Rieutord}{Reese
  et~al.}{2004}]{reese2004oscillations}
Reese D.,  Rincon F.,   Rieutord M.,  2004, Astronomy \& Astrophysics, 427, 279

\bibitem[\protect\citeauthoryear{Riabouchinsky}{Riabouchinsky}{1911}]{riabouchinsky1911methode}
Riabouchinsky D.,  1911, L’a{\'e}rophile, 1, 407

\bibitem[\protect\citeauthoryear{Rincon \& Rieutord}{Rincon \&
  Rieutord}{2003}]{rincon2003oscillations}
Rincon F.,  Rieutord M.,  2003, Astronomy \& Astrophysics, 398, 663

\bibitem[\protect\citeauthoryear{Rogers \& MacGregor}{Rogers \&
  MacGregor}{2010}]{rogers2010interaction}
Rogers T.,  MacGregor K.,  2010, Monthly Notices of the Royal Astronomical
  Society, 401, 191

\bibitem[\protect\citeauthoryear{Schneider, Ohlmann, Podsiadlowski, R{\"o}pke,
  Balbus, Pakmor  \& Springel}{Schneider et~al.}{2019}]{schneider2019stellar}
Schneider F.,  Ohlmann S.~T.,  Podsiadlowski P.,  R{\"o}pke F.~K.,  Balbus
  S.~A.,  Pakmor R.,   Springel V.,  2019, Nature, 574, 211

\bibitem[\protect\citeauthoryear{Shu}{Shu}{1991}]{shu1991physics}
Shu F.~H.,  1991, The Physics of Astrophysics: Gas Dynamics.
~ Vol. 2, University Science Books

\bibitem[\protect\citeauthoryear{Spruit}{Spruit}{2002}]{spruit2002dynamo}
Spruit H.,  2002, Astronomy \& Astrophysics, 381, 923

\bibitem[\protect\citeauthoryear{Stello, Cantiello, Fuller, Garcia  \&
  Huber}{Stello et~al.}{2016a}]{stello2016suppression}
Stello D.,  Cantiello M.,  Fuller J.,  Garcia R.~A.,   Huber D.,  2016a,
  Publications of the Astronomical Society of Australia, 33, e011

\bibitem[\protect\citeauthoryear{Stello, Cantiello, Fuller, Huber, Garc{\'\i}a,
  Bedding, Bildsten  \& Aguirre}{Stello et~al.}{2016b}]{stello2016prevalence}
Stello D.,  Cantiello M.,  Fuller J.,  Huber D.,  Garc{\'\i}a R.~A.,  Bedding
  T.~R.,  Bildsten L.,   Aguirre V.~S.,  2016b, Nature, 529, 364

\bibitem[\protect\citeauthoryear{Szary}{Szary}{2013}]{szary2013non}
Szary A.,  2013, arXiv preprint arXiv:1304.4203

\bibitem[\protect\citeauthoryear{Takata \& Saio}{Takata \&
  Saio}{2013}]{takata2013rosette}
Takata M.,  Saio H.,  2013, Publications of the Astronomical Society of Japan,
  65

\bibitem[\protect\citeauthoryear{Thompson \& Duncan}{Thompson \&
  Duncan}{1993}]{thompson1993neutron}
Thompson C.,  Duncan R.~C.,  1993, The Astrophysical Journal, 408, 194

\bibitem[\protect\citeauthoryear{Tout, Wickramasinghe  \& Ferrario}{Tout
  et~al.}{2004}]{tout2004magnetic}
Tout C.~A.,  Wickramasinghe D.~T.,   Ferrario L.,  2004, Monthly Notices of the
  Royal Astronomical Society, 355, L13

\bibitem[\protect\citeauthoryear{Townsend}{Townsend}{2003}]{townsend2003asymptotic}
Townsend R.,  2003, Monthly Notices of the Royal Astronomical Society, 340,
  1020

\bibitem[\protect\citeauthoryear{Townsend}{Townsend}{2020}]{townsend2020improved}
Townsend R.,  2020, Monthly Notices of the Royal Astronomical Society, 497,
  2670

\bibitem[\protect\citeauthoryear{Tutukov \& Fedorova}{Tutukov \&
  Fedorova}{2010}]{tutukov2010possible}
Tutukov A.,  Fedorova A.,  2010, Astronomy Reports, 54, 156

\bibitem[\protect\citeauthoryear{Unno, Osaki, Ando, Saio  \& Shibahashi}{Unno
  et~al.}{1989}]{unno1989nonradial}
Unno W.,  Osaki Y.,  Ando H.,  Saio H.,   Shibahashi H.,  1989, Unno, W.,
  Osaki, Y., Ando, H., Saio, H., \& Shibahashi, H, 755

\bibitem[\protect\citeauthoryear{Vaschy}{Vaschy}{1892}]{vaschy1892lois}
Vaschy A.,  1892, in Annales t{\'e}l{\'e}graphiques. pp 25--28

\bibitem[\protect\citeauthoryear{Vidotto et~al.,}{Vidotto
  et~al.}{2014}]{vidotto2014stellar}
Vidotto A.,  et~al., 2014, Monthly Notices of the Royal Astronomical Society,
  441, 2361

\bibitem[\protect\citeauthoryear{Wang, Boyd  \& Akmaev}{Wang
  et~al.}{2016}]{wang2016computation}
Wang H.,  Boyd J.~P.,   Akmaev R.~A.,  2016, Geoscientific Model Development,
  9, 1477

\bibitem[\protect\citeauthoryear{Whittaker}{Whittaker}{1903}]{whittaker1903expression}
Whittaker E.~T.,  1903

\bibitem[\protect\citeauthoryear{Wickramasinghe \& Ferrario}{Wickramasinghe \&
  Ferrario}{2000}]{wickramasinghe2000magnetism}
Wickramasinghe D.,  Ferrario L.,  2000, Publications of the Astronomical
  Society of the Pacific, 112, 873

\bibitem[\protect\citeauthoryear{Wickramasinghe, Tout  \&
  Ferrario}{Wickramasinghe et~al.}{2014}]{wickramasinghe2014most}
Wickramasinghe D.~T.,  Tout C.~A.,   Ferrario L.,  2014, Monthly Notices of the
  Royal Astronomical Society, 437, 675

\bibitem[\protect\citeauthoryear{Widdowson, Hurricane  \& Cowley}{Widdowson
  et~al.}{1998}]{widdowson1998continuum}
Widdowson S.,  Hurricane O.,   Cowley S.,  1998, Physics of Plasmas, 5, 1259

\makeatother
\end{thebibliography}

\appendix

\section{Magnetogravity eigenproblems in other geometries} \label{toys}

In this Appendix, we non-dimensionalize the fluid equations for the geometries considered by \citet{fuller2015asteroseismology} and \citet{lecoanet2016conversion}, and show that they can be interpreted as similar eigenvalue problems as considered in our work.

\subsection{\citet{fuller2015asteroseismology}: Uniform radial field model}

The model presented by \citet{fuller2015asteroseismology} can be precisely reproduced by adopting a purely uniform radial magnetic field,
\begin{equation} \label{monopole}
    \vec{B}_0=B_0(r)\,\hat{r}
\end{equation}

\nzraddone{Adopting a WKB approximation in the radial direction, w}\nzrremoveone{W}e can define
\begin{subequations} \label{fullerab}
    \begin{gather}
        b_{\mathrm{F}} = \frac{k_rv_A}{\omega} \\
        a_{\mathrm{F}} = \left(\frac{N}{\omega}\right)\left(\frac{v_A/r}{\omega}\right)
    \end{gather}
\end{subequations}

\noindent where \nzrremoveone{$v_{A,0}=B_0(r)/\sqrt{4\pi\rho_0}$}\nzraddone{$v_A=B_0(r)/\sqrt{4\pi\rho_0}$}.

While such a monopolar field is clearly unphysical, it is also a useful toy model because it retains the spherical symmetry of the zero-field problem.
Therefore, its horizontal eigenfunctions are simply spherical harmonics, and its dispersion relation is
\begin{equation} \label{fullerdisp}
    \frac{\ell(\ell+1)}{1-b_{\mathrm{F}}^2} = b_{\mathrm{F}}^2/a_{\mathrm{F}}^2
\end{equation}

\nzrremoveone{where this time $v_{A,0}$ is defined with respect to $B_0(r)$ in Equation \ref{monopole}.}
Equation \ref{fullerdisp} can be analytically solved for \nzraddone{$b_{\mathrm{F}}$}\nzrremoveone{$b$} to yield
\begin{equation} \label{fuller}
    b_{\mathrm{F}}^2 = \frac{1}{2} \pm \frac{1}{2}\sqrt{1 - 4a_F^2\ell(\ell+1)}
\end{equation}

\noindent where it can be seen that there are no real solutions for $b_{\mathrm{F}}$ for some critical $a_{\mathrm{F}}>a^\ell_c$ given by
\begin{equation}
    a^\ell_c = \frac{1}{2\sqrt{\ell(\ell+1)}}
\end{equation}

This is equivalent to the result originally presented by \citet{fuller2015asteroseismology} that there are no propagating solutions when the magnetogravity frequency (defined in Equation \ref{omegamg}) rises above the mode frequency.
At small buoyancy or Alfv\'en frequencies, the two solutions for \nzraddone{$b_{\mathrm{F}}^2$}\nzrremoveone{$b^2$} in Equation \ref{fuller} approach the usual internal gravity wave and Alfv\'en wave dispersion relations, and remain finite.
Equations \nzrremoveone{\ref{dipolegeo}}\nzraddone{\ref{monopole}}, \ref{fullerab}, and \ref{fullerdisp} are analogous to Equations \nzrremoveone{\ref{monopole}}\nzraddone{\ref{dipolegeo}}, \ref{b} and \ref{a}, and \ref{dispersionrelation} in the main text.

\subsection{\citet{lecoanet2016conversion}: Multipole Cartesian geometry} \label{lecoanet}

\citet{lecoanet2016conversion} consider a Cartesian geometry with the equilibrium magnetic field configuration
\begin{equation} \label{cart}
    \Vec{B}_0 = B_0e^{-k_Bz}\left[ \sin(k_Bx)\,\hat{x} + \cos(k_Bx)\,\hat{z} \right]
\end{equation}

\noindent where the oscillatory dependence in $x$ is chosen to closely mimic the $\theta$ dependence of a multipole magnetic field.

Define $k_P$ to the wavenumber which defines the periodicity of the domain, i.e., the solution is periodic in $x$ with a period $2\pi/k_P$ ($k_P$ is analogous to $1/r$ in the spherical problem).
We can first define
\begin{subequations} \label{lecoanetab}
    \begin{gather}
        b_{\mathrm{L}} = \frac{k_zv_A}{\omega} \\
        a_{\mathrm{L}} = \left(\frac{N}{\omega}\right)\left(\frac{k_Pv_A}{\omega}\right)
    \end{gather}
\end{subequations}
where $v_A=B_0e^{-k_Bz}/\sqrt{4\pi\rho_0}$.

Then, following a very similar procedure to the spherical dipole problem described in the main text, the three-dimensional Cartesian problem corresponding to the field in Equation \ref{cart} can be reduced to
\begin{equation} \label{lecoanetmte}
    \mathcal{L}^{m,b_{\mathrm{L}}}p'(\mu) + \frac{b_{\mathrm{L}}^2}{a_{\mathrm{L}}^2}p'(\mu) = 0
\end{equation}

\noindent where $\mu=\cos(k_Px)$.

In the special case that $k_B=k_P$ (i.e., a dipole field), the differential operator $\mathcal{L}^{m,b_{\mathrm{L}}}_{\mathrm{L}}$ is given by
\begin{equation} \label{lecoanetmto}
    \mathcal{L}^{m,b_{\mathrm{L}}}_{\mathrm{L}}p'(\mu) = \sqrt{1-\mu^2}\frac{\mathrm{d}}{\mathrm{d}\mu}\left(\frac{\sqrt{1-\mu^2}}{1 - b_{\mathrm{L}}^2\mu^2}\frac{\mathrm{d}p'(\mu)}{\mathrm{d}\mu}\right) - \frac{m^2}{1 - b_{\mathrm{L}}^2\mu^2}p'(\mu)
\end{equation}

\noindent where $m=k_y/k_P$.
Letting $\lambda_{\mathrm{L}}$ be a given (negative) eigenvalue of $\mathcal{L}^{m,b_{\mathrm{L}}}_{\mathrm{L}}$, the dispersion relation takes the form
\begin{equation} \label{lecoanetdisp}
    \lambda_{\mathrm{L}} = b_{\mathrm{L}}^2/a_{\mathrm{L}}^2
\end{equation}

\citet{lecoanet2016conversion} solve the problem described above in the two-dimensional zonal case (i.e., $m=0$), taking advantage of the fact that two-dimensional incompressibility defines a ``vector potential'' whose direction is everywhere orthogonal to the fluid motions.
They find the eigenfunctions to be Mathieu functions \citep{mathieu1868memoire}, with a branch of inward-traveling internal gravity waves \nzrremoveone{scattering}\nzraddone{refracting} up into a branch of slow \nzraddone{magnetic }
waves which diverge to infinite wavenumber at a finite \nzraddone{``cutoff''} radius.
We reproduce this behavior in the dipole geometry for the $m=0$ modes (see Section \ref{mequalszero}).

Equations \ref{cart}, \ref{lecoanetab}, \ref{lecoanetmte}, \ref{lecoanetmto}, and \ref{lecoanetdisp} are analogous to Equations \nzrremoveone{\ref{monopole}}\nzraddone{\ref{dipolegeo}}, \ref{b} and \ref{a}, \ref{mte}, \ref{mto}, and \ref{dispersionrelation}, in the main text.
We see that, aside from geometrical factors $\propto\sqrt{1-\mu^2}$ (which become irrelevant in the large $b_{\mathrm{L}}$ limit), $\mathcal{L}^{m,b}_{\mathrm{mag}}$ and $\mathcal{L}^{m,b_{\mathrm{L}}}_{\mathrm{L}}$ are identical.

\section{Numerically solving $\mathcal{L}^{\lowercase{m,b}}_{\mathrm{\lowercase{mag}}}$} \label{num}

We use a relaxation scheme to numerically solve Equations \ref{firstorder} and \ref{firstorder2}, following closely the procedure used by \citet{lee1997low} and \citet{fuller2014dynamical} to diagonalize the Laplace tidal equation.
\noindent using the \texttt{C++} implementation given by \textit{Numerical Recipes} \citep{press2007numerical}.

However, the procedures which we adopt vary somewhat for the dissipative and dissipationless cases.
The numerical solution procedure for Sections \ref{mequalszero} and \ref{mnotequalszero} is summarized in Section \ref{numa}, and the procedure for Section \ref{viscous} is summarized in Section \ref{numb}.

\subsection{Numerical solution without dissipation} \label{numa}

In the $m=0$ (Section \ref{mequalszero}) and $m\neq0,|b|\leq1$ (Section \ref{mnotequalszero}) cases without dissipation, \nzraddone{little} special care is \nzrremoveone{not }required to solve the requisite two first-order differential equations.
In the former case, the solutions (Hough functions) are known to be second-differentiable across the singularity.
In the latter case, the problem is a standard Sturm--Liouville problem with no internal singularities.
Moreover, because $\mathcal{L}_{\mathrm{mag}}^{m,b}$ (Equation \ref{mto}) is even with respect to $\mu$, its eigenfunctions can be partitioned into even and odd parity, which is given by \nzraddone{$(-1)^{\ell+m}$}\nzrremoveone{$(-1)^{k+m}$} for \nzraddone{$p'$}\nzrremoveone{$\mathcal{P}$} and $(-1)^{\ell+m+1}$ for \nzraddone{$\xi_\theta$}\nzrremoveone{$\mathcal{Z}_\theta$}.
This known parity \nzraddone{greatly} simplifies the problem, allowing us to solve for the eigenfunctions for only $\mu\in[0,+1)$ rather than over the full domain.

Then, defining $\mathcal{P}=p'/\rho_0\omega^2r^2$ and $\mathcal{Z}_\theta=\sqrt{1-\mu^2}\xi_\theta/r$ as (assumed real) non-dimensionalized versions of $\mathcal{P}$ and $\mathcal{Z}_\theta$, our equations become
\begin{subequations} \label{firstorder}
    \begin{gather}
        \frac{\mathrm{d}\mathcal{P}}{\mathrm{d}\mu} = -\frac{1-b^2\mu^2}{1-\mu^2}\mathcal{Z}_\theta \\
        \frac{\mathrm{d}\mathcal{Z}_\theta}{\mathrm{d}\mu} = \left(\frac{b^2}{a^2} - \frac{m^2}{\left(1-\mu^2\right)\left(1-b^2\mu^2\right)}\right)\mathcal{P}
    \end{gather}
\end{subequations}

\noindent where the first equation follows from the $\theta$ component of the momentum equation (Equation \ref{thetamtmdipole}) and the second equation follows from the continuity equation (Equation \ref{continuitydipole}).

Starting from Legendre polynomials as initial guess, we gradually increase $b$ ($>0$) from $0$, retaining lower $b$ solutions as initial guesses for higher $b$ relaxations.
Note that, in the numerical implementation, we promote $a=a(\mu)$ to a function of $\mu$, and additionally enforce $\mathrm{d}a/\mathrm{d}\mu=0$ \citep[this is the standard \nzrremoveone{implementation}\nzraddone{technique for solving such eigenproblems} in, e.g.,][]{press2007numerical}.
We impose boundary conditions on $\mathcal{P}$ and $\mathcal{Z}_\theta$ at $\mu=0$ depending on parity, setting one of these variables to $0$ and the other to $1$ for normalization.
Additionally, at the right boundary $\mu=1-\epsilon$, we enforce that $\mathcal{Z}_\theta$ must vanish (which can be seen in its definition).

In the $m\neq0,|b|<1$ case, we demonstrate in Section \ref{mnotequalszero} that eigenfunctions in the formal zero-dissipation limit are exactly localized to the band between the internal singularities at $\mu=\pm1/b$ (where we have taken $b>0$ without loss of generality).
Then we can rescale Equations \ref{firstorder} via $x=b\mu$ to
\begin{subequations} \label{firstorder2}
    \begin{gather}
        \frac{\mathrm{d}\mathcal{P}}{\mathrm{d}x} = -\frac{1-x^2}{b^2-x^2}b^3\mathcal{Z}_\theta \\
        \frac{\mathrm{d}\mathcal{Z}_\theta}{\mathrm{d}x} = \left(\frac{1}{a^2} - \frac{m^2}{\left(b^2-x^2\right)\left(1-x^2\right)}\right)b^3\mathcal{P}
    \end{gather}
\end{subequations}

\noindent over the range $x\in[0,+1)$ (i.e., $\mu\in[0,+1/b)$), enforcing $\mathcal{P}=0$ at $x=1-\epsilon$ (see Section \ref{mnotequalszero}).

\subsection{Numerical solution with dissipation} \label{numb}

When dissipation is considered, the internal singularities are ``softened'' in the sense that they are shifted off of the real line.
Therefore, a case-wise treatment of the singularity (as in Section \ref{numa}) is not necessary.
However, in general, both the perturbations and at least one of the quantities $k_r$ and $\omega$ are complex, doubling the number of equations to be solved.

Moreover, in the dissipative case (even for arbitrarily small viscosities/diffusivities), the delta function which appears in $\xi_\phi$ becomes softened to a sharp peak with a finite width.
Therefore, instead of solving for $\mathcal{P}\sim(1-b^2\mu^2)\xi_\phi$ as a perturbation, we probe this peak by solving the complex versions of the following equations,
\begin{subequations} \label{viscousfirstorder}
    \begin{gather}
        \frac{\mathrm{d}\mathcal{Z}_\phi}{\mathrm{d}\mu} = \frac{2b^2\mu}{1-b^2\mu^2-ic}\mathcal{Z}_\phi - \frac{1}{1-\mu^2}\mathcal{Z}_\theta \\
        \frac{\mathrm{d}\mathcal{Z}_\theta}{\mathrm{d}\mu} = \left(\lambda\left(1 - b^2\mu^2 - ic\right) - \frac{m^2}{1-\mu^2}\right)\mathcal{Z}_\phi
    \end{gather}
\end{subequations}

\noindent where $\mathcal{Z}_\phi=\sqrt{1-\mu^2}\xi_\phi/imr$.
In Section \ref{numerical1}, we pick $c=10^{-3}$ and take $\lambda=|\lambda|e^{i\zeta_1}$ and $b=|b|e^{i\zeta_1/2}$ and solve for $|\lambda|$ and $\zeta_1$, while varying $|b|$.
In Section \ref{numerical2}, we pick $c=10^{-2}$ and take $\lambda=|\lambda|e^{i\zeta_2}$ and $b=|b|$ (real) and solve for $|\lambda|$ and $\zeta_2$, while again varying $|b|$.
As in Section \ref{numa}, the first equation follows from the $\theta$ component of the momentum equation, and the second equation follows from the continuity equation (but in terms of different perturbations).

%In Equations \ref{viscousfirstorder}, we have assumed that $a$ is real but $b$ is complex---this corresponds to assuming that $\omega$ is real but $k_r$ is complex.
%This choice (compared to taking $\omega$ to be complex) is found to be much more numerically stable.
%As with $a$, we promote the phase $\zeta$ to a function $\zeta(\mu)$, and enforce a ``flatness'' condition $\mathrm{d}\zeta/\mathrm{d}\mu=0$.
The evenness and oddness conditions can be applied as in Section \ref{numa} to $\mathrm{Re}\left(\mathcal{Z}_\phi\right)$ (which has the same parity as $\mathcal{P}$) and $\mathrm{Re}\left(\mathcal{Z}_\theta\right)$ at the left boundary $\mu=0$.
At this same boundary, we enforce (due to overall phase invariance) $\mathrm{Im}\left(\mathcal{Z}_\phi\right)=\mathrm{Im}\left(\mathcal{Z}_\theta\right)=0$.
Finally, at $\mu=1-\epsilon$, we enforce $\mathrm{Re}\left(\mathcal{Z}_\theta\right)=\mathrm{Im}\left(\mathcal{Z}_\theta\right)=0$.
Equations \ref{viscousfirstorder} are then solved for increasingly large values of $|b|$, using associated Legendre polynomials as the initial $|b|=0$ guesses.

\section{Tightly confined equatorial magnetogravity waves} \label{mgextra}

In this Appendix, we investigate the behavior of magnetogravity waves very close to the equator (for a dipole field), where $v_{Ar}\approx0$.
\nzraddone{In the following, we demonstrate in this narrow equatorial band the existence of self-consistent, propagating solutions with wavenumbers enhanced in magnitude by a factor $N/\omega$.
These solutions are not captured by the analysis in the main text, which assumed that the vertical component of the Alfv\'en velocity dominates the mode structure.
However, the dynamics of these modes may play an important role in understanding the observable, asteroseismic consequences of strong core magnetic fields.}

In this region, the assumption (used throughout this work) that the Alfv\'en frequency $\omega_A=\Vec{k}\cdot\Vec{v}_A$ is dominated by the radial component is violated.
We expect this violation to be important when
\begin{equation}
    v_{Ar}/v_{Ah}\lesssim k_h/k_r\sim\omega/N
\end{equation}

\noindent or in a narrow band around the equator with angular extent $\delta\theta\sim\omega/N$.

Assuming a WKB approximation in all directions, the dispersion relation for magnetogravity waves near the equator becomes
\begin{equation}
    \omega^2 - \frac{k_h^2}{k_r^2}N^2 - k_h^2v_{Ah}^2 = 0 \, ,
\end{equation}

\noindent where now $\omega_A\approx k_hv_{Ah}$ is dominated by the horizontal component.
Then, solving for $k_r$, we have
\begin{equation} \label{horizontaldisp}
    k_r^2 = \frac{k_h^2N^2}{\omega^2-k_h^2v_{Ah}^2} \, ,
\end{equation}

\noindent where the criterion for radial propagation is
\begin{equation} \label{crit1}
    \omega^2 > k_h^2v_{Ah}^2 \, .
\end{equation}

Because these solutions are only accurate in an equatorial band $\delta\theta\lesssim\omega/N$, it follows that\nzrremoveone{,}
\begin{equation} \label{crit2}
    k_h \gtrsim \frac{N}{\omega r}\,\mathrm{,}
\end{equation}

\noindent i.e., at least one horizontal wavelength fits within this band.
The criteria in Equation \ref{crit1} and \ref{crit2} can be combined to obtain
\begin{equation}
    \omega \gtrsim \sqrt{Nv_{Ah}/r} \sim \omega_B
\end{equation}
\noindent where $\omega_B$ is the critical magnetic field strength from Equation \ref{omegab}, but now applied to the horizontal field rather than the radial field. 
Therefore, such confined magnetogravity waves would remain radially propagating in roughly the same regions that magnetogravity waves throughout the rest of the star would (within a small, order-unity factor in radius).

Interestingly, when Equation \ref{crit1} (for the minimum $k_h$) is combined with Equation \ref{horizontaldisp} for propagating waves, we obtain
\begin{equation}
    |k_r| \gtrsim \frac{N}{\sqrt{\omega^2/k_{h,\mathrm{min}}^2-v_{Ah}^2}} \sim \left(\frac{N}{\omega}\right)^2\frac{1}{r}
\end{equation}

\noindent where we have examined the limit where $v_{Ah}\ll r\omega^2/N$ (i.e., at fields much lower than the critical field, or where $\omega\gg\omega_B$).
This radial wavenumber is larger than the radial wavenumber of normal, low-$\ell$ gravity waves by a large factor $\sim N/\omega$, implying that magnetogravity waves near the equator will develop very small radial wavelengths.
This \nzrremoveone{behavior}\nzraddone{increased radial wavenumber} appears to be qualitatively consistent with numerical simulations conducted by \citet{lecoanet2016conversion}, which seem to exhibit such waves at locations where the radial magnetic field vanishes (see their Figure 6).
\nzraddone{However, their simulation also seems to show outgoing equatorially confined evanescent waves, whose driving we cannot explain.
Moreover, due to their small spatial scale, it is unclear to us whether these modes are numerically resolved.
Future work will be required to further elucidate the nature of these highly confined modes.}

Note that the dispersion relation in Equation \ref{horizontaldisp} implies that the group and phase velocities of these confined magnetogravity waves are in opposite directions (similar to normal, zero-field gravity waves).
Because the refracted, outgoing magnetogravity wave solutions at $b>1$ (described in the rest of this work) have aligned group and phase velocities, this implies that such outgoing magnetogravity waves couple most efficiently to \textit{ingoing}, equatorially confined magnetogravity waves (described above).
This poses a challenge for equatorially confined magnetogravity waves as a vehicle for bringing wave power out of the core.
Moreover, the very short wavelengths of the equatorially confined waves make them much more susceptible to damping processes.
Further work may elucidate the nature of these waves, and their role in wave power transport and dissipation.

\section{Structure of Alfv\'en resonances including horizontal-field contributions to the magnetic tension} \label{nonWKB}

In the solutions throughout \nzrremoveone{this}\nzraddone{the main} text, it has been assumed that the magnetic tension terms which appear in the momentum equations are dominated by the radial component (in, e.g., Equations \ref{subeq}).
However, very near to $\omega^2=k_r^2v_{Ar}^2$ (i.e., very near to a critical latitude), the horizontal components of the magnetic tension may become relevant.
As described throughout the text (e.g., Section \ref{wkb}), sharp horizontal fluid features may appear in the vicinity of Alfv\'en resonances.
To estimate the impact of such terms, we will make a WKB approximation in both the radial \textit{and} horizontal directions (i.e., $k_r,k_h\gg1/r$).
We further hypothesize that it still remains true that $k_h\ll k_r$ (this will set a condition \nzrremoveone{of}\nzraddone{for} validity).
For Alfv\'en \nzrremoveone{resonances}\nzraddone{waves}, we examine the horizontal momentum equation:

% To determine the behavior of the solution near the critical latitude, we consider the WKB dispersion relation for Alfv\'en waves (Section \ref{wkb}), which can be written as
\begin{equation} \label{mtmmtm}
    \left( k_r^2v_{Ar}^2 + \frac{2 i k_r v_{Ar} v_{A\theta}}{r} \frac{\mathrm{d}}{\mathrm{d}\theta} - \frac{v_{A\theta}^2}{r^2} \frac{\mathrm{d}^2}{\mathrm{d}\theta^2 } \right)  \Vec{\xi}_h =  \omega^2 \Vec{\xi}_h  \, ,
\end{equation}

\noindent where we have assumed a poloidal field ($B_\phi=0$) for simplicity.
Because we are interested in the solution in the vicinity of $\omega^2-k_r^2v_{Ar}^2\approx0$, we can perform a Taylor expansion:
\begin{equation}
    \omega^2-k_r^2v_{Ar}^2 \approx -k_r^2\frac{\mathrm{d}v_{Ar}^2}{\mathrm{d}\theta}\delta\theta
\end{equation}

\noindent where $\delta\theta=\theta-\theta_c$, and $\theta_c$ is the critical latitude.
Then Equation \ref{mtmmtm} becomes
\begin{equation}
    \frac{2 i k_r v_{Ar} v_{A\theta}}{r}\frac{\mathrm{d}\Vec{\xi}_h}{\mathrm{d}\theta} = -k_r^2\frac{\mathrm{d}v_{Ar}^2}{\mathrm{d}\theta}\delta\theta\,\Vec{\xi}_h = -2k_r^2v_{Ar}^2\frac{\mathrm{d}\ln v_{Ar}}{\mathrm{d}\theta}\delta\theta\,\Vec{\xi}_h
\end{equation}

\noindent where $k_h\gg k_r$ allowed us to drop the term $\propto\mathrm{d}^2\Vec{\xi}_h/\mathrm{d}\theta^2$.
This becomes
\begin{equation}
    \frac{\mathrm{d}\Vec{\xi}_h}{\mathrm{d}\theta} = ik_rr\frac{v_{Ar}}{v_{A\theta}}\frac{\mathrm{d}\ln v_{Ar}}{\mathrm{d}\theta}\delta\theta\,\Vec{\xi}_h
\end{equation}

Then, since $v_{Ar}\sim v_{A\theta}$ and $\mathrm{d}\ln v_{Ar}/\mathrm{d}\theta\simeq1$ (since $v_{Ar}$ varies horizontally roughly on the order of a radian), the prefactors involving $v_{Ar}$ and $v_{Ah}$ are order-unity.
Doing this more carefully for a dipole field (where $v_{Ar}\propto2\cos\theta$ and $v_{Ah}\propto\sin\theta$)\nzrremoveone{, this becomes} \nzraddone{yields}
\begin{equation} \label{alf}
    \frac{\mathrm{d}\Vec{\xi}_h}{\mathrm{d}\theta} = -2ik_rr\delta\theta\,\Vec{\xi}_h
\end{equation}

Equation \ref{alf} is straightforwardly solved by
\begin{equation}
    \Vec{\xi}_h \simeq \Vec{\xi}_{h,0}e^{-ik_rr\delta\theta^2}
\end{equation}
This complex Gaussian describes a wave whose wavelength decreases away from the critical latitude $\theta_c$. The first wavelength occurs where $r k_r \delta \theta^2 = 2 \pi$, yielding a characteristic angular scale of $\delta \theta \sim 1/\sqrt{k_rr}$, or characteristic angular wavenumber of $r k_\theta \sim \sqrt{k_rr}$ (compare Equation \ref{wkblimit}).
Note that this horizontal wavenumber satisfies our assumption that $1/r \ll k_\theta \ll k_r$.

\section{Magnetic tangling-induced shear stress} \label{tangling}

While realistic stars do not have shear-restorative forces, \citet{hoven2011magnetar} argue that small-scale disordered magnetic fields (``tangling'') may introduce an effective shear stress, with characteristic wave speed $c_s^2=B_{\mathrm{rms}}^2/4\pi\rho_0$.
To characterize the effect that this shear modulus term has on the Alfv\'en \nzrremoveone{resonances}\nzraddone{waves}, we can focus on the magnetic tension term in the horizontal momentum equation, which is the only other term capable of restoring torsional mode components \citep{loi2017torsional}:
\begin{equation} \label{schrod}
    \omega^2\vec{\xi}_h = -c_s^2r^{-2}{\bm \nabla}_h^2\vec{\xi}_h + \omega_{Ar}^2\vec{\xi}_h
\end{equation}

We see that Equation \ref{schrod} is mathematically identical to a two-dimensional Schr\"odinger equation, where $E=\omega^2$ plays the role of the total energy, $V=\omega_{Ar}^2$ plays the role of the potential, and the small shear speed $c_s$ plays the role of $\hbar$.
As noted by \citet{hoven2011magnetar}, the effect of a small shear modulus is to transform the continuous spectrum of Alfv\'en \nzrremoveone{resonances}\nzraddone{waves} into a discrete one (as in standard bound-state spectra of the Schr\"odinger equation), whose mode spacings decrease to zero in the limit where $c_s\rightarrow0$.
If a WKB approximation is adopted for Equation \ref{schrod} in the horizontal direction, the coupling of these discrete \nzrremoveone{resonances}\nzraddone{waves} to a mode is similar to that of the continuum \nzrremoveone{resonances}\nzraddone{waves} in the $c_s=0$ case.
In the region where $\omega^2\gg\omega_{Ar}^2$, the discrete Alfv\'en \nzrremoveone{resonance}\nzraddone{wave} oscillates spatially very rapidly, and its overlap with a global-scale $g$ mode averages to zero.
Similarly, in the region where $\omega^2\ll\omega_{Ar}^2$, the \nzrremoveone{resonance}\nzraddone{wave} decays very rapidly, and therefore is very close to zero.
However, the solution very close to $\omega^2=\omega_{Ar}^2$ (the ``classical turning point'') is known to be an Airy function of angular width
\begin{equation}
    \delta\theta = \sqrt[3]{\frac{c_s^2}{r^2}\left\lvert\frac{\partial\omega_A^2}{\partial\theta}\right\rvert^{-1}}
\end{equation}

\noindent which sets the scale at which an interaction with a mode and Alfv\'en \nzrremoveone{resonance}\nzraddone{wave} will be ``smeared'' in the angular direction, due to shear stress.
Physically, the shear modulus-induced discretization of the Alfv\'en \nzrremoveone{resonances}\nzraddone{waves} occurs because shear adds an isotropic contribution to the wave speed (so that it is not exactly zero perpendicular to the field lines), and thus couples fluid motions across field lines.
\nzrremoveone{We note, however, that t}\nzraddone{T}his effect is likely to be small relative to similar effects associated \nzrremoveone{to}\nzraddone{with} the horizontal component of the mean magnetic field (Section \ref{wkb}).

% Don't change these lines
\bsp	% typesetting comment
\label{lastpage}
\end{document}